\documentclass[a4paper,11pt]{article}
\pdfoutput=1 
\usepackage{jheppub}
\usepackage{amsmath}
\usepackage{amsfonts}
\usepackage{amssymb}
\usepackage{slashed}
\usepackage{bm}
\usepackage{graphicx}%
\usepackage[T1]{fontenc} 
\usepackage{mathrsfs}
\usepackage{yfonts}

\DeclareMathAlphabet{\mathpzc}{OT1}{pzc}{m}{it}

\newcommand{\beq}{\begin{equation}} 
\newcommand{\eeq}{\end{equation}} 
\newcommand{\bega}{\begin{eqnarray}} 
\newcommand{\ega}{\end{eqnarray}}

\newcommand{\ie}{i\epsilon}

\newcommand{\dhd}{{\textstyle d}
\lower.03ex\hbox{\kern-0.38em$^{\scriptstyle-}$}\kern-0.05em{}}
\newcommand{\dbar}{{\textstyle \delta}
\lower.03ex\hbox{\kern-0.38em$^{\scriptstyle-}$}\kern-0.05em{}}
\newcommand{\half}{{1\over 2}}
\newcommand{\bu}{{\bullet}}

\newcommand{\bare}{{\bar e}}
\newcommand{\barf}{{\bar f}}
\newcommand{\barg}{{\bar g}}
\newcommand{\barh}{{\bar h}}

\newcommand{\barq}{{\bar q}}

\newcommand{\bsi}{{\bar \psi}}

\newcommand{\bhi}{{\bar \chi}}

\newcommand{\barB}{{\bar B}}

\newcommand{\Bsi}{{\bar \Psi}}
\newcommand{\Bxi}{{\bar \Xi}}

\newcommand{\Bigma}{{\bar \Sigma}}
\newcommand{\Byps}{{\bar\Upsilon}}

\newcommand{\cald}{{\cal D}}

\newcommand{\calo}{{\cal O}}

\newcommand{\calW}{{\cal W}} 
\newcommand{\calw}{{\cal W}}

\newcommand{\hatp}{{\hat p}}

\newcommand{\hatA}{{\hat A}}

\newcommand{\tile}{{\tilde e}}

\newcommand{\tilq}{{\tilde q}}

\newcommand{\tilA}{{\tilde A}}
\newcommand{\tilB}{{\tilde B}}

\newcommand{\tilF}{{\tilde F}}

\newcommand{\tigma}{\tilde {\sigma}}

\newcommand{\bref}{\breve {f}}

\newcommand{\ace}{\acute {e}} 
\newcommand{\acf}{\acute {f}} 
 
\newcommand{\acA}{\acute {A}} 
\newcommand{\acB}{\acute {B}}

\newcommand{\graf}{\grave {f}} 
 
\newcommand{\grA}{\grave {A}} 
\newcommand{\graB}{\grave {B}}

\newcommand{\chU}{\check {U}} 
\newcommand{\cheV}{\check {V}} 
\newcommand{\cheW}{\check {W}}

\newcommand{\pizb}{\mathpzc{B}}

\newcommand{\pizg}{\mathpzc{G}}

\newcommand{\pizp}{\mathpzc{P}}

\newcommand{\notp}{{\not\! p}}

\newcommand{\notA}{{\not\! \!A}}
\newcommand{\notB}{{\not\! \!B}}

\newcommand{\slp}{{\slashed{p}}}

\newcommand{\slA}{{\slashed{A}}}
\newcommand{\slB}{{\slashed{B}}}

\newcommand{\slP}{{\slashed{P}}}

\newcommand{\slpart}{{\slashed{\partial}}}

\pdfoutput=1

\abstract{
I calculate ${1\over Q^2}$  power corrections to unpolarized Drell-Yan hadronic tensor for electromagnetic (EM) current   at large $N_c$ and demonstrate 
the EM gauge invariance at this level.}
\arxivnumber{}
\affiliation{ Physics Department, Old Dominion University, Norfolk, VA 23529, USA and Thomas Jefferson National Accelerator Facility, Newport News, VA 23606, USA}

\emailAdd{balitsky@jlab.org}
\begin{document}

\title{\boldmath $1/Q^2$ power corrections to TMD factorization for Drell-Yan hadronic tensor}
\author{I. Balitsky }
\preprint{CERN-TH-2024-051,~JLAB-THY-24-4014}
\maketitle

\flushbottom

\section{Introduction\label{aba:sec1}}
Particle production  in hadron-hadron scattering with transverse momentum of produced particle(s)  much smaller 
than the invariant mass is described in the framework of TMD factorization \cite{Collins:2011zzd, Collins:1981uw, Collins:1984kg, Ji:2004wu, GarciaEchevarria:2011rb}.
The typical factorization formula for particle production in hadron-hadron scattering  looks like \cite{Collins:2011zzd, Collins:2014jpa}
\begin{eqnarray}
&&\hspace{-2mm}
{d\sigma\over  d\eta d^2q_\perp}~=~\sum_f\!\int\! d^2b_\perp e^{i(q,b)_\perp}
\cald_{f/A}(x_A,b_\perp,\eta)\cald_{f/B}(x_B,b_\perp,\eta)\sigma(ff\rightarrow X)
\nonumber\\
&&\hspace{-2mm}
+~{\rm power ~corrections}~+~{\rm Y-terms}
\label{TMDf}
\end{eqnarray}
where $\eta$ is the rapidity, $\cald_{f/A}(x,z_\perp,\eta)$ is the TMD density of  a parton $f$  in hadron $A$, and $\sigma(ff\rightarrow X)$ is the cross section of production of particle(s) $X$ 
of invariant mass $m_X^2=Q^2$ by the fusion of two partons.

Typically, leading first term in Eq. (\ref{TMDf}) is given by quark-antiquark TMDs 
(or two-gluon TMDs in the case of Higgs boson production). The second term stands for the power corrections given 
by a series in $q_\perp^2/Q^2$ while the third describes transition to the regime $q_\perp^2\sim Q^2$ governed by the
collinear factorization.

The significance  of power corrections is twofold. First, they show up to what $q_\perp^2$
 the differential cross section is given by the first term 
in the formula (\ref{TMDf}) with controlled accuracy. For example, the estimate  for
$Z$-boson production in DY process gives power corrections reaching  order of few per cent at 
${q_\perp\over Q}\sim {1\over 4}$  \cite{Balitsky:2017gis}. 

The second use of power corrections is due to the fact that for certain characteristics 
of a scattering the power corrections are actually the
leading terms.  It turns out that some  angular distributions of produced particle(s) are defined by 
quark-quark-gluon TMDs forming power corrections $\sim{q_\perp^2\over Q^2}$. 
For example,
the symmetric DY hadronic tensor $W_{\mu\nu}$, defined as 
\footnote{Here $p_A,p_B$ are hadron momenta, $q$ is the momentum of DY pair, $\sum_X$ denotes the sum over full set of ``out''  states and 
$J_\mu=\sum e_f\bsi^f\gamma_\mu\psi_f$ is an electromagnetic current. We take into account only $u,d,s,c$ quarks and consider them massless. In principle, one can include ``massless'' $b$-quark for $q_\perp^2\gg m_b^2$}
\begin{eqnarray}
\hspace{-1mm}
W_{\mu\nu}(q)~&\stackrel{\rm def}{=}&~{1\over (2\pi)^4}\sum_X\!\int\! d^4x~e^{-iqx}
\half\Big(\langle p_A,p_B|J_\mu(x)|X\rangle\langle X|J_\nu(0) |p_A,p_B\rangle+\mu\leftrightarrow\nu\Big)
\nonumber\\
~&=&~{1\over (2\pi)^4}\!\int\! d^4x~e^{-iqx}
\half\langle p_A,p_B|J_\mu(x)J_\nu(0) +\mu\leftrightarrow\nu|p_A,p_B\rangle  
\label{W},
\end{eqnarray}
has  4 tensor structures for unpolarized hadrons.  Two of them are determined by 
leading-twist quark TMDs while two other ones  start from terms ${q_\perp\over Q}$ and $\sim{q_\perp^2\over Q^2}$ described
by quark-quark-gluon TMDs. Note that while ${q_\perp\over Q}$ power corrections were known for more than two decades  \cite{Mulders:1995dh}, there
was no calculations of ${q_\perp^2\over Q^2}$ until recently, starting from the paper \cite{Balitsky:2017gis}.

In two previous papers \cite{Balitsky:2020jzt, Balitsky:2021fer} I calculated such ${q_\perp^2\over Q^2}$ power corrections and found DY angular distributions   at small Bjorken $x_B$ in the leading order 
in ${1\over N_c}$. In this paper I generalize the results of Ref. \cite{Balitsky:2020jzt}
to arbitrary values of $x_B$.  As a result,  the number of relevant TMDs increases: for unpolarized protons, in addition to eight quark-antiquark TMDs, there are about twenty quark-antiquark-gluon TMDs on ${1\over Q^2}$, leading-$N_c$ level.

The paper is organized as follows.  
In section  \ref{sec:rapfak}   I outline the derivation of  TMD factorization by
 rapidity factorization  of  the double functional integral for a cross section of particle production. 
Also, I  briefly remind the method of
calculation of power corrections based on approximate solution of classical Yang-Mills equations \cite{Balitsky:2017gis}.
In Sect. \ref{sec:lhtc} I present the leading-twist result and discuss the types of ${1\over Q^2}$ power corrections calculated in this paper. In the next three Sections I calculate different types of ${1\over Q^2}$ power corrections using the aforementioned
method. The result is presented In Sect. \ref{sec:result} and discussed in Sect.  \ref{sec:coutlook}. The necessary technical formulas and parametrizations of relevant TMDs can be found in appendices.

\section{TMD factorization from rapidity factorization \label{sec:rapfak}}

We use 
Sudakov variables $p=\alpha p_1+\beta p_2+p_\perp$, where $p_1$ and $p_2$ are light-like vectors close to $p_A$ and $p_B$ so that 
$p_A=p_1+{m^2\over s}p_2$ and $p_B=p_2+{m^2\over s}p_1$ with $m$ being the proton mass.
Also, we use the notations $x_\bu\equiv x_\mu p_1^\mu$ and $x_\star\equiv x_\mu p_2^\mu$ 
for the dimensionless light-cone ``Ioffe times'' $x_\star=\sqrt{s\over 2}x_+$ and $x_\bu=\sqrt{s\over 2}x_-$. Our metric is $g^{\mu\nu}~=~(1,-1,-1,-1)$ 
which we will frequently rewrite as a sum of longitudinal part and transverse part: 
\begin{equation}
g^{\mu\nu}~=~g_\parallel^{\mu\nu}+g^{\mu\nu}_\perp~=~{2\over s}\big(p_1^\mu p_2^\nu+p_2^\mu p_1^\nu)+g_\perp^{\mu\nu}
\label{delta}
\end{equation}
Consequently,  $p\cdot q~=~(\alpha_p\beta_q+\alpha_q\beta_p){s\over 2}-(p,q)_\perp$ where $(p,q)_\perp\equiv -p_iq^i$. 
Throughout the paper, the sum over the Latin indices $i$, $j$, ... runs over two transverse components while the sum over Greek indices $\mu$, $\nu$, ... runs over four components as usual.

\begin{figure}[htb]
\begin{center}

\hspace{-5mm}
\vspace{-4mm}
\includegraphics[width=155mm]{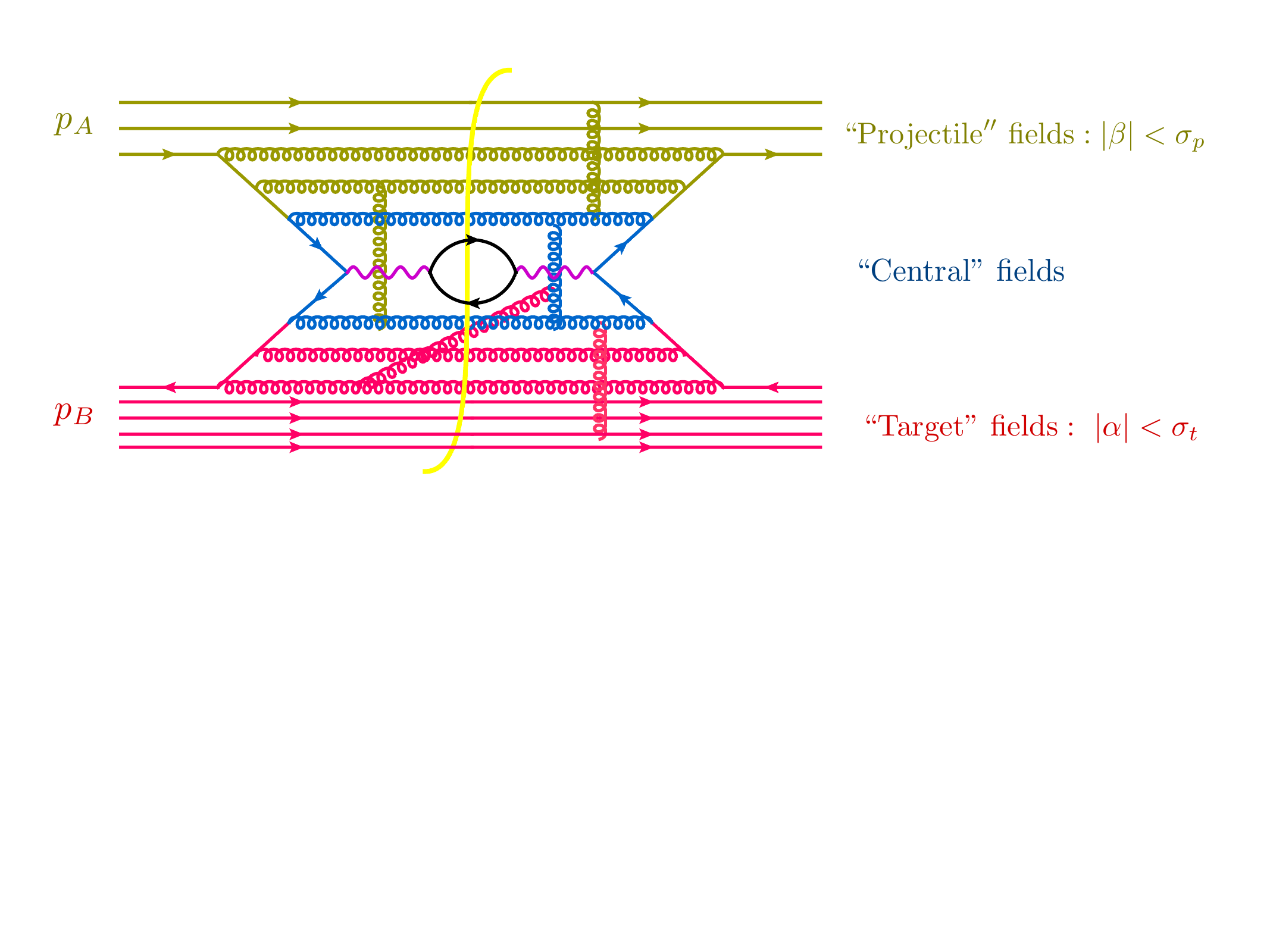}
\end{center}

\vspace{-64mm}
\caption{Rapidity factorization for DY particle production \label{fig:2}}
\end{figure}

Following Ref. \cite{Balitsky:2017flc}  we separate quark and gluon fields into three sectors (see figure \ref{fig:2}): 
``projectile'' fields $A_\mu, \psi_A$ 
with $|\beta|<\sigma_p$, 
``target'' fields $B_\mu, \psi_B$ with $|\alpha|<\sigma_t$ and ``central rapidity'' fields $C_\mu,\psi_C$ with $|\alpha|>\sigma_t$ and $|\beta|>\sigma_p$, 
see Fig. \ref{fig:2}. ( For convenience, I call hadron  $A$  by the name ``projectile''  and  hadron $B$ 
by the name ``target'').
Our goal is to integrate over central fields and get
the amplitude in the factorized form, i.e. as a product of functional integrals over $A$ fields representing projectile matrix elements (TMDs of the projectile) 
and functional integrals over $B$ fields representing target matrix elements (TMDs of the target).
In the spirit of background-field method, we ``freeze'' projectile and target fields and get a sum of diagrams in these external fields.  As we shall see below, for the purpose of calculation of most of the power corrections
we can set  $\beta=0$ for the  projectile fields and $\alpha=0$ for the target fields.  The corrections to this approximation
 are $O\big({m^2\over\sigma_p s}\big)$ and  $O\big({m^2\over\sigma_t s}\big)$ and can be neglected almost everywhere,  see the discussion in Sect. \ref{sec:qqpc}. 

In the coordinate space, the $\beta=0$ approximation means that projectile fields do not depend on $x_\ast$ and 
$\alpha=0$ means that target ones do not depend on $x_\bu$. 
\footnote{Beyond the tree level, the integration over $C$ fields produces
 logarithms of the cutoffs $\sigma_p$ and $\sigma_t$ which match  the corresponding
logs in TMDs of the projectile and the target, see the discussion in Ref. \cite{Balitsky:2023hmh} }
In this case, 
as discussed in Ref. \cite{Balitsky:2020jzt},  central fields at the tree level are given by a set of Feynman diagrams with {\it retarded} propagators 
in background field $A + B$ and $\psi_A+\psi_B$.The set of such ``retarded'' diagrams represent the solution of QCD equations of motion with sources being projectile and target fields.
After summation of these diagrams the hadronic tensor (\ref{W}) can  be represented as
\begin{equation}
\hspace{-1mm}  
W_{\mu\nu}~=~\frac{1}{(2\pi)^4}\!\int \! d^4x  e^{-iqx}
\sum_{m,n}\! \int\! dz_m c_{m,n}(q,x)
\langle p_A|\hat\Phi_A(z_m)|p_A\rangle\!\int\! dz'_n\langle p_B| \hat\Phi_B(z'_n)|p_B\rangle.   
\label{W5}
\end{equation}
where $c_{m,n}$ are coefficients and $\Phi$ can be any of  the background fields promoted to operators after integration over
projectile and target fields.

In general, solution of classical QCD equations with projectile and target sources  is a formidable task which still awaits its solution. Fortunately, as demonstrated in Ref. \cite{Balitsky:2020jzt},
at our kinematics we have a small parameter ${q_\perp^2\over Q^2}\ll 1$ and it is possible to expand classical solution for central fields in powers of this parameter.
It is convenient to choose a gauge where $A_\star=0$ for projectile fields and $B_\bu=0$ for target fields.
\footnote{Throughout the paper, we will keep different notations $A_i$ and  $B_i$ for the projectile and target gluon fields because 
of different gauge choices, see e.g. Eqs.
(\ref{ai}) and (\ref{bi}).
}
(The existence of such gauge was proved in appendix B of Ref. \cite{Balitsky:2017flc} by explicit construction). 
Also, since we are dealing with tree approximation and quark equations of motion, it is convenient to include coupling constant $g$
in the definition of gluon fields so that $D_\mu\psi=\partial_\mu\psi-iA_\mu\psi$,
 $\bsi\stackrel{\leftarrow}{D}_\mu=\partial_\mu\bsi+i\bsi A_\mu$ and 
$F_{\mu\nu}=\partial_\mu A_\nu-\partial_\nu A_\mu-i[A_\mu,A_\nu]$. \label{oboz}

As demonstrated in Ref. \cite{Balitsky:2017gis}, the
expansion of classical quark fields 
in powers of $p^2_\perp/p_\parallel^2$ has the form
\footnote{The corresponding expansion of classical gluon fields is presented in  
Ref. \cite{Balitsky:2017flc}, but we do not need it here.
}
\begin{eqnarray}
&&\hspace{-1mm}
\Psi(x)~=~\Psi_1(x)+\Psi_2(x)
+\dots,
\label{klfildz}
\end{eqnarray}
where ($P_\perp\equiv\partial_\perp+A_\perp+B_\perp$ )

\begin{eqnarray}
&&\hspace{-5mm}
\Psi_1~=~\psi_A+\Upsilon_1,~~~\Upsilon_1~=~\Xi_{1}+\Xi'_{1},~~~~
\Xi_{1}~=~-{\slashed{p}_2\over s}\slB_\perp{1\over \alpha+i\epsilon}\psi_A,
\nonumber\\
&&\hspace{5mm}
\Xi'_{1}~=~-{\slashed{p}_1\over s}\Big({1\over\beta+\ie}\slB_\perp\Big)\psi_A
+{1\over s^2}\Big({\slp_1\over\beta+\ie}\slP_\perp{\slp_2\over\alpha+\ie}+{\slp_2\over\alpha+\ie}\slP_\perp{\slp_1\over\beta+\ie}\Big)\slB_\perp\psi_A
\nonumber\\
&&\hspace{-5mm}
\Bsi_1~=~\bar\psi_A+\Byps_{1},~~~~\Byps_{1}~=~\Bigma_{1}+\Bigma'_{1},~~~~
\Bigma_{1}~=~-\big(\bar\psi_A{1\over\alpha-i\epsilon}\big)\slB_\perp{\slashed{p}_2\over s}
\nonumber\\
&&\hspace{5mm}
\Bxi'_{1}~=~-\bsi_A\Big({\slB\over\beta-\ie}\Big){\slp_1\over s}
+{1\over s^2}\bsi_A\slB\Big({\slp_1\over\beta-\ie}\slP_\perp{\slp_2\over\alpha-\ie}
+{\slp_2\over\alpha-\ie}\slP_\perp{\slp_1\over\beta-\ie}\Big)
\nonumber\\
&&\hspace{-5mm}
\Psi_2~=~\psi_B+\Upsilon_2,~~~\Upsilon_2~=~\Xi_{2}+\Xi'_{2},~~~~
\Xi_{2}~=~-{\slashed{p}_1\over s}\slA_\perp{1\over\beta+i\epsilon}\psi_B
\nonumber\\
&&\hspace{5mm}
\Xi'_{2}~=~-{\slashed{p}_2\over s}\Big({1\over\alpha+\ie}\slA\Big)\psi_B
+{1\over s^2}\Big({\slp_1\over\beta+\ie}\slP_\perp{\slp_2\over\alpha+\ie}
+{\slp_2\over\alpha+\ie}\slP_\perp{\slp_1\over\beta+\ie}\Big)\slA_\perp\psi_B
\nonumber\\
&&\hspace{-5mm}
\Bsi_2~=~\bar\psi_A+\Byps_{1},~~~~\Byps_{2}~=~\Bigma_{2}+\Bigma'_{2},~~~~
\Bxi_{2}~=~-\big(\bar\psi_B{1\over \beta-i\epsilon}\big)\slA_\perp{\slashed{p}_1\over s}
\nonumber\\
&&\hspace{5mm}
\Bxi'_{2}~=~-\bsi_B\Big({\slA\over \alpha-\ie)}\Big){\slp_2\over s}
+{1\over s^2}\bsi_B\slA\Big({\slp_1\over\beta-\ie}\slP_\perp{\slp_2\over\alpha-\ie}
+{\slp_2\over\alpha-\ie}\slP_\perp{\slp_1\over\beta-\ie}\Big)
\label{fildz0}
\end{eqnarray}
and the dots stand for higher-order power corrections.
\footnote{The relevant expressions for  $\Xi'_i,\Bxi'_i$ from Ref. \cite{Balitsky:2017gis} are more complicated than those of Eq. (\ref{fildz0}) but the additional terms 
are shown in Sect. \ref{sec:primed} to be negligible.}
In the above formulas 
\begin{eqnarray}
&&\hspace{-1mm}
{1\over \alpha+i\epsilon}\psi_A(x_\bu,x_\perp)~\equiv~-i\!\int_{-\infty}^{x_\bu}\! dx'_\bu~\psi_A(x'_\bu,x_\perp),
\nonumber\\
&&\hspace{-1mm}
\Big(\bsi_A{1\over \alpha-i\epsilon}\Big)(x_\bu,x_\perp)~\equiv~i\!\int_{-\infty}^{x_\bu}\! dx'_\bu~\bsi_A(x'_\bu,x_\perp)
\label{3.25}
\end{eqnarray}
and similarly for ${1\over\beta\pm i\epsilon}$. 
For brevity,  in what follows we denote $\big(\bar\psi_A{1\over\alpha}\big)(x)\equiv \big(\bsi_A{1\over \alpha-i\epsilon}\big)(x)$ and
$\big(\bar\psi_B{1\over\beta}\big)(x)\equiv \big(\bsi_B{1\over \beta-i\epsilon}\big)(x)$. 
Similarly to Eq. (\ref{3.25}), more complicated  expressions for $\Bsi$ should be read from right to left, for example
\begin{eqnarray}
&&\hspace{-5mm}
\bsi_A\slB\Big({\slp_1\over\beta}\slP_\perp{\slp_2\over\alpha}+{\slp_2\over\alpha}\slP_\perp{\slp_1\over\beta}\Big)(x)~
=~\int\! dz~\bsi_A(z)\slB(z)(z|{\slp_1\over\beta}\slP_\perp{\slp_2\over\alpha}+{\slp_2\over\alpha}\slP_\perp{\slp_1\over\beta}|x)
\label{forexa}
\end{eqnarray}
with $\alpha-\ie$ and $\beta-\ie$ in the denominators. Here $(x|f(p)|y)\equiv (2\pi)^{-d}\int\! d^dp e^{-ip(x-y)}f(p)$ are Schwinger's notations for propagators.

The contributions from the terms $\Xi_i,\Bxi_i$ were calculated in Ref. \cite{Balitsky:2020jzt} in the 
$\alpha_q,\beta_q\ll 1$ approximation and in this paper we will repeat the calculation relaxing the aforementioned  condition. 
The  contributions from the terms $\Xi'_i,\Bxi'_i$ are new and will be calculated in Sect. \ref{sec:primed}.

\section{Hadronic tensor at  $Q^2\gg q_\perp^2$: leading twist and power corrections\label{sec:lhtc}}

As we noted above, we take into account only hadronic tensor due to electromagnetic currents of $u,d,s,c$ quarks and consider these quarks to be massless. 
It is convenient to define coordinate-space hadronic tensor multiplied by $N_c{2\over s}$ (and denoted by extra ``check'' mark) as follows
\begin{eqnarray}
&&\hspace{-1mm}
\cheW_{\mu\nu}(x)~\equiv~
N_c{1\over s}\langle A,B|J_\mu(x)J_\nu(0)+\mu\leftrightarrow\nu|A,B\rangle
\label{defW}
\end{eqnarray}
so that
\begin{eqnarray}
&&\hspace{-1mm}
W_{\mu\nu}(q)~=~{s/2\over(2\pi)^4N_c}\int\!d^4x ~e^{-iqx} \cheW_{\mu\nu}(x).
\nonumber\\
&&\hspace{26mm}
=~{1\over(2\pi)^4N_c}\!\int\! dx_\bu dx_\star d^2x_\perp ~e^{-i\alpha_qx_\bu-i\beta_q x_\star+i(q,x)_\perp}
\cheW_{\mu\nu}(x).
\label{furie}
\end{eqnarray}
Hereafter we use notation $|A,B\rangle\equiv|p_A,p_B\rangle$ for brevity.

For future use, let us also define the hadronic tensor in mixed representation: in the momentum longitudinal 
space but in the transverse coordinate space
\begin{eqnarray}
&&\hspace{-1mm}
W_{\mu\nu}(q)~=~\int\!d^2x_\perp ~e^{i(q,x)_\perp} W_{\mu\nu}(\alpha_q,\beta_q, x_\perp),
\label{defWcoord}\\
&&\hspace{-1mm}
W_{\mu\nu}(\alpha_q,\beta_q,x_\perp)~\equiv~
{1\over(2\pi)^4s}\!\int\! dx_\bu dx_\star ~e^{-i\alpha_qx_\bu-i\beta_q x_\star}
\langle A,B|J_\mu(x_\bu,x_\star,x_\perp)J_\nu(0)+\mu\leftrightarrow\nu|A,B\rangle.
\nonumber
\end{eqnarray}

After integration over central fields in the tree approximation we obtain
\begin{equation}
\hspace{-1mm}
\cheW_{\mu\nu}(x)~\equiv~
N_c{1\over s}\langle A,B|J_\mu(x_\bu,x_\star,x_\perp)J_\nu(0)+\mu\leftrightarrow\nu|A,B\rangle
\label{4.3}
\end{equation}
where
\begin{eqnarray}
&&\hspace{-1mm}
J^\mu~=~J^\mu_A+J^\mu_B+J^\mu_{AB}+J^\mu_{BA},
\nonumber\\
&&\hspace{-1mm}
J^\mu_A~=~\sum_f e_f\bar\Psi_1^f\gamma^\mu\Psi_1^f,~~~
J^\mu_{AB}~=~\sum_f e_f\bar\Psi_1^f\gamma^\mu\Psi_2^f
\label{jeiz}
\end{eqnarray}
and similarly for $J^\mu_B$ and $J^\mu_{BA}$.  Here $\langle A,B|\calo(\psi_A,A_\mu,\psi_B,B_\mu)|A,B\rangle$ denotes 
double functional integral over $A$ and $B$ fields  which gives matrix elements between projectile and target states of Eq. (\ref{W5}) type.

The leading-twist contribution to $W_{\mu\nu}(q)$  comes only from product $J_{AB}^\mu(x)J_{BA}^\nu(0)$ (or $J_{BA}^\mu(x)J_{AB}^\nu(0)$), 
while power corrections may come also from other terms like $J_A^\mu(x)J_{B}^\nu(0)$. 
However, as demonstrated in Refs. \cite{Balitsky:2017gis,Balitsky:2020jzt}, at leading-$N_c$ power corrections 
come only from $J_{AB}^\mu(x)J_{BA}^\nu(0)$ or $J_{BA}^\mu(x)J_{AB}^\nu(0)$. Since these contributions are 
diagonal in flavor, we will perform the calculations for one flavor of quarks (with $J_\mu=\bsi\gamma_\mu\psi$) and will write down sum over flavors only in the final result 
(\ref{finalresult}).

\subsection{Leading-$N_c$ terms from $J_{AB}^\mu(x)J_{BA}^\nu(0)$}

With our ${1\over Q^2}$, leading-$N_c$ accuracy we get from Eq.  (\ref{fildz0}):
\begin{eqnarray}
&&\hspace{-1mm}
J_{AB}^\mu(x)J_{BA}^\nu(0)
~+~x\leftrightarrow 0~=~\Bsi_1(x)\gamma^\mu\Psi_2(x)\Bsi_2(0)\gamma^\nu\Psi_1(0)~+~x\leftrightarrow 0
~+~...
\nonumber\\
&&\hspace{-1mm}
=~\big[\big(\bar\psi_A +\Byps_{1}\big)(x)\gamma_\mu\big(\psi_B+\Upsilon_{2}\big)(x)\big]
[\big(\bar\psi_B+\Byps_{2}\big)(0)\gamma_\nu\big(\psi_A+\Upsilon_{1}\big)(0)\big]~+~x\leftrightarrow 0
\nonumber\\
&&\hspace{-1mm}
=~[\bar\psi_A(x)\gamma_\mu\psi_B(x)\big]\big[\bar\psi_B(0)\gamma_\nu\psi_A(0)\big]
\label{7lines}\\
&&\hspace{-1mm}
+~[(\Bxi_{1}+\Bxi'_1)(x)\gamma_\mu\psi_B(x)\big]\big[\bar\psi_B(0)\gamma_\nu\psi_A(0)\big]
+[\bar\psi_A(x)\gamma_\mu\psi_B(x)\big]\big[\bar\psi_B(0)\gamma_\nu(\Xi_{1}+\Xi'_1)(0)\big]
\nonumber\\
&&\hspace{-1mm}
+~
[\bar\psi_A(x)\gamma_\mu\psi_B(x)\big]\big[(\Bxi_{2}+\Bxi'_2)(0)\gamma_\nu\psi_A(0)\big]
+[\bar\psi_A(x)\gamma_\mu(\Xi_{2}+\Xi'_2)(x)\big]\big[\bar\psi_B(0)\gamma_\nu\psi_A(0)\big]
\nonumber\\
&&\hspace{-1mm}
+~[\Bxi_{1}(x)\gamma_\mu\psi_B(x)\big]\big[\bar\psi_B(0)\gamma_\nu\Xi_{1}(0)\big]
+[\bar\psi_A(x)\gamma_\mu\Xi_{2}(x)\big]\big[\Bxi_{2}(0)\gamma_\nu\psi_A(0)\big]
\nonumber\\
&&\hspace{-1mm}
+~[\Bxi_{1}(x)\gamma_\mu\psi_B(x)\big]\big[\Bxi_{2}(0)\gamma_\nu\psi_A(0)\big]
+[\bar\psi_A(x)\gamma_\mu\Xi_{2}(x)\big]\big[\bar\psi_B(0)\gamma_\nu\Xi_{1}(0)\big]
\nonumber\\
&&\hspace{-1mm}
+~[\Bxi_{1}(x)\gamma_\mu\Xi_{2}(x)\big]\big[\bar\psi_B(0)\gamma_\nu\psi_A(0)\big]
+[\bar\psi_A(x)\gamma_\mu\psi_{B}(x)\big]\big[\Bxi_{2}(0)\gamma_\nu\Xi_{1}(0)\big]
~+~x\leftrightarrow 0.
\nonumber
\end{eqnarray}
where the square brackets mean trace over Lorentz and color indices. 
\footnote{As demonstrated in Sect. \ref{sec:twoperators}, the terms coming from expressions with one $\Xi_i,\Bxi_i$ and one $\Xi'_i,\Bxi'_i$ are negligible in our approximation)
}

First, let us consider the leading-twist term and power corrections coming from the first term in the r.h.s. of this equation.

\subsection{Contribution of quark-antiquark TMDs }

\subsubsection{Leading-twist contribution}
As we mentioned, the leading-twist term comes from from the first term in the r.h.s. of Eq. (\ref{7lines}). Using Fierz transformation (\ref{fierz}) one obtains the quark-antiquark
contribution in the form
\begin{eqnarray}
&&\hspace{-1mm}
\cheW^{qq}_{\mu\nu}(x)~=~{N_c\over s}\big(\big[\bar\psi_A(x_\bu,x_\perp)\gamma_\mu\psi_B(x_\star,x_\perp)\big]\big[\bar\psi_B(0)\gamma_\nu\psi_A(0)\big]+\mu\leftrightarrow\nu\big)
~+~x\leftrightarrow 0
\nonumber\\
&&\hspace{-1mm}
=~{g_{\mu\nu}\over 2s}\big(-[\bsi_A\psi_A][\bsi_B\psi_B)+[\bsi_A\gamma_5\psi_A][\bsi_B\gamma_5\psi_B]
+[\bsi_A\gamma_\alpha\psi_A][\bsi_B\gamma^\alpha\psi_B]
\nonumber\\
&&\hspace{-1mm}
+~[\bsi_A\gamma_\alpha\gamma_5\psi_A][\bsi_B\gamma^\alpha\gamma_5\psi_B]
-\half[\bsi_A\sigma^{\alpha\beta}\psi_A][\bsi_B\sigma_{\alpha\beta}\psi_B]\big)
\nonumber\\
&&\hspace{-1mm}
-~{1\over 2s}\big([\bsi_A\gamma_\mu\psi_A][\bsi_B\gamma_\nu\psi_B]+\mu\leftrightarrow\nu\big)
-{1\over 2s}\big([\bsi_A\gamma_\mu\gamma_5\psi_A][\bsi_B\gamma_\nu\gamma_5\psi_B]+\mu\leftrightarrow\nu\big)
\nonumber\\
&&\hspace{-1mm}
+~{1\over 2s}\big([\bsi_A\sigma_{\nu\alpha}\psi_A][\bsi_B\sigma_{\mu\alpha}\psi_B]
+[\bsi_A\sigma_{\mu\alpha}\psi_A][\bsi_B\sigma_{\nu\alpha}\psi_B]\big)~+~x\leftrightarrow 0
\label{ltfierz}
\end{eqnarray}
where the arguments of the fields are the same as in the l.h.s..
From the parametrization of two-quark operators in section \ref{sec:paramlt},
it is clear that the leading-twist contribution to $W_{\mu\nu}(q)$  comes from 
\begin{eqnarray}
&&\hspace{-1mm}
{1\over 2s}(g_{\mu\nu}g^{\alpha\beta}-\delta_\mu^\alpha\delta_\nu^\beta-\delta_\nu^\alpha\delta_\mu^\beta)
[\bsi_A(x)\gamma_\alpha\psi_A(0)]\bsi_B(0)\gamma_\beta\psi_B(x)]
\label{cheklt}\\
&&\hspace{-1mm}
+~{1\over 2s}\big(\delta_\mu^\alpha\delta_\nu^\beta+\delta_\nu^\alpha\delta_\mu^\beta-\half g_{\mu\nu}g^{\alpha\beta}\big)
[\bsi_A(x)\sigma_{\alpha\xi}\psi_A(0)][\bsi_B(0)\sigma_\beta^{~\xi}\psi_B(x)]
\nonumber\\
&&\hspace{-2mm}
+~{1\over 2s}(g_{\mu\nu}g^{\alpha\beta}-\delta_\mu^\alpha\delta_\nu^\beta-\delta_\nu^\alpha\delta_\mu^\beta)
[\bsi_A(x)\gamma_\alpha\gamma_5\psi_A(0)][\bsi_B(0)\gamma_\beta\gamma_5\psi_B(x)]
\nonumber\\
&&\hspace{-2mm}
-~{g_{\mu\nu}\over 2s}[\bsi_A(x)\psi_A(0)][\bsi_B(0)\psi_B(x)]
~+~x\leftrightarrow 0
\nonumber
\end{eqnarray}
With the leading-twist accuracy we can replace $\delta_\mu^\alpha\rightarrow{2\over s}p_{1\mu}p_2^\alpha$, 
$\delta_\nu^\beta\rightarrow{2\over s}p_{2\nu}p_1^\beta$, $g^{\alpha\beta}\rightarrow {2\over s}p_2^\alpha p_1^\beta$, and get
\begin{eqnarray}
&&\hspace{-1mm}
\cheW^{\rm l}_{\mu\nu}~=~
{1\over s^2}g^\perp_{\mu\nu}\big(
[\bsi_A(x)\slp_2\psi_A(0)][\bsi_B(0)\slp_1\psi_B(x)]
+[\bsi_A(x)\slp_2\gamma_5\psi_A(0)][\bsi_B(0)\slp_1\gamma_5\psi_B(x)]\big)
\nonumber\\
&&\hspace{-1mm}
+~{1\over s^2}\big(g^\perp_{\mu\xi}g^\perp_{\nu\eta}+g^\perp_{\nu\xi}g^\perp_{\mu\eta}-g^\perp_{\mu\nu}g^\perp_{\xi\eta}\big)
[\bsi_A(x)\sigma_{\star\xi}\psi_A(0)][\bsi_B(0)\sigma_\bu^{~\xi}\psi_B(x)]
~+~x\leftrightarrow 0
\label{cheklt2}
\end{eqnarray}
where $g^\parallel_{\mu\nu}\equiv {2\over s}\big(p_{1\mu}p_{2\nu}+\mu\leftrightarrow\nu\big)$ and
$g^\perp_{\mu\nu}\equiv g_{\mu\nu}-g^\parallel_{\mu\nu}$.

As mentioned above, the dependence of $\psi_A$ on $x_\star$ and $\psi_B$ on $x_\bu$ is very slow so
at the leading-twist order we can replace  $\psi_A(x)\rightarrow\psi_A(x_\bu,x_\perp)$ and 
 $\psi_B(x)\rightarrow\psi_B(x_\bu,x_\perp)$ (the corrections will be considered in next Section). 
 
 Next, after integration over background fields $A$ and $B$ we promote $A$, $\psi_A$ and $B$, $\psi_B$ to operators $\hatA$, $\hat\psi$.
A subtle point is that our operators are not under T-product ordering so one should be careful while changing the order of operators in
 formulas like Fierz transformation. Fortunately, all operators in the r.h.s of Eq. (\ref{cheklt2}) and in
 similar formulas for power corrections are separated either by space-like intervals or light-like intervals so they commute with each other.
We get
\footnote{
In a general gauge for
projectile and target fields these matrix elements read 
\begin{eqnarray}
&&\hspace{-2mm}
\langle p_A|\bsi(x)\gamma_\mu \psi(0)|p_A\rangle
~=~\langle p_A|\psi(x)\gamma_\mu[x,x-\infty p_2][x-\infty p_2,-\infty p_2][-\infty p_2,0]\psi(0)|p_A\rangle,
\nonumber\\
&&\hspace{-2mm}
\langle p_B|\psi(x)\gamma_\mu \psi(0)|p_B\rangle
~=~\langle p_B|\psi(x)\gamma_\mu[x,x-\infty p_1][x-\infty p_1,-\infty p_1][-\infty p_1,0]\psi(0)|p_B\rangle
\label{gaugelinks}
\end{eqnarray}
where $[x,y]\equiv {\rm Pexp}\Big\{i\!\int_0^1\! du ~x^\mu A_\mu((ux+(1-u)y)\Big\}$,
and similarly for other matrix elements.}
\begin{eqnarray}
&&\hspace{-7mm}
\cheW^{\rm lt}_{\mu\nu}~=~
{1\over s^2}g^\perp_{\mu\nu}
\langle\bsi(x_\bu,x_\perp)\slp_2\psi(0)\rangle_A\langle\bsi(0)\slp_1\psi(x_\star,x_\perp)\rangle_B
+~(\slp_2\otimes\slp_1\leftrightarrow\slp_2\gamma_5\otimes\slp_1\gamma_5)
\label{cheklt3}\\
&&\hspace{-7mm}
+~{1\over s^2}\big(g^\perp_{\mu\xi}g^\perp_{\nu\eta}+g^\perp_{\nu\xi}g^\perp_{\mu\eta}-g^\perp_{\mu\nu}g^\perp_{\xi\eta}\big)
\langle\bsi(x_\bu,x_\perp)\sigma_\star^{~\xi}\psi(0)\rangle_A\langle\bsi(0)\sigma_\bu^{~\eta}\psi(x_\star,x_\perp)\rangle_B~+~x\leftrightarrow 0
\nonumber
\end{eqnarray}

Hereafter we use notations $\langle\calo\rangle_A\equiv\langle p_A|\calo|p_A\rangle$ and 
$\langle\calo\rangle_A\equiv\langle p_B|\calo|p_B\rangle$ for brevity.
The corresponding leading-twist contribution to  to $W_{\mu\nu}(q)$  
has the form \cite{Tangerman:1994eh}
\begin{eqnarray}
&&\hspace{-1mm}W_{\mu\nu}^{\rm lt}(\alpha_q,\beta_q,q_\perp)
~=~{1\over 16\pi^4N_c}\!\int\! dx_\bu dx_\star d^2x_\perp~e^{-i\alpha_qx_\bu-i\beta_qx_\star+i(q,x)_\perp}\cheW^{\rm lt}_{\mu\nu}(x)
\nonumber\\
&&\hspace{-1mm}
~=~{1\over N_c}\!\int\! d^2k_\perp\Big(-g_{\mu\nu}^\perp\big[f_1(\alpha_q,k_\perp)\barf_1(\beta_q,q_\perp-k_\perp) 
+\barf_1(\alpha_q,k_\perp)f_1(\beta_q,q_\perp-k_\perp)\big]
\nonumber\\
&&\hspace{17mm}
-~{1\over m^2}\big[k^\perp_\mu(q-k)^\perp_\nu+k^\perp_\nu(q-k)^\perp_\mu+g_{\mu\nu}^\perp(k,q-k)_\perp\big]
\nonumber\\
&&\hspace{22mm}
\times~\big[h^\perp_1(\alpha_q,k_\perp)\barh^\perp_1(\beta_q,q_\perp-k_\perp)
+\barh^\perp_1(\alpha_q,k_\perp)h^\perp_1(\beta_q,q_\perp-k_\perp)\big]\Big)
\label{WLT}
\end{eqnarray}
To compare with general formula (\ref{TMDf}) we need to identify $\alpha_q\equiv x_A$ and $\beta_q\equiv x_B$. 
To avoid confusion with coordinates, throughout the paper we will keep notations $\alpha_q$ and $\beta_q$.

\subsubsection{Types of ${1\over Q^2}$ power corrections}
Let us outline power corrections calculated in this paper. As I mentioned in the Introduction, 
only leading-$N_c$ power corrections up to ${1\over Q^2}$ will be taken into account. Specifically, this means
the PCs 
\begin{eqnarray}
&&\hspace{-1mm}
N_cW_{\mu\nu}(q)~\sim~g_\perp^{\mu\nu}\Big[1+{q_\perp^2\over Q^2}\Big],~~ {q_\perp^\mu q_\perp^\nu\over m_\perp^2}
\Big[1+{m_\perp^2\over Q^2}\Big],~~g_\parallel^{\mu\nu}\Big[0+{m_\perp^2\over Q^2}\Big],~~
\label{pc}\\
&&\hspace{-1mm}
{1\over Q^2}\big(p_2^\mu q_\perp^\nu+\mu\leftrightarrow\nu\big),~~
{1\over Q^2}\big(p_2^\mu q_\perp^\nu+\mu\leftrightarrow\nu\big),~~
{1\over Q^2}\big(p_1^\mu q_\perp^\nu+\mu\leftrightarrow\nu\big),~~
{p_1^\mu p_1^\nu \over Q^4}, ~~
{p_2^\mu p_2^\nu\over Q^4}
\nonumber
\end{eqnarray}
Among those, corrections $\sim {p_{1,2}^\mu q_\perp^\nu}$ are of order ${1\over Q}$ while the rest is $\sim {1\over Q^2}$.
Here $m_\perp^2\sim q_\perp^2,m^2$. When counting powers of ${1\over Q}$ we do not distinguish between $q_\perp^2, k_\perp^2$ and $m^2$ but in concrete formulas we keep them different so we can consider, for example, case $q_\perp^2\gg m^2$.
Similarly,  parametrically we do not distinguish between $s$ and $Q^2=\alpha_q\beta_qs-q_\perp^2$ but  keep track
in our formulas so they are correct both at $s\sim Q^2$ and
$s\gg Q^2$.

Let us also specify the terms which are not calculated in this paper.  Roughly speaking, they correspond to terms in Eq. (\ref{pc}) multiplied by extra power(s) of
${m_\perp\over Q}$ or by extra ${1\over N_c}$. Our strategy in the next sections is to compare a certain  term in 
$\cheW_{\mu\nu}$ to terms in Eq. (\ref{pc}), and, if it is smaller, neglect, and if it is of the same size, calculate. 

\subsubsection{Power corrections due to quark-antiquark TMDs \label{sec:qqpc}}
As one can see from parametrization in Sect. \ref{sec:paramlt}, the r.h.s. of Eq. (\ref{ltfierz}) contains not only the leading-twist 
contributions (\ref{WLT}), but also a number of power corrections. 

 We start from the corrections obtained by expansions
\begin{eqnarray}
&&\hspace{-2mm}
\psi(x)~=~\psi(x_\perp,x_\bu,0)+x_\star {2\over s}D_\bu\psi(x_\perp,x_\bu,0)~+...,
\nonumber\\
&&\hspace{-2mm}
\bsi(x)~=~\bsi(x_\perp,x_\bu,0)+x_\star {2\over s}\bsi\stackrel{\leftarrow} D_\bu(x_\perp,x_\bu,0)~+...  
\end{eqnarray}
for projectile matrix elements and
\begin{eqnarray}
&&\hspace{-2mm}
\psi(x)~=~\psi(x_\perp,0,x_\star)+x_\bu {2\over s}D_\star\psi(x_\perp,0,x_\star)~+...
\nonumber\\
&&\hspace{-2mm}
\bsi(x)~=~\bsi(x_\perp,0,x_\star)+x_\star {2\over s}\bsi\stackrel{\leftarrow} D_\bu(x_\perp,0,x_\star)~+...  
\label{xpanshens}
\end{eqnarray}
for the target ones. 

Let us show that second terms in these expansions are $\sim {1\over Q^2}$. To this end,
note that $ D_\bu\sim \beta_{\rm proj}s\leq \sigma_p s$. As discussed in Ref. \cite{Balitsky:2023hmh}, the natural scales for rapidity 
factorization outlined in Sect. \ref{sec:rapfak} are  $\sigma_p\sim{q_\perp^2\over \alpha_q s}$ 
and $\sigma_t\sim{q_\perp^2\over \beta_q s}$.  Adding   estimates $x_\bu\sim {1\over\alpha_q}$, 
 $x_\star\sim {1\over\beta_q}$
we get
\begin{equation}
x_\star {2\over s}D_\bu~\sim~ {1\over\alpha_q}\sigma_ps~\sim~{q_\perp^2\over\alpha_q\beta_q s}~\sim~{q_\perp^2\over Q^2}
\label{estimproj}
\end{equation}
 for the projectile and 
\begin{equation}
x_\bu {2\over s}D_\star~\sim~ {1\over\beta_q}\sigma_ts~\sim~{q_\perp^2\over\alpha_q\beta_q s}~\sim~{q_\perp^2\over Q^2}
\label{estimtar}
\end{equation}
for the target matrix elements. This means that no further terms in expansions (\ref{estimproj}), (\ref{estimtar})
are necessary and moreover, the only place where we need these corrections is the leading contributions (\ref{cheklt2}).
\footnote{The author is indebted to A. Vladimirov for clarifying this point.}

Before expansions (\ref{xpanshens}) it is convenient to use translational invariance and make a shift 
in Eq. (\ref{cheklt2})
\begin{eqnarray}
&&\hspace{-1mm}
\cheW^{\rm l}_{\mu\nu}~=~
{1\over s^2}g^\perp_{\mu\nu}
\langle\bsi\big(x_\bu,x_\perp,{x_\ast\over 2}\big)\slp_2\psi\big(-{x_\ast\over 2}\big)\rangle_A
\langle\bsi\big(-{x_\bu\over 2}\big)\slp_1\psi\big(x_\star,x_\perp,{x_\bu\over 2}\big)\rangle_B
\nonumber\\
&&\hspace{-1mm}
+~(\slp_2\otimes\slp_1\leftrightarrow\slp_2\gamma_5\otimes\slp_1\gamma_5)
+~{1\over s^2}\big(g^\perp_{\mu\xi}g^\perp_{\nu\eta}+g^\perp_{\nu\xi}g^\perp_{\mu\eta}-g^\perp_{\mu\nu}g^\perp_{\xi\eta}\big)
\nonumber\\
&&\hspace{-1mm}
\times~
\langle\bsi\big(x_\bu,x_\perp,{x_\ast\over 2}\big)\sigma_\star^{~\xi}\psi\big(-{x_\ast\over 2}\big)\rangle_A\langle\bsi\big(-{x_\bu\over 2}\big)\sigma_\bu^{~\eta}\psi\big(x_\star,x_\perp,{x_\bu\over 2}\big)\rangle_B
\nonumber\\
&&\hspace{-1mm}
+~\Big(\{x_\bu,x_\perp,{x_\ast\over 2}\}\leftrightarrow \{-{x_\ast\over 2}\}, \{x_\ast,x_\perp,{x_\bu\over 2}\}\leftrightarrow \{-{x_\bu\over 2}\}\Big) 
\nonumber\\
&&\hspace{-1mm}
=~\cheW_{\mu\nu}^{\rm l.t.}-~{g^\perp_{\mu\nu}\over 2s^2}\Big[x_\star
\langle\bsi(x_\bu,x_\perp)\slp_2\stackrel{\leftrightarrow}D_\bu\psi(0)\rangle_A
\langle\bsi(0)\slp_1\psi(x_\star,x_\perp)\rangle_B
\nonumber\\
&&\hspace{-1mm}
+~x_\bu
\langle\bsi(x_\bu,x_\perp)\slp_2\psi(0)\rangle_A
\langle\bsi(0)\slp_1\stackrel{\leftrightarrow}D_\star\psi(x_\star,x_\perp)\rangle_B
+(\slp_2\otimes\slp_1\leftrightarrow\slp_2\gamma_5\otimes\slp_1\gamma_5)\Big]
\nonumber\\
&&\hspace{-1mm}
-~{1\over 2s^2}\big(g^\perp_{\mu\xi}g^\perp_{\nu\eta}+g^\perp_{\nu\xi}g^\perp_{\mu\eta}
-g^\perp_{\mu\nu}g^\perp_{\xi\eta}\big)\Big[x_\star
\langle\bsi(x_\bu,x_\perp)\sigma_\star^{~\xi}\stackrel{\leftrightarrow}D_\bu\psi(0)\rangle_A
\langle\bsi(0)\sigma_\bu^{~\eta}\psi(x_\star,x_\perp)\rangle_B
\nonumber\\
&&\hspace{-1mm}
+~x_\bu
\langle\bsi(x_\bu,x_\perp)\sigma_\star^{~\xi}\psi(0)\rangle_A
\langle\bsi(0)\sigma_\bu^{~\eta}\stackrel{\leftrightarrow}D_\star\psi(x_\star,x_\perp)\rangle_B
\Big]~=~\cheW_{\mu\nu}^{\rm l.t.}+\cheW_{\mu\nu}^D
\end{eqnarray}
Here 
\beq
\bsi(x_\bu,x_\perp)\slp_2\stackrel{\leftrightarrow}D_\bu\psi(0)
\equiv\bsi(x_\bu,x_\perp)\slp_2D_\bu\psi(0)\rangle_A-\bsi\stackrel{\leftarrow}D_\bu(x_\bu,x_\perp)\slp_2\psi(0),
\eeq
and similarly for other terms. Using parametrizations (\ref{Dmaelp}) and (\ref{Dmaelt}), one easily obtains
\begin{eqnarray}
&&\hspace{-1mm}W_{\mu\nu}^D(\alpha_q,\beta_q,q_\perp)
~=~{1\over 16\pi^4N_c}\!\int\! dx_\bu dx_\star d^2x_\perp~e^{-i\alpha_qx_\bu-i\beta_qx_\star+i(q,x)_\perp}\cheW^{\rm lt}_{\mu\nu}(x)
\label{Wder}\\
&&\hspace{-1mm}
~=~{2\over sN_c}\!\int\! d^2k_\perp\Big\{m^2g_{\mu\nu}^\perp\big[
\Re f_D(\alpha_q,k_\perp)\barf'_1(\beta_q,q_\perp-k_\perp) 
+\Re \barf_D(\alpha_q,k_\perp)f'_1(\beta_q,q_\perp-k_\perp) 
\nonumber\\
&&\hspace{33mm}
+~f'_1(\alpha_q,k_\perp) \Re \barf_D(\beta_q,q_\perp-k_\perp)
+\barf'_1(\alpha_q,k_\perp) \Re f_D(\beta_q,q_\perp-k_\perp)
\big]
\nonumber\\
&&\hspace{17mm}
+~\big[k^\perp_\mu(q-k)^\perp_\nu+k^\perp_\nu(q-k)^\perp_\mu+g_{\mu\nu}^\perp(k,q-k)_\perp\big]
\nonumber\\
&&\hspace{22mm}
\times~\big[\Re h_D(\alpha_q,k_\perp)\bar{h'}_1^\perp(\beta_q,q_\perp-k_\perp)
+\Re\barh_D(\alpha_q,k_\perp){h'}^\perp_1(\beta_q,q_\perp-k_\perp)
\nonumber\\
&&\hspace{25mm}
+~{h'}^\perp_1(\alpha_q,k_\perp)\Re\barh_D(\beta_q,q_\perp-k_\perp)
+\bar{h'}_1^\perp(\alpha_q,k_\perp)\Re h_D(\beta_q,q_\perp-k_\perp)
\big]\Big\}
\nonumber
\end{eqnarray}
where $\barf'_1(\beta_q,q_\perp-k_\perp) \equiv {\partial\over\partial\beta_q}\barf_1(\beta_q,q_\perp-k_\perp)$ etc.

As was mentioned above, for the rest of ${1\over Q^2}$ corrections one can neglect the dependence of
projective fields on $x_\ast$ and target ones on $x_\bu$. Using parametrizations in Sect. \ref{sec:paramlt}, one obtains for quark-antiquark contribution (\ref{ltfierz})
\begin{eqnarray}
&&\hspace{-1mm}
W^{q\barq}(\alpha_q,\beta_q,q_\perp)~=~
{2\over N_cs}\!\int\! d^2k_\perp\Big\{-{s\over 2}g^\perp_{\mu\nu}\{f_1\barf_1+\barf_1 f_1\}
\label{ququ}\\
&&\hspace{-1mm}
+[k^\perp_\mu(q-k)^\perp_\nu+k^\perp_\nu(q-k)^\perp_\mu+g_{\mu\nu}(k,q-k)_\perp]\{f_\perp\barf_\perp+\barf_\perp f_\perp\}
\nonumber\\
&&\hspace{-1mm}
+~(k^\perp_\mu p_{2\nu}+k^\perp_\nu p_{2\mu})\{f_\perp\barf_1+\barf_\perp f_1\}
+[(q-k)^\perp_\mu p_{1\nu}+(q-k)^\perp_\nu p_{1\mu}]\{f_1\barf_\perp+\barf_1 f_\perp)\}
\nonumber\\
&&\hspace{-1mm}
+~{4m^2\over s}p_{1\mu}p_{1\nu}\{f_1\barf_3+\barf_1 f_3\}+{4m^2\over s}p_{2\mu}p_{2\nu}\{f_3\barf_1+\barf_3 f_1\}\Big\}
\nonumber\\
&&\hspace{-1mm}
+~
[(g_{\mu\nu}^\parallel-g_{\mu\nu}^\perp)(k,q-k)_\perp-k^\perp_\mu(q-k)^\perp_\nu-k^\perp_\nu(q-k)^\perp_\mu]
\{g^\perp\barg^\perp+\barg^\perp g^\perp\}-m^2g_{\mu\nu}\{e\bare+\bare e\}
\nonumber\\
&&\hspace{-1mm}
-{s\over 2m^2}\big(g^\perp_{\mu\nu}+k^\perp_\mu(q-k)^\perp_\nu+k^\perp_\nu(q-k)^\perp_\mu\big)
\{h_1^\perp\barh_1^\perp+\barh_1^\perp h_1^\perp\}+m^2\big(g^\perp_{\mu\nu}-g^\parallel_{\mu\nu}\big)\{h\barh+\barh h\}
\nonumber\\
&&\hspace{-1mm}
+~\big(k^\perp_\mu p_{1\nu}+ k^\perp_\nu p_{1\mu}\big)\{h_1^\perp\barh+\barh_1^\perp h\}
+\big[p_{2\mu}(q-k)^\perp_\nu+p_{2\nu}(q-k)^\perp_\mu\big]\{h\barh_1^\perp+\barh h_1^\perp\}
\nonumber\\
&&\hspace{-1mm}
+~{4\over s}(k,q-k)_\perp\big[p_{1\mu} p_{1\nu} \{h_1^\perp\barh_3^\perp+\barh_1^\perp h_3^\perp\}
+p_{2\mu} p_{2\nu} \{h_3^\perp\barh_1^\perp+\barh_3^\perp h_1^\perp\}\big]
\nonumber\\
&&\hspace{-1mm}
-~{2\over s}m^2\big[(k,q-k)_\perp(g^\perp_{\mu\nu}+k^\perp_\mu(q-k)^\perp_\nu+k^\perp_\nu(q-k)^\perp_\mu\big]
\{h_3^\perp\barh_3^\perp+\barh_3^\perp h_3^\perp\}\Big]
\nonumber
\end{eqnarray}
Hereafter we use the notation
\beq
\{f_1\barf_2+\barf_1f_2\}~\equiv ~f_1(\alpha_q,k_\perp)\barf_2(\beta_q, q_\perp-k_\perp)
+\barf_1(\alpha_q,k_\perp)f_2(\beta_q, q_\perp-k_\perp)
\label{glavnotation}
\eeq
so that the argument of the first function is always $(\alpha_q,k_\perp)$ and that of the second  is $(\beta_q, q_\perp-k_\perp)$, similarly to the leading-twist contribution (\ref{WLT}) which we included for completeness.

 \section{Terms with one quark-quark-gluon operator $\Xi_i$ or $\Bxi_i$ \label{1qqG}}
We separate terms in Eq. (\ref{7lines}) according to the number of gluon fields (contained in  $\Xi$'s ).
\begin{equation}
\cheW_{\mu\nu}~\stackrel{{\rm sym}~\mu,\nu}{=}~\cheW_{\mu\nu}^{\rm lt}+\cheW_{\mu\nu}^{(1)}
+\cheW_{\mu\nu}^{(1')}+\cheW_{\mu\nu}^{(2a)}+\cheW_{\mu\nu}^{(2b)}
+\cheW_{\mu\nu}^{(2c)}
\end{equation}
where leading-twist terms without gluons (quark-antiquark TMDs) were considered in previous Section, and
\begin{eqnarray}
&&\hspace{-1mm}
\cheW_{\mu\nu}^{(1)}(x)~=~
{N_c\over s}\langle A,B|
\big[\bar\psi_A(x)\gamma_\mu\psi_B(x)\big]\big[\bar\psi_B(0)\gamma_\nu\Xi_{1}(0)\big]
\nonumber\\
&&\hspace{-1mm}
+~\big[\Bxi_{1}(x)\gamma_\mu\psi_B(x)\big]\big[\bar\psi_B(0)\gamma_\nu\psi_A(0)\big]
+\big[\bar\psi_A(x)\gamma_\mu\Xi_{2}(x)\big]\big[\bar\psi_B(0)\gamma_\nu\psi_A(0)\big]
\nonumber\\
&&\hspace{-1mm}
+~\big[\bar\psi_A(x)\gamma_\mu\psi_B(x)\big]\big[\Bxi_{2}(0)\gamma_\nu\psi_A(0)\big]
+\mu\leftrightarrow\nu|A,B\rangle~+~x\leftrightarrow 0
\label{kalw1}
\end{eqnarray}
\begin{eqnarray}
&&\hspace{-1mm}
\cheW_{\mu\nu}^{(1')}(x)~=~
{N_c\over s}\langle A,B|
\big[\bar\psi_A(x)\gamma_\mu\psi_B(x)\big]\big[\bar\psi_B(0)\gamma_\nu\Xi'_1(0)\big]
\nonumber\\
&&\hspace{-1mm}
+~\big[\Bxi'_1(x)\gamma_\mu\psi_B(x)\big]\big[\bar\psi_B(0)\gamma_\nu\psi_A(0)\big]
+\big[\bar\psi_A(x)\gamma_\mu\Xi'_2(x)\big]\big[\bar\psi_B(0)\gamma_\nu\psi_A(0)\big]
\nonumber\\
&&\hspace{-1mm}
+~\big[\bar\psi_A(x)\gamma_\mu\psi_B(x)\big]\big[\Bxi'_2(0)\gamma_\nu\psi_A(0)\big]
+\mu\leftrightarrow\nu|A,B\rangle~+~x\leftrightarrow 0
\label{kalw1shtrix}
\end{eqnarray}
and
\begin{eqnarray}
&&\hspace{-1mm}
\cheW_{\mu\nu}^{(2a)}(x)~=~
{N_c\over s}\langle A,B|\big[\bar\psi_A(x)\gamma_\mu\Xi_{2}(x)\big]\big[\bar\psi_B(0)\gamma_\nu\Xi_{1}(0)\big]
\nonumber\\
&&\hspace{-1mm}
+~[\Bxi_{1}(x)\gamma_\mu\psi_B(x)\big]\big[\Bxi_{2}(0)\gamma_\nu\psi_A(0)\big]
+\mu\leftrightarrow\nu|A,B\rangle~+~x\leftrightarrow 0
\label{kalw2a}
\end{eqnarray}
\begin{eqnarray}
&&\hspace{-1mm}
\cheW_{\mu\nu}^{(2b)}(x)~=~
{N_c\over s}\langle A,B|
\big[\bar\psi_A(x)\gamma_\mu\Xi_{2}(x)\big]\big[\Bxi_{2}(0)\gamma_\nu\psi_A(0)\big]
\nonumber\\
&&\hspace{-1mm}
+~
\big[\Bxi_{1}(x)\gamma_\mu\psi_B(x)\big]\big[\bar\psi_B(0)\gamma_\nu\Xi_{1}(0)\big]
+\mu\leftrightarrow\nu|A,B\rangle~+~x\leftrightarrow 0
\label{kalw2b}
\end{eqnarray}
\begin{eqnarray}
&&\hspace{-1mm}
\cheW_{\mu\nu}^{(2c)}(x)~=~
{N_c\over s}\langle A,B|\big[\Bxi_{1}(x)\gamma_\mu\Xi_{2}(x)\big]\big[\bar\psi_B(0)\gamma_\nu\psi_A(0)\big]
\nonumber\\
&&\hspace{-1mm}
+~\big[\bar\psi_A(x)\gamma_\mu\psi_B(x)\big]\big[\Bxi_{2}(0)\gamma_\nu\Xi_{1}(0)\big]
+\mu\leftrightarrow\nu|A,B\rangle~+~x\leftrightarrow 0
\label{kalw2c}
\end{eqnarray}
The corresponding contributions to $W_{\mu\nu}(q)$ will be denoted $W_{\mu\nu}^{(1)}$,  $W_{\mu\nu}^{(1')}$, $W_{\mu\nu}^{(2)a}$,  $W_{\mu\nu}^{(2)b}$, and 
 $W_{\mu\nu}^{(2)c}$, respectively. In this and  next two Sections, 
I will consider these contributions in turn.  
Whenever possible, I will refer to calculations in Ref.  \cite{Balitsky:2020jzt}
to pinpoint terms which can be safely neglected. 
The calculations in this paper are very similar to those of Ref.  \cite{Balitsky:2020jzt}  but much more  lengthly. 
The result is presented in Sect. \ref{sec:result}. 

\subsection{Terms with $\Xi_{1}$ \label{sec:onexifirst}}
Let us start with the first term in Eq. (\ref{kalw1}). The Fierz transformation (\ref{fierz}) 
yields
\begin{eqnarray}
&&\hspace{-1mm}
\half[\bar\psi_A(x)\gamma_\mu\psi_B(x)\big]\big[\bar\psi_B(0)\gamma_\nu\Xi_{1}(0)\big]~+~\mu\leftrightarrow\nu
\nonumber\\
&&\hspace{-1mm}
=~{g_{\mu\nu}\over 4}\big\{\big[\bar\psi_A^{m}(x){\notp_2\over s}\gamma^i{1\over \alpha}\psi_A^k(0)\big]\big[\bar\psi_B^{n}\barB_i^{nk}(0)\psi_{B}^{m}(x)\big]~-~(\psi_A^k\otimes\psi_B^n\leftrightarrow\gamma_5\psi_A^k\otimes\gamma_5\psi_B^n\big\}
\nonumber\\
&&\hspace{-1mm}
+~{1\over 4}(\delta_\mu^\alpha\delta_\nu^\beta+\delta_\nu^\alpha\delta_\mu^\beta-g_{\mu\nu}g^{\alpha\beta})
\nonumber\\
&&\hspace{-1mm}
\times~\big\{\big[\bar\psi_A^{m}(x)\gamma_\alpha{\notp_2\over s}\gamma^i{1\over \alpha}\psi_A^k(0)\big]
\big[\bar\psi_B^{n}\barB_i^{nk}(0)\gamma_\beta\psi^{m}_{B}(x)\big]
~+~(\gamma_\alpha\otimes\gamma_\beta\leftrightarrow\gamma_\alpha\gamma_5\otimes\gamma_\beta\gamma_5)\big\}
\nonumber\\
&&\hspace{-1mm}
-~{1\over 4}(\delta_\mu^\alpha\delta_\nu^\beta+\delta_\nu^\alpha\delta_\mu^\beta-\half g_{\mu\nu}g^{\alpha\beta})
\big[\bar\psi_A^{m}(x)\sigma_{\alpha\xi}{\notp_2\over s}\gamma^i{1\over \alpha}\psi_A^k(0)\big]
\big[\bar\psi_B^{n}\barB_i^{nk}(0)\sigma_\beta^{~\xi}\psi^{m}_{B}(x)\big]
\label{onexi1}
\end{eqnarray}
where we used Eq. (\ref{fildz0}) $\Xi_{1}(0)=-{\notp_2\over s}\gamma^i\barB_i{1\over \alpha}\psi_A(0)$. 
To save space, from now on we use the notations $A\psi(x)\equiv A(x)\psi(x)$ and $\bsi A(x)\equiv\bsi(x)A(x)$. Note that all
colors are in the fundamental representation so e.g. $B^{mn}(x)\equiv (t_a)^{mn}B^a(x)$.

Promoting $A$ and $B$ fields to operators and sorting out the color-singlet contributions we get
\begin{eqnarray}
&&\hspace{-3mm}
\cheW_{\mu\nu}^{(1\Xi_1)}(x)~=~{N_c\over s}\langle A,B|[\bsi_A(x)\gamma_\mu\psi_B(x)\big]\big[\bsi_B(0)\gamma_\nu\Xi_{1}(0)\big]~+~\mu\leftrightarrow\nu
|A,B\rangle~+~x\leftrightarrow 0
\label{chus}\\
&&\hspace{-1mm}
=~g_{\mu\nu}\chU^{(1a)}(x)
+(\delta_\mu^\alpha\delta_\nu^\beta+\delta_\nu^\alpha\delta_\mu^\beta-g_{\mu\nu}g^{\alpha\beta})
\chU_{\alpha\beta}^{(1b)}(x)
+(\delta_\mu^\alpha\delta_\nu^\beta+\delta_\nu^\alpha\delta_\mu^\beta-\half g_{\mu\nu}g^{\alpha\beta})
\chU_{\alpha\beta}^{(1c)}(x)
\nonumber
\end{eqnarray}
where
\begin{eqnarray}
&&\hspace{-3mm}
\chU^{(1a)}(x)~=~{1\over 2s^2}\big[\langle\bsi(x)\notp_2\gamma^i{1\over \alpha}\psi(0)\rangle_A\langle\bsi\barB_i(0)\psi(x)]\rangle_B
~-~(\psi(0)\otimes\psi(x)\leftrightarrow\gamma_5\psi(0)\otimes\gamma_5\psi(x)\big]
\nonumber\\
&&\hspace{15mm}
+~x\leftrightarrow 0
\nonumber\\
&&\hspace{-3mm}
\chU_{\alpha\beta}^{(1b)}(x)~=~{1\over 4s^2}
\big[\langle\bsi(x)\gamma_\alpha\notp_2\gamma_i{1\over \alpha}\psi(0)\rangle_A
\langle\bsi B^i(0)\gamma_\beta\psi(x)\rangle_B
\nonumber\\
&&\hspace{13mm}
~+~(\psi(0)\otimes\psi(x)\leftrightarrow\gamma_5\psi(0)\otimes\gamma_5\psi(x)+\alpha\leftrightarrow\beta\big]
~+~x\leftrightarrow 0
\label{fla49}\\
&&\hspace{-3mm}
\chU_{\alpha\beta}^{(1c)}(x)~=~-\big[{1\over 4s^2}\langle\bsi(x)\sigma_{\alpha\xi}\notp_2\gamma^i{1\over \alpha}\psi(0)\rangle_A
\langle\bsi B^i(0)\sigma_\beta^{~\xi}\psi(x)\rangle_B+\alpha\leftrightarrow\beta\big]~+~x\leftrightarrow 0
\nonumber
\end{eqnarray}

\subsubsection{Term $~\chU^{(1a)}$}
It is easy to see that 
\begin{eqnarray}
&&\hspace{-1mm}
\chU^{(1a)}(x)~=~{1\over 2s^2}\big\{\langle\bar\psi(x)\notp_2\gamma^i{1\over \alpha}\psi(0)\rangle_A\langle\bar\psi\barB_i(0)\psi(x)]\rangle_A
\nonumber\\
&&\hspace{15mm}
-~(\psi(0)\otimes\psi(x)\leftrightarrow\gamma_5\psi(0)\otimes\gamma_5\psi(x)\big\}
~+~x\leftrightarrow 0
\nonumber\\
&&\hspace{-1mm}
=~-i{1\over 2s^2}\langle\bar\psi(x)\sigma_{\star i}{1\over \alpha}\psi(0)\rangle_A\langle\bsi[B_i(0)+i\tilB_i\gamma_5(0)]\psi(x)\rangle_A
~+~x\leftrightarrow 0
\end{eqnarray}
where we used formula
\begin{equation}
\sigma_{\star i}\otimes B^i-\sigma_{\star i}\gamma_5\otimes B^i\gamma_5~=~\sigma_{\star i}\otimes \acB^i
\end{equation}
Throughout the paper we will use the notations
\begin{equation}
\acA_i\equiv A_i+i\tilA_i\gamma_5,~~~~\grA\equiv A_i-i\tilA_i\gamma_5,~~~~~
\acB_i\equiv B_i+i\tilB_i\gamma_5,~~~~\graB\equiv B_i-i\tilB_i\gamma_5
\label{agrebi}
\end{equation}

From formula (\ref{maelqg3}) and parametrization (\ref{bres}) from Appendix \ref{sec:qqgparam} we get 
\begin{eqnarray}
&&\hspace{-1mm}
U^{(1a)}(q)~=~{1\over 16\pi^4N_c}\!\int\! dx_\star dx_\bu d^2x_\perp~e^{-i\alpha_qx_\bu-i\beta_qx_\star+i(q,x)_\perp}
\chU^{(1a)}(x)
\label{w1aq}\\
&&\hspace{-1mm}
=~-i{1\over\alpha_qsN_c}\!\int\! d^2k_\perp~ (k,q-k)_\perp 
\{h_1^\perp\bar\ace_G
+\barh_1^\perp\ace_G\}
\nonumber
\end{eqnarray}

\subsubsection{Term $~\chU_{\alpha\beta}^{(1b)}$}
In this section we consider
\begin{eqnarray}
&&\hspace{-1mm}
\chU_{\alpha\beta}^{(1b)}(x)~=~~{1\over 4s^2}
\big\{\langle\bar\psi(x)\gamma_\alpha\notp_2\gamma_i{1\over \alpha}\psi(0)\rangle_A
\langle\bar\psi B^i(0)\gamma_\beta\psi(x)\rangle_B
\label{w1b1}\\
&&\hspace{-1mm}
+~\big(\psi(0)\otimes\psi(x)\leftrightarrow\gamma_5\psi(0)\otimes\gamma_5\psi(x)\big)
~+~\alpha\leftrightarrow\beta\big\}~+~x\leftrightarrow 0~=~
\chU_{1\alpha\beta}^{(1b)}(x)+\chU_{1\alpha\beta}^{(1b)}(x\leftrightarrow 0)
\nonumber
\end{eqnarray}
Let us start from the first term in the r.h.s. of this equation. From Eq. (\ref{fformula1})
\begin{eqnarray}
&&\hspace{-1mm}
\chU_{1\alpha\beta}^{(1b)}(x)~=~{1\over 4s^2}
\big\{\langle\bar\psi(x)\gamma_\alpha\notp_2\gamma_i{1\over \alpha}\psi(0)\rangle_A
\label{w1bodd}\\
&&\hspace{14mm}
\times~
\langle\bar\psi B^i(0)\gamma_\beta\psi(x)\rangle_B
~+~(\psi(0)\otimes\psi(x)\leftrightarrow\gamma_5\psi(0)\otimes\gamma_5\psi(x)\big\}
 ~+~\alpha\leftrightarrow\beta
\nonumber\\
&&\hspace{-1mm}
=~{1\over 4s^2}
\Big\{
-\langle\bar\psi(x)\slp_2{1\over \alpha}\psi(0)\rangle_A\langle\bar\psi \gamma_{\beta_\perp}\graB_\alpha(0)\psi(x)\rangle_B
-~{2\over s}p_{2\alpha}p_{2\beta}\langle\bar\psi(x)\gamma_i{1\over \alpha}\psi(0)\rangle_A
\nonumber\\
&&\hspace{14mm}
\times~\langle\bar\psi \slB(0)\slp_1\gamma^i\psi(x)\rangle_B
-~{2\over s}p_{2\alpha}p_{2\beta}\langle\bar\psi(x)\gamma_i\gamma_5
{1\over \alpha}\psi(0)\rangle_A\langle\bar\psi \slB(0)\slp_1\gamma^i\gamma_5\psi(x)\rangle_B
\nonumber\\
&&\hspace{22mm}
+~{2\over s}p_{2\beta}\langle\bar\psi(x)\slp_2{1\over \alpha}\psi(0)\rangle_A
\langle\bar\psi \slB(0)\slp_1\gamma_{\alpha_\perp}\psi(x)\rangle_B~+~\alpha\leftrightarrow\beta\Big\},
\nonumber
\end{eqnarray}
From Eqs. (\ref{tarmaels}) and (\ref{bresgi})
we get
\begin{eqnarray}
&&\hspace{-1mm}
{U}_{1\alpha\beta}^{(1b)}(q)~=~{1\over 16\pi^4N_c}\!\int\! dx_\star dx_\bu d^2x_\perp~e^{-i\alpha_qx_\bu-i\beta_qx_\star+i(q,x)_\perp}
\chU_{1\alpha\beta}^{(1b)}(x)
\label{wf1b}\\
&&\hspace{-1mm}
=~{1\over 2\alpha_qs}\!\int\! d^2k_\perp\Big[
(q-k)_{\alpha_\perp}p_{2\beta}f_1(\alpha_q,k_\perp)[\barf_1^\perp-\beta_q(\barf_\perp-i\barg_\perp)](\beta_q,q_\perp-k_\perp)
\nonumber\\
&&\hspace{-1mm}
-~f_1(\alpha_q,k_\perp)\bar\bref_{1G}(\beta_q,q_\perp-k_\perp)\big[(q-k)_\alpha(q-k)_\beta
+(q-k)_\perp^2{g_{\alpha\beta}^\perp\over 2}\big]
\nonumber\\
&&\hspace{-1mm}
-~{g_{\alpha\beta}^\perp\over 2}f_1(\alpha_q,k_\perp)\big[(q-k)_\perp^2(\barf+i\barg)-2\beta_qm^2\barf_3\big](\beta_q,q_\perp-k_\perp)
\nonumber\\
&&\hspace{-1mm}
+~{2\over s}p_{2\alpha}p_{2\beta}(k,q-k)_\perp [f_\perp+ig_\perp](\alpha_q,k_\perp)
\big(\barf_{1}^\perp 
-\beta_q[\barf_\perp-i\barg_\perp)]\big)(\beta_q,q_\perp-k_\perp\Big]
~+~\alpha\leftrightarrow\beta
\nonumber
\end{eqnarray}
where for brevity 
\beq
[\barf_1^\perp-\beta_q(\barf_\perp-i\barg_\perp)](\beta_q,q_\perp-k_\perp)\equiv \barf_1^\perp
-\beta_q\Big(\barf_\perp(\beta_q,q_\perp-k_\perp)-i\barg_\perp(\beta_q,q_\perp-k_\perp)\Big)
\eeq
and similarly for other terms here and throughout the paper.

Next,
\begin{eqnarray}
&&\hspace{-1mm}
\chU_{2\alpha\beta}^{(1b)}(x)~\equiv~\chU_{1\alpha\beta}^{(1b)}(x\leftrightarrow 0)
\label{w2bodd}\\
&&\hspace{-1mm}
=~{1\over 4s^2}
\big\{\langle\bar\psi(0)\gamma_\alpha\notp_2\gamma_i{1\over \alpha}\psi(x)\rangle_A
\langle\bar\psi B^i(x)\gamma_\beta\psi(0)\rangle_B
\nonumber\\
&&\hspace{22mm}
+~(\psi(x)\otimes\psi(0)\leftrightarrow\gamma_5\psi(x)\otimes\gamma_5\psi(0)\big\}
 ~+~\alpha\leftrightarrow\beta
\nonumber\\
&&\hspace{-1mm}
=~
\Big\{
\langle\bar\psi(0)\slp_2\psi(x)\rangle_A\langle\bar\psi\gamma_{\beta_\perp} \graB_\alpha(x)\psi(0)\rangle_B
+~{2\over s}p_{2\alpha}p_{2\beta}\langle\bar\psi(0)\gamma_i\psi(x)\rangle_A
\langle\bar\psi \slB(x)\slp_1\gamma^i\psi(0)\rangle_B
\nonumber\\
&&\hspace{13mm}
+~{2\over s}p_{2\alpha}p_{2\beta}\langle\bar\psi(0)\gamma_i\gamma_5
\psi(x)\rangle_A\langle\bar\psi \slB(x)\slp_1\gamma^i\gamma_5\psi(0)\rangle_B
\nonumber\\
&&\hspace{22mm}
-~{2\over s}p_{2\beta}\langle\bar\psi(0)\slp_2\psi(x)\rangle_A
\langle\bar\psi \slB(x)\slp_1\gamma_{\alpha_\perp}\psi(0)\rangle_B~
\Big\}{1\over 4s^2\alpha_q}+~\alpha\leftrightarrow\beta
\nonumber
\end{eqnarray}
Now, from Eqs. (\ref{bresgi}) and (\ref{tarmaels5})
we get
\begin{eqnarray}
&&\hspace{-11mm}
{U}_{2\alpha\beta}^{(1b)}(q)~\equiv~{1\over 16\pi^4N_c}\!\int\! dx_\star dx_\bu d^2x_\perp~e^{-i\alpha_qx_\bu-i\beta_qx_\star+i(q,x)_\perp}
\chU_{2\alpha\beta}^{(1b)}(x)
\nonumber\\
&&\hspace{-11mm}
=~{1\over 2\alpha_qsN_c}\!\int\! d^2k_\perp\Big[-\barf_1^\perp(\alpha_q,k_\perp)\graf_{1G}(\beta_q,q_\perp-k_\perp)\big[(q-k)^\perp_\alpha(q-k)^\perp_\beta
+(q-k)_\perp^2{g^\perp_{\alpha\beta}\over 2}\big]
\nonumber\\
&&\hspace{-11mm}
-~{g^\perp_{\alpha\beta}\over 2}\barf_1^\perp(\alpha_q,k_\perp)
\big[(q-k)_\perp^2(f-ig)-2\beta_qm^2f_3\big](\beta_q,q_\perp-k_\perp)
\nonumber\\
&&\hspace{-11mm}
+~{2\over s}p_{2\alpha}p_{2\beta}(k,q-k)_\perp \big(\barf_\perp-i\barg_\perp\big)(\alpha_q,k_\perp)
\big(f_{1}
-\beta_q[f_\perp+ig_\perp](\beta_q,q_\perp-k_\perp)\big)
\nonumber\\
&&\hspace{-11mm}
+~(q-k)_{\alpha_\perp}p_{2\beta}\barf_1(\alpha_q,k_\perp)[f_1-\beta_q(f_\perp+ig_\perp)](\beta_q,q_\perp-k_\perp)\Big]
~+~\alpha\leftrightarrow\beta
\label{w1b2q}
\end{eqnarray}
Finally,
\begin{eqnarray}
&&\hspace{-1mm}
U_{\alpha\beta}^{(1b)}(q)~=~{1\over 16\pi^4N_c}\!\int\! dx_\star dx_\bu d^2x_\perp~e^{-i\alpha_qx_\bu-i\beta_qx_\star+i(q,x)_\perp}
\big(\chU_{1\alpha\beta}^{(1b)}(x)+\chU_{2\alpha\beta}^{(1b)}(x)\big)
\label{w1bq}\\
&&\hspace{-1mm}
=~{1\over 2\alpha_qsN_c}\!\int\! d^2k_\perp\Big\{
(q-k)_{\alpha_\perp}p_{2\beta}\{f_1[\barf_1-\beta_q(\barf_\perp-i\barg_\perp)]
+~\barf_1[f_1-\beta_q(f_\perp+ig_\perp)]\}
\nonumber\\
&&\hspace{-1mm}
-~\big[(q-k)_\alpha(q-k)_\beta+(q-k)_\perp^2{g_{\alpha\beta}^\perp\over 2}\big]
\{f_1\bar\graf_{1G}
+~\barf_1\graf_{1G}\}
\nonumber\\
&&\hspace{-1mm}
-~{g_{\alpha\beta}^\perp\over 2}\Big((q-k)_\perp^2\{f_1[\barf+i\barg]+\barf_1[f-ig]\}
-2\beta_qm^2\{f_1\barf_3+\barf_1 f_3\}\Big)
\nonumber\\
&&\hspace{-1mm}
+~{2\over s}p_{2\alpha}p_{2\beta}(k,q-k)_\perp \Big[\{[f_\perp+ig_\perp]\barf_1+[\barf_\perp-i\barg_\perp] f_1\}
\nonumber\\
&&\hspace{-1mm}
-~\beta_q\{[f_\perp+ig_\perp][\barf_\perp-i\barg_\perp]+[\barf_\perp-i\barg_\perp][f_\perp+ig_\perp]\}\Big]
\Big\}
~+~\alpha\leftrightarrow\beta
\nonumber
\end{eqnarray}
where we used notation (\ref{glavnotation}).

\subsubsection{Term $\chU_{\alpha\beta}^{(1c)}$}
Next, consider
\beq
\chU_{\alpha\beta}^{(1c)}(x)~=-~{1\over 4s^2}\big[
\langle\bar\psi(x)\sigma_{\alpha\xi}\notp_2\gamma^i{1\over \alpha}\psi(0)\rangle_A
\langle\bar\psi B_i(0)\sigma_\beta^{~\xi}\psi(x)\rangle_B~+~\alpha\leftrightarrow\beta\big]~+~x\leftrightarrow 0
\label{onexi9}
\eeq

From Eq. (\ref{hformulas}) we get
\begin{eqnarray}
&&\hspace{-1mm}
\chU_{1\alpha\beta}^{(1c)}(x)~=~{1\over 4s^2}
\langle\bar\psi(x)i\sigma_{\alpha\xi}\sigma_{\star i}{1\over \alpha}\psi(0)\rangle_A
\langle\bar\psi B^i(0)\sigma_\beta^{~\xi}\psi(x)\rangle_B~+~\alpha\leftrightarrow\beta
\nonumber\\
&&\hspace{-1mm}
=~{1\over 4s^2}
\Big\{\langle\bar\psi(x)\sigma_{\star j}{1\over \alpha}\psi(0)\rangle_A\langle\bar\psi B_\alpha(0)\sigma_{\beta_\perp}^{~ j}\psi(x)\rangle_B
+i\langle\bar\psi(x)\sigma_{\star \alpha_\perp}{1\over \alpha}\psi(0)\rangle_A
\nonumber\\
&&\hspace{-1mm}
\times~\langle\bar\psi (\slB\gamma_{\beta_\perp}-B_\beta)(0)\psi(x)\rangle_B
+~{2i\over s}p_{2\beta}\langle\bar\psi(x)\sigma_{\star \alpha_\perp}{1\over \alpha}\psi(0)\rangle_A\langle\bar\psi  \slB(0)\slp_1\psi(x)\rangle_B
\nonumber\\
&&\hspace{-1mm}
+~{2\over s}p_{2\alpha}p_{2\beta}\langle\bar\psi(x){1\over \alpha}\psi(0)\rangle_A\langle\bsi \slB(0)\slp_1\psi(x)\rangle_B
-~{4i\over s^2}p_{2\alpha}p_{2\beta}\langle\bar\psi(x)\sigma_{\bu\ast}{1\over \alpha}\psi(0)\rangle_A\langle\bsi \slB(0)\slp_1\psi(x)\rangle_B
\nonumber\\
&&\hspace{-1mm}
+~{4\over s^2}\big(p_{1\alpha}p_{2\beta}+p_{2\alpha}p_{1\beta}\big)\langle\bar\psi(x)\sigma_{\star i}{1\over \alpha}\psi(0)\rangle_A\langle\bsi B^i(0)\sigma_{\star\bu}\psi(x)\rangle_B
\Big\}~+~\alpha\leftrightarrow\beta
\label{w1c1x}
\end{eqnarray}
where we used the fact that
\begin{equation}
\langle\bar\psi(0) [B_i(0)\sigma_{\bu j}-B_j(0)\sigma_{\bu i}]\psi(x)\rangle_A=0
\label{flanpolarized}
\end{equation}
for unpolarized hadrons.
Next, from Eq. (\ref{besigmat})
and Eq. (\ref{hbes})
we obtain
\begin{eqnarray}
&&\hspace{-1mm}
{U}_{1\alpha\beta}^{(1c)}(q)~\equiv~{1\over 16\pi^4N_c}\!\int\! dx_\star dx_\bu d^2x_\perp~e^{-i\alpha_qx_\bu-i\beta_qx_\star+i(q,x)_\perp}
\chU_{1\alpha\beta}^{(1c)}(x)
\label{w1c1q}\\
&&\hspace{-1mm}
=~{1\over 2\alpha_qsN_c}. 
\!\int\! d^2k_\perp\Big[(k,q-k)_\perp g^\perp_{\alpha\beta}h_1^\perp(\alpha_q,k_\perp)
[i\bare-i\bare_G+\beta_q\barh_3^\perp+\barh_D](\beta_q,q_\perp-k_\perp)
\nonumber\\
&&\hspace{-1mm}
-~p_{2\beta}k^\perp_\alpha {1\over m^2}h_1^\perp(\alpha_q,k_\perp)
\big[(q-k)_\perp^2\barh_1^\perp
+m^2\beta_q(i\bare+\barh)\big](\beta_q,q_\perp-k_\perp)
\nonumber\\
&&\hspace{-1mm}
+~{2\over s}p_{2\alpha}p_{2\beta}~e(\alpha_q,k_\perp)\big[-i(q-k)_\perp^2\barh_{1}^\perp
+\beta_q m^2(\bare-i\barh)\big](\beta_q,q_\perp-k_\perp)
\nonumber\\
&&\hspace{-1mm}
+~{2\over s}p_{2\alpha}p_{2\beta}h(\alpha_q,k_\perp)\big[(q-k)_\perp^2\barh_{1}^\perp+\beta_q m^2(i\bare+\barh)\big](\beta_q,q_\perp-k_\perp)
\nonumber\\
&&\hspace{-1mm}
-~{2\over s}\big(p_{1\alpha}p_{2\beta}+p_{2\alpha}p_{1\beta}\big)(k,q-k)_\perp h_1^\perp(\alpha_q,k_\perp)
\big[\beta \barh_3^\perp-\barh-\barh_D-i\bar\tile_G\big](\beta_q,q_\perp-k_\perp)
+\alpha\leftrightarrow\beta
\nonumber
\end{eqnarray}

Now consider
\begin{eqnarray}
&&\hspace{-1mm}
\chU_{2\alpha\beta}^{(1c)}(x)~=~\chU_{1\alpha\beta}^{(1c)}(x\leftrightarrow 0)~=~{1\over 4s^2}
\langle\bar\psi(0)i\sigma_{\alpha\xi}\sigma_{\star i}{1\over \alpha}\psi(x)\rangle_A
\langle\bar\psi B^i(x)\sigma_\beta^{~\xi}\psi(0)\rangle_B~+~\alpha\leftrightarrow\beta
\nonumber\\
&&\hspace{-1mm}
=~{1\over 4s^2}
\Big\{\langle\bar\psi(0)\sigma_{\star j}{1\over \alpha}\psi(x)\rangle_A\langle\bar\psi B_\alpha(x)\sigma_{\beta_\perp}^{~ j}\psi(0)\rangle_B
+i\langle\bar\psi(0)\sigma_{\star \alpha_\perp}{1\over \alpha}\psi(x)\rangle_A
\nonumber\\
&&\hspace{-1mm}
\times~\langle\bar\psi (\slB\gamma_{\beta_\perp}-B_\beta)(x)\psi(0)\rangle_B
+~{2i\over s}p_{2\beta}\langle\bar\psi(0)\sigma_{\star \alpha_\perp}{1\over \alpha}\psi(x)\rangle_A\langle\bar\psi  \slB(x)\slp_1\psi(0)\rangle_B
\nonumber\\
&&\hspace{-1mm}
+~{2\over s}p_{2\alpha}p_{2\beta}\langle\bar\psi(0){1\over \alpha}\psi(x)\rangle_A\langle\bsi \slB(x)\slp_1\psi(0)\rangle_B
-~{4i\over s^2}p_{2\alpha}p_{2\beta}\langle\bar\psi(0)\sigma_{\bu\ast}{1\over \alpha}\psi(x)\rangle_A\langle\bsi \slB(x)\slp_1\psi(0)\rangle_B
\nonumber\\
&&\hspace{-1mm}
+~{4\over s^2}\big(p_{1\alpha}p_{2\beta}+p_{2\alpha}p_{1\beta}\big)\langle\bar\psi(0)\sigma_{\star i}{1\over \alpha}\psi(x)\rangle_A\langle\bsi B^i(x)\sigma_{\star\bu}\psi(0)\rangle_B
\Big\}~+~\alpha\leftrightarrow\beta
\nonumber
\end{eqnarray}
where we  again used Eq. (\ref{flanpolarized}) for
unpolarized hadrons.

Using Eqs. (\ref{besigmat}), (\ref{10.88}), and (\ref{hbes})
we get
\begin{eqnarray}
&&\hspace{-1mm}
{U}_{2\alpha\beta}^{(1c)}(q)~=~{1\over 16\pi^4N_c}\!\int\! dx_\star dx_\bu d^2x_\perp~e^{-i\alpha_qx_\bu-i\beta_qx_\star+i(q,x)_\perp}
\cheW_{2\alpha\beta}^{(1c)}(x)
\label{w1c2q}\\
&&\hspace{-1mm}
=~{1\over2\alpha_qsN_c}
\!\int\! d^2k_\perp\Big[-i(k,q-k)_\perp g^\perp_{\alpha\beta}\barh_1^\perp(\alpha_q,k_\perp)
(e+e_G+i\beta_qh_3^\perp-ih_D)(\beta_q,q_\perp-k_\perp)
\nonumber\\
&&\hspace{-1mm}
-~k^\perp_\alpha p_{2\beta}{1\over m^2}\barh_1^\perp(\alpha_q,k_\perp)\big[(q-k)_\perp^2h_1^\perp+\beta_qm^2(h-ie)\big](\beta_q,q_\perp-k_\perp)
\nonumber\\
&&\hspace{-1mm}
+~{p_{2\alpha}p_{2\beta}\over s} i\bare(\alpha_q,k_\perp)
\big[(q-k)_\perp^2h_1^\perp+m^2\beta_q(h-ie)\big](\beta_q,q_\perp-k_\perp)
\nonumber\\
&&\hspace{-1mm}
+~{2\over s}p_{2\alpha}p_{2\beta} \barh(\alpha_q,k_\perp)
\big[(q-k)_\perp^2h_1^\perp+m^2\beta_q(h-ie)\big](\beta_q,q_\perp-k_\perp)
\nonumber\\
&&\hspace{-1mm}
-~{2\over s}\big(p_{1\alpha}p_{2\beta}+p_{2\alpha}p_{1\beta}\big)\barh_1^\perp(\alpha_q,k_\perp)
\big[\beta h_3^\perp-h-i\tile_G+h_D\big](\beta_q,q_\perp-k_\perp)
~+~\alpha\leftrightarrow\beta
\nonumber
\end{eqnarray}
The sum of Eqs. (\ref{w1c1q}) and  (\ref{w1c2q}) is
\begin{eqnarray}
&&\hspace{-1mm}
{U}_{\alpha\beta}^{(1c)}(q)~=~{1\over 16\pi^4N_c}\!\int\! dx_\star dx_\bu d^2x_\perp~e^{-i\alpha_qx_\bu-i\beta_qx_\star+i(q,x)_\perp}
\chU_{\alpha\beta}^{(1c)}(x)~=~{1\over 2\alpha_qsN_c}
\label{w1cq}\\
&&\hspace{-1mm}
\times~
\!\int\! d^2k_\perp\Big\{ (k,q-k)_\perp g^\perp_{\alpha\beta}\Big(\{h_1^\perp
[i\bare-i\bare_G+\beta_q\barh_3^\perp+\barh_D]
+~\barh_1^\perp[-ie-ie_G+\beta_qh_3^\perp-h_D]\}\Big)
\nonumber\\
&&\hspace{-1mm}
-~p_{2\alpha}k^\perp_\beta \Big({(q-k)_\perp^2\over m^2}\{h_1^\perp\barh_1^\perp+\barh_1^\perp h_1^\perp\}
+\beta_q\{h_1[\barh+i\bare]+\barh_1[h-ie]\}\Big)+~{2\over s}p_{2\alpha}p_{2\beta}
\nonumber\\
&&\hspace{-1mm}
\times~
\Big(
(q-k)_\perp^2\{[h-ie]\barh_1^\perp+[\barh+i\bare]h_1^\perp\}+\beta_qm^2\{[h-ie][\barh+i\bare]+[\barh+i\bare][h-ie]\}\Big)
\nonumber\\
&&\hspace{-1mm}
-~g^\parallel_{\alpha\beta}(k,q-k)_\perp 
\{h_1^\perp[\beta \barh_3^\perp-\barh-i\bar\tile_G-\barh_D]
+~\barh_1^\perp[\beta h_3^\perp-h-i\tile_G+h_D]\}\Big\}
~+~\alpha\leftrightarrow\beta
\nonumber
\end{eqnarray}
Term with $\cheW_{\mu\nu}^{(1\Xi_1)}(q)$ is given by the sum of Eqs.  (\ref{w1aq}), (\ref{w1bq}), and (\ref{w1cq}).

\subsection{Terms with $\Bxi_1$, $\Xi_2$ and $\Bxi_2$ }
Let us consider now the second term in Eq. (\ref{kalw1}). from
\begin{eqnarray}
&&\hspace{-1mm}
\Big\{\int\! dx~e^{-iqx}\big[\Bxi_1(x)\gamma_\mu\psi_B(x)\big]\big[\bsi_B(0)\gamma_\nu\psi_A(0)\big]\Big\}^\ast
\nonumber\\
&&\hspace{-1mm}
=~\int\! dx~e^{-iqx}\big[\bsi_A(x)\gamma_\nu\psi_B(x)\big]\big[\bsi_B(0)\gamma_\mu\Xi_1(0)\big]
\label{procc}
\end{eqnarray}
we see that effectively addition of the term  $\cheW_{\mu\nu}^{(1\Bxi_1)}(q)$ doubles the real part of  $\cheW_{\mu\nu}^{(1\Xi_1)}(q)$
so one obtains
\begin{eqnarray}
&&\hspace{-1mm}
W^{(1\Xi_1+1\Bxi_1)}_{\mu\nu}(q)~=~{1\over 16\pi^4N_c}\!\int\! dx_\star dx_\bu d^2x_\perp~e^{-i\alpha_qx_\bu-i\beta_qx_\star+i(q,x)_\perp}
\big[\cheW^{(1\Xi_1)}_{\mu\nu}(x)+\cheW^{(1\Bxi_1)}_{\mu\nu}(x)\big]
\nonumber\\
&&\hspace{-1mm}
=~{1\over 16\pi^4N_c}\!\int\! dx_\star dx_\bu d^2x_\perp~e^{-i\alpha_qx_\bu-i\beta_qx_\star+i(q,x)_\perp}{N_c\over s}
\Big(\langle A,B|
\big[\bar\psi_A(x)\gamma_\mu\psi_B(x)\big]\big[\bar\psi_B(0)\gamma_\nu\Xi_{1}(0)\big]
\nonumber\\
&&\hspace{-1mm}
+~\big[\Bxi_{1}(x)\gamma_\mu\psi_B(x)\big]\big[\bar\psi_B(0)\gamma_\nu\psi_A(0)\big]|A,B\rangle~+~\mu\leftrightarrow\nu~+~x\leftrightarrow 0\Big)
\label{w1qtru}\\
&&\hspace{-1mm}
=~g_{\mu\nu}\big[\chU^{(1a)}(q)+\chU^{(2a)}(q)\big]
+(\delta_\mu^\alpha\delta_\nu^\beta+\delta_\nu^\alpha\delta_\mu^\beta-g_{\mu\nu}g^{\alpha\beta})
\big[\chU_{\alpha\beta}^{(1b)}(q)+\chU_{\alpha\beta}^{(2b)}(q)\big]
\nonumber\\
&&\hspace{-1mm}
+~(\delta_\mu^\alpha\delta_\nu^\beta+\delta_\nu^\alpha\delta_\mu^\beta-\half g_{\mu\nu}g^{\alpha\beta})
\big[\chU_{\alpha\beta}^{(1c)}(q)+\chU_{\alpha\beta}^{(2c)}(q)\big]
\nonumber\\
&&\hspace{-1mm}
=~{2\over\alpha_qsN_c}\!\int\! d^2k_\perp~ 
\Big\{(k,q-k)_\perp g_{\mu\nu}\{h_1^\perp \Im\bar\ace_G
+\barh_1^\perp\Im\ace_G\}
\nonumber\\
&&\hspace{-1mm}
+~
[(q-k)_{\mu_\perp}p_{2\nu}+\mu\leftrightarrow\nu]\big(\{f_1\barf_1+\barf_1f_1\}
-\beta_q\{f_1\barf_\perp+\barf_1f_\perp\}\big)
\nonumber\\
&&\hspace{-1mm}
-~\big[(q-k)^\perp_\mu(q-k)^\perp_\nu+\mu\leftrightarrow\nu+g^\perp_{\mu\nu}(q-k)_\perp^2\big]
\{f_1\Re\bar\graf_{1G}
+\barf_1\Re\graf_{1G}\}
\nonumber\\
&&\hspace{-1mm}
+~g^\parallel_{\mu\nu}\big[(q-k)_\perp^2\{f_1\barf_\perp+\barf_1f_\perp\}-2\beta_q m^2\{f_1\barf_3+\barf_1f_3\}
+m^2(f_1\Re \barf_D+\barf_1\Re f_D)\big]
\nonumber\\
&&\hspace{-1mm}
+~{4\over s}p_{2\mu}p_{2\nu}
(k,q-k)_\perp\big(
\{ f_\perp\barf_1+\barf_\perp f_1\}-\beta_q\{ f_\perp\barf_\perp+\barf_\perp f_\perp\}
-\beta_q\{g_\perp\barg_\perp+\barg_\perp g_\perp\}\big)
\nonumber\\
&&\hspace{-1mm}
+~
[g^\perp_{\mu\nu}-g^\parallel_{\mu\nu}](k,q-k)_\perp\big(2\beta_q\{h_1^\perp\barh_3^\perp+\barh_1^\perp h_3^\perp\}-\{h_1^\perp\barh+\barh_1^\perp h\}
+\{h_1^\perp\Im\bar\ace_G+\barh_1^\perp\Im \ace_G\}\big)
\nonumber\\
&&\hspace{-1mm}
-~[p_{2\mu}k^\perp_\nu+p_{2\nu}k^\perp_\mu] \Big({(q-k)_\perp^2\over m^2}\{h_1^\perp\barh_1^\perp+\barh_1^\perp h_1^\perp\}+\beta_q \{h_1^\perp\barh +\barh_1^\perp h\}
\Big)
\nonumber\\
&&\hspace{-1mm}
+~{4\over s}p_{2\mu}p_{2\nu}\big[m^2\beta_q\{e\bare+\bare e\}
+(q-k)_\perp^2\{h\barh_1^\perp+\barh h_1^\perp\}+\beta_qm^2\{h\barh+\barh h\}
\big]\Big\}
\nonumber
\end{eqnarray}
where we used $e_G+\tile_G=\ace_G$, see parametrizations (\ref{bes}) and (\ref{bres})

Next, the term with $\Xi_{2}$
can be obtained by projectile$\leftrightarrow$target replacement
\beq
p_1\leftrightarrow p_2, ~~~~x_\bu\leftrightarrow x_\star,~~~
\alpha_q\leftrightarrow\beta_q,~~~k^\perp\leftrightarrow (q-k)^\perp 
\label{protareplace}
\eeq
so we get
\begin{eqnarray}
&&\hspace{-1mm}
W^{(1\Xi_2+1\Bxi_2)}_{\mu\nu}(q)~=~{1\over 16\pi^4N_c}\!\int\! dx_\star dx_\bu d^2x_\perp~e^{-i\alpha_qx_\bu-i\beta_qx_\star+i(q,x)_\perp}
[\cheW^{(1\Xi_2)}_{\mu\nu}(x)+\cheW^{(1\Bxi_2)}_{\mu\nu}(x)]~
\nonumber\\
&&\hspace{-1mm}
=~{1\over 16\pi^4N_c}\!\int\! dx_\star dx_\bu d^2x_\perp~e^{-i\alpha_qx_\bu-i\beta_qx_\star+i(q,x)_\perp}{N_c\over s}
\Big(\langle A,B|
\big[\bar\psi_A(x)\gamma_\mu\Xi_2(x)\big]\big[\bar\psi_B(0)\gamma_\nu\psi_A(0)\big]
\nonumber\\
&&\hspace{-1mm}
+~\big[\bar\psi_A(x)\gamma_\nu\psi_B(x)\big]\big[\Bxi_{2}(0)\gamma_\mu\psi_B(0)\big]|A,B\rangle~+~\mu\leftrightarrow\nu~+~x\leftrightarrow 0\Big)
\label{w2qtru}\\
&&\hspace{-1mm}
=~{2\over\beta_qsN_c}\!\int\! d^2k_\perp~ 
\bigg[(k,q-k)_\perp g_{\mu\nu}\{\Im\bar\ace_Gh_1^\perp+\Im\ace_G\barh_1^\perp\}
\nonumber\\
&&\hspace{-1mm}
+~
\Big\{[k_{\mu_\perp}p_{1\nu}+\mu\leftrightarrow\nu]\big(\{f_1\barf_1+\barf_1f_1\}
-\alpha_q\{f_\perp\barf_1+\barf_\perp f_1\}\big)
\nonumber\\
&&\hspace{-1mm}
-~\big[k^\perp_\mu k^\perp_\nu+\mu\leftrightarrow\nu+g^\perp_{\mu\nu}k_\perp^2\big]
\{\Re\bar\graf_{1G}f_1+\Re\graf_{1G}\barf_1\}
\nonumber\\
&&\hspace{-1mm}
+~g^\parallel_{\mu\nu}\big[k_\perp^2\{\barf_\perp f_1+f_\perp\barf_1\}-2\alpha_q m^2\{\barf_3f_1+f_3\barf_1\}
+m^2\{\Re \barf_Df_1+\Re f_D\barf_1\}\big]
\nonumber\\
&&\hspace{-1mm}
+~{4\over s}p_{1\mu}p_{1\nu}
(k,q-k)_\perp\big[
\{\barf_1 f_\perp+f_1\barf_\perp \}-\alpha_q\{\barf_\perp f_\perp+ f_\perp\barf_\perp\}
-\alpha_q\{g_\perp\barg_\perp+\barg_\perp g_\perp\}\big]
\nonumber\\
&&\hspace{-1mm}
+~
[g^\perp_{\mu\nu}-g^\parallel_{\mu\nu}](k,q-k)_\perp\big(2\alpha_q\{\barh_3^\perp h_1^\perp+ h_3^\perp\barh_1^\perp\}
-\{\barh h_1^\perp+ h\barh_1^\perp\}
+\{\Im\bar\ace_Gh_1^\perp+\Im \ace_G\barh_1^\perp\}\big)
\nonumber\\
&&\hspace{-1mm}
-~[p_{1\mu}(q-k)^\perp_\nu+p_{1\nu}(q-k)^\perp_\mu]  \Big({k_\perp^2\over m^2}\{h_1^\perp\barh_1^\perp+\barh_1^\perp h_1^\perp\}
+\alpha_q \{\barh h_1^\perp + h\barh_1^\perp \}\Big)
\nonumber\\
&&\hspace{-1mm}
+~{4\over s}p_{1\mu}p_{1\nu}\big[m^2\alpha_q\{e\bare+\bare e\}
+k_\perp^2\{\barh_1^\perp h+h_1^\perp \barh \}+\alpha_qm^2\{h\barh+\barh h\}
\big]\Big\}
\nonumber
\end{eqnarray}
%

\section{Terms with $\Xi_i'$ or $\Bxi'_i$ \label{sec:primed}}
These  terms were not considered in Ref. \cite{Balitsky:2020jzt} so the analysis below will be more detailed than in 
previous Section.

\subsection{Terms with $\Xi'_1$}
We start from
\begin{eqnarray}
&&\hspace{-7mm}
\cheW_{\mu\nu}^{(1\Xi'_1)}(x)~=~
{N_c\over s}\langle A,B|[\bsi_A(x)\gamma_\mu\psi_B(x)\big]\big[\bsi_B(0)\gamma_\nu\Xi'_{1}(0)\big]~+~\mu\leftrightarrow\nu
|A,B\rangle~+~x\leftrightarrow 0
\label{dob1}
\end{eqnarray}
where $\Xi'_{1}$ can be taken from Ref.  \cite{Balitsky:2017gis}
\begin{eqnarray}
&&\hspace{-5mm}
\Xi'_{1}~=~-{\slashed{p}_1\over s}\gamma^i{B_i\over\beta}\psi_A
+{1\over s^2}\Big({\slp_1\over\beta}\slP_\perp{\slp_2\over\alpha}+{\slp_2\over\alpha}\slP_\perp{\slp_1\over\beta}\Big)\gamma_iB^i\psi_A
\nonumber\\
&&\hspace{-1mm}
-~{2\over s^2}\slp_2\slp_1B_\star {1\over\alpha}\psi_A
+{2\over s^2}\Big(B_\star{\slp_2\over\alpha^2}+A_\bu{\slp_1\over\beta^2}\Big)\slB_\perp\psi_A
\label{dob2}
\end{eqnarray}
First, note that the term $\slA_\perp$ in $\slP_\perp$ and $A_\bu$ in the last term can be neglected at large $N_c$. Indeed, 
after separation of color singlet contributions
\begin{eqnarray}
&&\hspace{-1mm}
\langle A,B|(\bsi_A^k \psi_B^k)(\bsi_B^m A_\mu^{ml}B_\nu^{ln} \psi_A^n)|A,B\rangle
~=~
\langle A,B|(\bsi_A^k A_\mu^{ml}\psi_A^n)(\bsi_B^m B^{ln}_\nu \psi_B^k)|A,B\rangle
\nonumber\\
&&\hspace{-1mm}
=~\langle A,B|(\bsi_A^k A_\mu^{ln}\psi_A^n)(\bsi_B^m B^{ml}_\nu \psi_B^k)|A,B\rangle
+if^{abc}\langle A,B|A^a_\mu B^b_\nu(\bsi_A^k \psi_A^n)(\bsi_B^m (t^c)_{mn}\psi_B^k)|A,B\rangle
\nonumber\\
&&\hspace{-1mm}
=~{1\over N_c}\langle \bsi_A A_\mu\psi_A\rangle_A\langle\bsi_B B_\nu\psi_B\rangle_B
+2if^{abc}\langle\bsi_A^k t^d_{kn}A_\mu^a\psi_A^n\rangle_A\langle\bsi_B^m t^c_{ms} t^d_{sl}B_\nu^b \psi_B^l\rangle_B
\nonumber\\
&&\hspace{-1mm}
+~{i\over N_c}f^{abc}\langle\bsi_A^k A_\mu^a\psi_A^k\rangle_A\langle\bsi_B^m t^c_{ml} B_\nu^b \psi_B^l\rangle_B
\nonumber\\
&&\hspace{-1mm}
=~{1\over N_c}\langle \bsi_A A_\mu\psi_A\rangle_A\langle\bsi_B B_\nu\psi_B\rangle_B
+2if^{abc}\langle A^a_mu(\bsi_A t^d\psi_A)\rangle_A\langle B^b_\nu(\bsi_B t^ct^d\psi_B)\rangle_B
\nonumber\\
&&\hspace{44mm}
+~{i\over N_c}f^{abc}\langle A^a_\mu(\bsi_A \psi_A)\rangle_A\langle B^b_\nu(\bsi_Bt^c \psi_B)\rangle_B
\nonumber\\
&&\hspace{-1mm}
=~{1\over N_c}\langle \bsi_A A_\mu\psi_A\rangle_A\langle\bsi_B B_\nu\psi_B\rangle_A
+2i{f^{abc}\over N_c^2-1}\langle (\bsi_A A_j\psi_A)\rangle_A\langle \bsi_B t^ct^aB^b_j\psi_B\rangle_B
\nonumber\\
&&\hspace{-1mm}
=~-{1\over N_c(N_c^2-1)}\langle\bsi_A A_\mu\psi_A\rangle_A\langle\bsi_B B_\nu\psi_B\rangle_B  
\label{dob3}
\end{eqnarray}
so effectively  $\Xi_1'$ reduces to
\begin{eqnarray}
&&\hspace{-1mm}
\Xi'_{1}~
\nonumber\\
&&\hspace{-1mm}
=~-{\slashed{p}_1\over s}\gamma^i{B_i\over\beta}\psi_A
+{1\over s^2}\Big({\slp_1\over\beta}\slP^B_\perp{\slp_2\over\alpha}+{\slp_2\over\alpha}\slP^B_\perp{\slp_1\over\beta}\Big)\gamma_iB^i\psi_A
-~{2\over s^2}\slp_2\slp_1B_\star \psi_A+{2\over s^2}\slp_2B_\star\slB_\perp{1\over\alpha^2}\psi_A
\nonumber\\
&&\hspace{22mm}
=~-{\slashed{p}_1\over s}\gamma^i{B_i\over\beta}\psi_A-{1\over s\beta}\slP^B_\perp\slB_\perp{1\over\alpha}\psi_A
+{2\over s^2}\slp_2B_\star\slB_\perp{1\over\alpha^2}\psi_A
\nonumber\\
&&\hspace{44mm}
+~{1\over s^2}\slp_2\slp_1\big[{1\over\beta},\slP^B_\perp\big]_-{1\over\alpha}\psi_A
-~{2\over s^2}\slp_2\slp_1B_\star \psi_A
\nonumber\\
\label{dob4}
\end{eqnarray}
where $\big[{1\over\beta},\slP^B_\perp\big]_-$ is a commutator and $P^B_i=i\partial_i+B_i$.

The Fierz transformation (\ref{fierz}) yields (cf. Eq. (\ref{chus}))
\begin{eqnarray}
&&\hspace{-1mm}
\cheW_{\mu\nu}^{(1\Xi'_1)}(x)~=~{N_c\over s}\langle A,B|[\bsi_A(x)\gamma_\mu\psi_B(x)\big]\big[\bsi_B(0)\gamma_\nu\Xi'_{1}(0)\big]~+~\mu\leftrightarrow\nu
|A,B\rangle~+~x\leftrightarrow 0
\label{dob5}\\
&&\hspace{-1mm}
=~g_{\mu\nu}\chU^{(1'a)}(x)
+(\delta_\mu^\alpha\delta_\nu^\beta+\delta_\nu^\alpha\delta_\mu^\beta-g_{\mu\nu}g^{\alpha\beta})
\chU_{\alpha\beta}^{(1'b)}(x)
+(\delta_\mu^\alpha\delta_\nu^\beta+\delta_\nu^\alpha\delta_\mu^\beta-\half g_{\mu\nu}g^{\alpha\beta})
\chU_{\alpha\beta}^{(1'c)}(x)
\nonumber
\end{eqnarray}
where
\begin{eqnarray}
&&\hspace{-1mm}
\chU^{(1'a)}(x)~=~\langle A,B|-{N_c\over 2s}\big[\bar\psi_A(x)\Xi'_{1}(0)\big]\big[\bar\psi_B(0)\psi_B(x)\big]
\label{dob6}\\
&&\hspace{22mm}
+~
{1\over 2s}\big[\bar\psi_A(x)\gamma_5\Xi'_{1}(0)\big]\big[\bar\psi_B(0)\gamma_5\psi_B(x)\big]|A,B\rangle~+~x\leftrightarrow 0
\nonumber\\
&&\hspace{-1mm}
\chU^{(1'b)}_{\alpha\beta}(x)~=-~{N_c\over 4s}
\langle A,B|\Big(\big[\bar\psi_A(x)\gamma_\alpha\Xi'_{1}(0)\big]\big[\bar\psi_B(0)\gamma_\beta\psi_B(x)\big]
\nonumber\\
&&\hspace{22mm}
+~\big[\bar\psi_A(x)\gamma_\alpha\gamma_5\Xi'_{1}(0)\big]\big[\bar\psi_B(0)\gamma_\beta\gamma_5\psi_B(x)\big]
\Big)|A,B\rangle~+~\alpha\leftrightarrow\beta~+~x\leftrightarrow 0
\nonumber\\
&&\hspace{-1mm}
\chU^{(1'c)}_{\alpha\beta}(x)~=~{N_c\over 4s}\langle A,B|\big[\bar\psi_A(x)\sigma_{\alpha\xi}\Xi'_{1}(0)\big]
\big[\bar\psi_B(0)\sigma_\beta^{~\xi}\psi_B(x)\big]|A,B\rangle~+~\alpha\leftrightarrow\beta~+~x\leftrightarrow 0
\nonumber
 \end{eqnarray}

First, it is easy to see that $\chU^{(1'a)}(x)$ is $\sim g_{\mu\nu}{m_\perp^4\over s^2}$ or less.

\subsubsection{Term $\chU^{(1'b)}$}
From Eq. (\ref{dob4}) we get
\begin{eqnarray}
&&\hspace{-1mm}
\chU^{(1'b)}_{\alpha\beta}~=~-{N_c\over 4s}\langle A,B|\big[\bsi_A(x)\gamma_\alpha\Xi_{1}(0)\big]\big[\bsi_B(0)\gamma_\beta\psi_B(x)\big]
\label{dob7}\\
&&\hspace{-1mm}
+~\big(\psi(0)\otimes\psi(x)\leftrightarrow\gamma_5\psi(0)\otimes\gamma_5\psi(x)\big)|A,B\rangle+\alpha\leftrightarrow\beta+x\leftrightarrow 0
\nonumber\\
&&\hspace{-1mm}
=~{N_c\over 4s^2}\langle A,B|\big[
\bsi_A(x)\gamma_\alpha\Big(\slashed{p}_1\gamma^i{B_i\over\beta}
+{1\over \alpha_q}{1\over \beta}\slP_\perp\slB
\nonumber\\
&&\hspace{-1mm}
-~{1\over \alpha_qs}\slashed{p}_2\slashed{p}_1\big(\big[{1\over \beta},\slP^B_\perp\big]_-\slB-2B_\star\big)
-{2\over \alpha_q^2s}\slp_2B_\star\slB\Big)(0)\psi_A(0)\big]
\big[\bsi_B(0)\gamma_\beta\psi_B(x)\big]]|A,B\rangle
\nonumber\\
&&\hspace{-1mm}
+~\big(\psi(0)\otimes\psi(x)\leftrightarrow\gamma_5\psi(0)\otimes\gamma_5\psi(x)\big)+\alpha\leftrightarrow\beta
+\big(x\leftrightarrow 0,\alpha_q\leftrightarrow -\alpha_q\big)
\nonumber
\end{eqnarray}
(in the few fromulas involving $B_\star$ the notation $\slB$ still means $B^i\gamma_i$). 
Let us demonstrate that the terms in the fourth line of the r.h.s are negligible. First,
\begin{eqnarray}
&&\hspace{-1mm}
{1\over s^3}\big[\bsi_A(x)\gamma_\alpha\slashed{p}_2\slashed{p}_1\big(\big[{1\over \beta},\slP^B_\perp\big]_-\slB
-2B_\star\big)(0)\psi_A(0)\big]
\big[\bsi_B(0)\gamma_\beta\psi_B(x)\big]
\label{dob8}\\
&&\hspace{-1mm}
=~{1\over 2\alpha_qs^3}\Big(\big[\bsi_A(x)\gamma_{\alpha_\perp}\slashed{p}_2\slashed{p}_1\big(\big[{1\over \beta},\slP^B_\perp\big]_-\slB-2B_\star\big)(0)\psi_A(0)\big]
\nonumber\\
&&\hspace{-1mm}
+2p_{2\alpha} \big[\bsi_A(x)\slashed{p}_1\big(\big[{1\over \beta},\slP^B_\perp\big]_-\slB-2B_\star\big)(0)\psi_A(0)\big]\Big)\big[\bsi_B(0)\gamma_\beta\psi_B(x)\big]
\nonumber
\end{eqnarray}
If index $\beta$ is transverse, the first term in the r.h.s. is $\sim q^\perp_\alpha q^\perp_\beta {m_\perp^2\over s^2}$ 
(after Fourier transformation (\ref{furie})) 
and the second is $\sim p_{2\alpha}q^\perp_\beta {m_\perp^4\over s^3}$. If the index $\beta$ is longitudinal, the first term
is $\sim q^\perp_\alpha p_{2\beta} {m_\perp^2\over s^2}$ and the second is $\sim p_{2\alpha} p_{2\beta}{m_\perp^4\over s^3}$
so all these terms are negligible in comparison to those in Eq. (\ref{pc}).
Second,
\begin{eqnarray}
&&\hspace{-1mm}
{1\over s^3}\big[\bsi_A(x)\gamma_\alpha\slashed{p}_2B_\star\slB(0)\psi_A(0)\big]
\big[\bsi_B(0)\gamma_\beta\psi_B(x)\big]
~
=~{1\over 2\alpha_qs^3}\Big(\big[\bsi_A(x)\gamma_{\alpha_\perp}\slashed{p}_2B_\star\slB(0)\psi_A(0)\big]
\nonumber\\
&&\hspace{-1mm}
+{2\over s}p_{2\alpha} \big[\bsi_A(x)\slashed{p}_1\slashed{p}_2B_\star\slB(0)\psi_A(0)\big]\Big)
\big[\bsi_B(0)\gamma_\beta\psi_B(x)\big]
\label{dob9}
\end{eqnarray}
Again, if index $\beta$ is transverse, the first term in the r.h.s. is $\sim q^\perp_\alpha q^\perp_\beta {m_\perp^2\over s^2}$ 
and the second is $\sim p_{2\alpha}q^\perp_\beta {m_\perp^4\over s^3}$, and if  the index $\beta$ is longitudinal, the first term
is $\sim q^\perp_\alpha p_{2\beta} {m_\perp^2\over s^2}$ and the second is $\sim p_{2\alpha} p_{2\beta}{m_\perp^4\over s^3}$,
so all these terms are negligible in comparison to those in Eq. (\ref{pc}).

We get
\begin{eqnarray}
&&\hspace{-1mm}
\chU^{(1'b)}_{\alpha\beta}(x)~=~{N_c\over 4s^2}\langle A,B|
\Big(\big[
\bsi_A(x)\gamma_\alpha\slashed{p}_1\gamma^i{1\over\beta}B_i\psi_A(0)\big]
\label{dob10}\\
&&\hspace{16mm}
+~{1\over \alpha_q}\big[\bsi_A(x)\gamma_\alpha{1\over \beta}\slP_\perp\slB\psi_A(0)\big]
\Big)\big[\bsi_B(0)\gamma_\beta\psi_B(x)\big]|A,B\rangle
\nonumber\\
&&\hspace{16mm}
+~\big(\psi(0)\otimes\psi(x)\leftrightarrow\gamma_5\psi(0)\otimes\gamma_5\psi(x)\big)+\alpha\leftrightarrow\beta
+\big(x\leftrightarrow 0,\alpha_q\leftrightarrow -\alpha_q\big)
\nonumber
\end{eqnarray}
Similar analysis shows that  the only non-negligible contribution is $\sim{p_{1\alpha}p_{2\beta}\over s^2}$ so separating color-singlet contributions
we obtain
\begin{eqnarray}
&&\hspace{-1mm}
\chU^{(1'b)}_{\alpha\beta}~
=~{p_{1\alpha}p_{2\beta}\over s^4}\Big\{\Big(\langle
\bsi_A(x)\slp_2\slashed{p}_1\gamma^i\psi_A(0)\rangle\langle\bsi_B\big({1\over\beta}B_i\big)(0)\slp_1\psi_B(x)\rangle
\label{dob11}\\
&&\hspace{10mm}
+~{1\over \alpha_q}\langle\bsi_A(x)\slp_2\slp_\perp\gamma^i\psi_A(0)\rangle\Big)
\langle\bsi_B\big({1\over\beta}B_i\big)(0)\slp_1\psi_B(x)\rangle
\nonumber\\
&&\hspace{10mm}+~{1\over \alpha_q}\langle\bsi_A(x)\slp_2\gamma^j\gamma^i\psi_A(0)\rangle
\times~\langle\bsi_B\big({1\over\beta}P^B_jB_i\big)(0)\slp_1\psi_B(x)
\rangle\Big\}
\nonumber\\
&&\hspace{10mm}
+~\big(\psi(0)\otimes\psi(x)\leftrightarrow\gamma_5\psi(0)\otimes\gamma_5\psi(x)\big)+\alpha\leftrightarrow\beta
+\big(x\leftrightarrow 0,\alpha_q\leftrightarrow -\alpha_q\big)
\nonumber
\end{eqnarray}
Using Eqs. (\ref{gammas1fild}) and (\ref{14.19}),  
we get
\begin{eqnarray}
&&\hspace{-11mm}
\chU^{(1'b)}_{\alpha\beta}~
=~{p_{1\alpha}p_{2\beta}\over s^4}\Big\{ {s\over 2}\langle
\bsi(x)\gamma^i\psi(0)\rangle_A\langle\bsi\grave\pizb_i(0)\slp_1\psi(x)\rangle_B
+{1\over \alpha_q}\langle\bsi(x)\slp_2p_i\psi(0)\rangle_A
\nonumber\\
&&\hspace{-11mm}
\times~\langle\bsi\acute\pizb_i(0)\slp_1\psi(x)\rangle_B
+~{1\over \alpha_q}\langle\bsi(x)\slp_2\psi(0)\rangle_A
\langle\bsi\slp_1\big(\pizp\pizb+{\epsilon^{ij}\over 2}\pizb_{ij}\gamma_5)(0)\psi(x)
\rangle_B\Big\}
\nonumber\\
&&\hspace{-11mm}
+~\big(\psi(0)\otimes\psi(x)\leftrightarrow\gamma_5\psi(0)\otimes\gamma_5\psi(x)\big)+\alpha\leftrightarrow\beta
+\big(x\leftrightarrow 0,\alpha_q\leftrightarrow -\alpha_q\big)
\label{dob13}
\end{eqnarray}
where we introduced the notations 
\begin{eqnarray}
&&\hspace{-1mm}
\pizb_i(x_\star,x_\perp)~\equiv~{1\over \beta+i\epsilon}B_i(x_\star,x_\perp)~\equiv~-i\!\int_{-\infty}^{x_\star}\! dx'_\star~B_i(x'_\star,x_\perp)
\nonumber\\
&&\hspace{-1mm} 
\pizp\pizb(x_\star,x_\perp)~\equiv~{1\over \beta+i\epsilon}iD^iB_i(x_\star,x_\perp)~=~\!\int_{-\infty}^{x_\star}\! dx'_\star~(\partial^i-iB^i)B_i(x'_\star,x_\perp)
\nonumber\\
&&\hspace{-1mm}
\pizb_{ij}(x_\star,x_\perp)~\equiv~{1\over \beta+\ie}F^{(B)}_{ij}(x_\star,x_\perp)
~=~-i\!\int_{-\infty}^{x_\star}\! dx'_\star~F^{(B)}_{ij}(x'_\star,x_\perp)
\label{pizbez}
\end{eqnarray}
and  $\grave\pizb_i\equiv\pizb_i-i\gamma_5\tilde\pizb_i,~\acute\pizb_i\equiv\pizb_i+i\gamma_5\tilde\pizb_i$  similarly to  Eq. (\ref{agrebi}).  Using parametizations (\ref{frabs1}), (\ref{frabs2}), (\ref{frabs3}) and formula $\graB_i\gamma_5=-i\epsilon_{ij}\graB^j$
we get
\begin{eqnarray}
&&\hspace{-1mm}
\chU^{(1'b)}_{\alpha\beta}~
=~{g^\parallel_{\alpha\beta}\over 2\alpha_qsN_ c}\!\int\! d^2k_\perp
\Big[(k,q-k)_\perp\Big(\{f_1\bar\acf_{1\pizg}+\barf_1\acf_{1\pizg}\}\bar\acf_{1\pizg}
\nonumber\\
&&\hspace{-1mm}
+\alpha_q\{[f_\perp-ig_\perp]\bar\graf_{1\pizg}+[\barf_\perp+i\barg_\perp)]\graf_{1\pizg}\}\Big)
+~m^2\{f_1(\alpha_q,k_\perp)
+\barf_1[f_{2\pizg}+f_{3\pizg}]\}\Big],
\label{dob15}
\end{eqnarray}

cf. Eq. (\ref{w1bq}).

\subsubsection{Term $\chU^{(1'c)}$}
Next,
\begin{eqnarray}
&&\hspace{-5mm}
\chU^{(1'c)}(x)~=~{N_c\over 4s}\langle A,B|\big[\bar\psi_A(x)\sigma_{\alpha\xi}\Xi'_{1}(0)\big]\big[\bar\psi_B(0)\sigma_\beta^{~\xi}\psi_B(x)|A,B\rangle~+~\alpha\leftrightarrow\beta~+~x\leftrightarrow 0
\nonumber\\
&&\hspace{7mm}
=~-{N_c\over 4s^2}\langle A,B|\big[
\bsi_A(x)\sigma_{\alpha\xi}\Big(\slashed{p}_1\gamma^i{B_i\over\beta}
+{1\over \alpha_q}{1\over \beta}\slP_\perp\slB
-~{1\over \alpha_qs}\slashed{p}_2\slashed{p}_1\big(\big[{1\over \beta},\slP^B_\perp\big]_-\slB
-2B_\star\big)
\nonumber\\
&&\hspace{7mm}
-~{2\over \alpha_q^2s}\slp_2B_\star\slB_\perp\Big)(0)\psi_A(0)\big]
\big[\bsi_B(0)\sigma_\beta^{~\xi}\psi_B(x)\big]|A,B\rangle
+~\alpha\leftrightarrow\beta
+\big(x\leftrightarrow 0,\alpha_q\leftrightarrow -\alpha_q\big)
\nonumber\\
\label{dob16}
\end{eqnarray}
Let us again demonstrate that the last two terms in curly brackets are negligible. Ignoring the transverse factors $p_i$, $B_i$ 
and $B_\star$ which cannot produce factor $s$, we obtain the following estimates
\begin{eqnarray}
&&\hspace{-1mm}
{1\over s^3}\sigma_{\alpha_\perp\xi}\slp_2(\slp_1)\otimes\sigma_{\beta_\perp}^{~\xi}
~=~O\big({g^\perp_{\alpha\beta}\over s^2}\big),~~~~
\label{dob17}\\
&&\hspace{-1mm}
{p_{2\beta}\over s^4}\sigma_{\alpha_\perp\xi}\slp_2(\slp_1)\otimes\sigma_{\bu}^{~\xi}
~=~O\big({p_{2\beta}q^\perp_\alpha\over s^2}\big),~~~~
{p_{1\beta}\over s^4}\sigma_{\alpha_\perp\xi}\slp_2(\slp_1)\otimes\sigma_{\star}^{~\xi}
~=~O\big({p_{1\beta}q^\perp_\alpha\over s^3}\big)
\nonumber\\
&&\hspace{-1mm}
{p_{1\alpha}\over s^4}\sigma_{\star\xi}\slp_2(\slp_1)\otimes\sigma_{\beta_\perp}^{~\xi}~=~O\big({p_{1\alpha}q^\perp_\beta\over s^2}\big),~~~~
{p_{2\alpha}\over s^4}\sigma_{\bu\xi}\slp_2(\slp_1)\otimes\sigma_{\beta_\perp}^{~\xi}~=~O\big({p_{2\alpha}q^\perp_\beta\over s^2}\big),
\nonumber\\
&&\hspace{-1mm}
{p_{1\alpha}p_{1\beta}\over s^5}\sigma_{\star\xi}\slp_2(\slp_1)\otimes\sigma_{\star}^{~\xi}~=~0,~~~~
{p_{1\alpha}p_{2\beta}\over s^5}\sigma_{\star\xi}\slp_2(\slp_1)\otimes\sigma_{\bu}^{~\xi}~=~O\big({p_{1\alpha}p_{2\beta}\over s^3}\big),~~~~
\nonumber\\
&&\hspace{-1mm}
{p_{2\alpha}p_{1\beta}\over s^5}\sigma_{\bu\xi}\slp_2(\slp_1)\otimes\sigma_{\star}^{~\xi}~=~O\big({p_{2\alpha}p_{1\beta}\over s^3}\big),~~~~
{p_{2\alpha}p_{2\beta}\over s^5}\sigma_{\bu\xi}\slp_2(\slp_1)\otimes\sigma_{\bu}^{~\xi}~=~O\big({p_{2\alpha}p_{2\beta}\over s^3}\big)
\nonumber
\end{eqnarray}
where the factors $(\slp_1)$ means that inclusion (or non-inclusion) of $\slp_1$ does not change the power of $s$ in the projectile TMDs. 
Similarly to Eq. (\ref{dob10}) we get
\begin{eqnarray}
&&\hspace{-1mm}
\chU^{(1'c)}_{\alpha\beta}(x)~
\nonumber\\
&&\hspace{-1mm}
=~-{N_c\over 4s^2}\langle A,B|\Big(\big[
\bsi_A(x)\sigma_{\alpha\xi}\slashed{p}_1\gamma^i\pizb_i\psi_A(0)\big]
+{1\over \alpha_q}\big[\bsi_A(x)\sigma_{\alpha\xi}{1\over \beta}\slP_\perp\slB\psi_A(0)\big]
\Big)
\nonumber\\
&&\hspace{-1mm}
\times~\big[\bsi_B(0)\sigma_\beta^{~\xi}\psi_B(x)\big]|A,B\rangle
+\alpha\leftrightarrow\beta
+\big(x\leftrightarrow 0,\alpha_q\leftrightarrow -\alpha_q\big)
\label{dob18}
\end{eqnarray}

For the first term, 
from Eq. (\ref{mainfla3}) we get 
\begin{eqnarray}
&&\hspace{-1mm}
{1\over 4s^2}i\sigma_{\alpha\xi}\sigma_{\bu i}\otimes\sigma_\beta^{~\xi}\pizb^i~+~{\alpha\leftrightarrow\beta}
\label{dob19}\\
&&\hspace{-1mm}
=~{1\over 4s^2}\Big[-{2\over s}\sigma_{\bu\ast}\otimes
\big(\sigma_{\bu\beta_\perp} \pizb_\alpha-{g^\perp_{\alpha\beta}\over 2}\sigma_{\bu i}\pizb^i\big)
+{i\over 2}g_{\alpha\beta}\otimes\sigma_{\bu i}\pizb^i+{g_{\alpha\beta}\over s}\sigma_{\star\bu} \otimes\sigma_{\bu i} \pizb^i
\nonumber\\
&&\hspace{22mm}
+~i\otimes\big(\sigma_{\bu\beta_\perp}\pizb_\alpha-{g_{\alpha\beta}^\perp\over 2}\sigma_{\bu i}\pizb^i\big)
~+~{\alpha\leftrightarrow\beta}\Big]
\nonumber
\end{eqnarray}
Next, due to the definition (\ref{dob5}), $U^{1'c}_{\alpha\beta}$ will be multiplied by the traceless tensor 
$(\delta_\mu^\alpha\delta_\nu^\beta+\delta_\nu^\alpha\delta_\mu^\beta-\half g_{\mu\nu}g^{\alpha\beta})$ so 
the contributions to $U^{1'c}_{\alpha\beta}$ proportional to $g_{\alpha\beta}$ can be ignored. Effectively, 
\begin{eqnarray}
&&\hspace{-1mm}
{1\over 4s^2}i\sigma_{\alpha\xi}\sigma_{\bu i}\otimes\sigma_\beta^{~\xi}\pizb^i~+~{\alpha\leftrightarrow\beta}
\label{dob20}\\
&&\hspace{-1mm}
=~{1\over 4s^2}\Big[-{2\over s}\sigma_{\bu\ast}\otimes\big(\sigma_{\bu\beta_\perp} \pizb_\alpha-{g^\perp_{\alpha\beta}\over 2}\sigma_{\bu i}\pizb^i\big)
+i\otimes\big(\sigma_{\bu\beta_\perp}\pizb_\alpha-{g_{\alpha\beta}^\perp\over 2}\sigma_{\bu i}\pizb^i\big)
~+~{\alpha\leftrightarrow\beta}\Big]
\nonumber
\end{eqnarray}
and we get the first term in Eq. (\ref{dob18}) in the form
\begin{eqnarray}
&&\hspace{-1mm}
\chU^{(1'c)}_{\alpha\beta(1)}(x)~
\equiv~-{N_c\over 4s^2}\langle A,B|\big[
\bsi_A(x)\sigma_{\alpha\xi}\slashed{p}_1\gamma^i{B_i\over\beta}\psi_A(0)\big]
\label{dob21}\\
&&\hspace{16mm}
\times~\big[\bsi_B(0)\sigma_\beta^{~\xi}\psi_B(x)\big]|A,B\rangle
+\alpha\leftrightarrow\beta+x\leftrightarrow 0,
+\big(x\leftrightarrow 0,\alpha_q\leftrightarrow -\alpha_q\big)
\nonumber\\
&&\hspace{5mm}
=~\Big({2\over s^3}\langle\bsi(x)\sigma_{\star\bu}\psi(0)\rangle_A+{i\over 4s^2}\langle\bsi(x)\psi(0)\rangle_A\Big)
\nonumber\\
&&\hspace{16mm}
\times~\langle\bsi
\big(\sigma_{\bu\beta_\perp} \pizb_\alpha-{g^\perp_{\alpha\beta}\over 2}\sigma_{\bu i}\pizb^i\big)(0)\psi(x)\rangle_B
+\alpha\leftrightarrow\beta+x\leftrightarrow 0
\nonumber
\end{eqnarray}
Using parametrization (\ref{hrabs1}) we get
\begin{eqnarray}
&&\hspace{-1mm}
U^{(1'c)}_{\alpha\beta(1)}(q)~=~{1\over 16\pi^4N_c}\!\int\! dx_\star dx_\bu d^2x_\perp~e^{-i\alpha_qx_\bu-i\beta_qx_\star+i(q,x)_\perp}
U^{(1'c)}_{\alpha\beta(1)}~
\label{dob22}\\
&&\hspace{1mm}
=~-{1\over sN_c}\!\int\! d^2k_\perp~\big[(q-k)^\perp_\alpha(q-k)^\perp_\beta+{g^\perp_{\alpha\beta}\over 2}(q-k)_\perp^2\big]\{[h+ie]\barh_{1\pizg}+[\barh-i\bare]h_{1\pizg}\}
\nonumber\end{eqnarray}

To get the second term in Eq. (\ref{dob18}) we need a table of estimates similar to Eq. (\ref{dob17})
\begin{eqnarray}
&&\hspace{-1mm}
{1\over s^2}\sigma_{\alpha_\perp\xi}\otimes\sigma_{\beta_\perp}^{~\xi}
~=~{2\over s^3}\sigma_{\alpha_\perp\ast}\otimes\sigma_{\beta_\perp\bu},~~~~
\nonumber\\
&&\hspace{-1mm}
{p_{2\beta}\over s^4}\sigma_{\alpha_\perp\xi}\slp_1\slp_2\otimes\sigma_{\bu}^{~\xi}
~=~O\big({p_{2\beta}q^\perp_\alpha\over s^2}\big),~~~~
{p_{1\beta}\over s^4}\sigma_{\alpha_\perp\xi}\slp_1\slp_2\otimes\sigma_{\star}^{~\xi}
~=~O\big({p_{1\beta}q^\perp_\alpha\over s^2}\big)
\nonumber\\
&&\hspace{-1mm}
{p_{1\alpha}\over s^4}\sigma_{\star\xi}\slp_1\slp_2\otimes\sigma_{\beta_\perp}^{~\xi}~=~O\big({p_{1\alpha}q^\perp_\beta\over s^2}\big),~~~~
{p_{2\alpha}\over s^4}\sigma_{\bu\xi}\slp_1\slp_2\otimes\sigma_{\beta_\perp}^{~\xi}~=~O\big({p_{2\alpha}q^\perp_\beta\over s^2}\big),
\nonumber\\
&&\hspace{-1mm}
{p_{1\alpha}p_{1\beta}\over s^5}\sigma_{\star\xi}\slp_1\slp_2\otimes\sigma_{\star}^{~\xi}~=~O\big({p_{1\alpha}p_{1\beta}\over s^3}\big),~~~~
{p_{1\alpha}p_{2\beta}\over s^5}\sigma_{\star\xi}\slp_1\slp_2\otimes\sigma_{\bu}^{~\xi}
~=~{p_{1\alpha}p_{2\beta}\over s^4}\sigma_{\star i}\otimes\sigma_{\bu}^{~i}
,~~~~
\nonumber\\
&&\hspace{-1mm}
{p_{2\alpha}p_{1\beta}\over s^5}\sigma_{\bu\xi}\slp_1\slp_2\otimes\sigma_{\star}^{~\xi}~=~O\big({p_{2\alpha}p_{1\beta}\over s^3}\big),~~~~
{p_{2\alpha}p_{2\beta}\over s^5}\sigma_{\bu\xi}\slp_1\slp_2\otimes\sigma_{\bu}^{~\xi}~=~O\big({p_{2\alpha}p_{2\beta}\over s^3}\big)
\label{dob23}
\end{eqnarray}
Combining these equations and equations from  table (\ref{dob17}) (with $\slp_1$ included) we see that 
\begin{eqnarray}
&&\hspace{-1mm}
{1\over s^2}\sigma_{\alpha\xi}\otimes\sigma_{\beta}^{~\xi}+\alpha\leftrightarrow\beta
~=~{2\over s^3}\sigma_{\alpha_\perp\ast}\otimes\sigma_{\beta_\perp\bu}
+{4p_{1\alpha}p_{2\beta}\over s^4}\sigma_{\star i}\otimes\sigma_{\bu}^{~i}+\alpha\leftrightarrow\beta
\label{dob24}\\
&&\hspace{-1mm}
=~{2\over s^3}\sigma_\star^{~ i}\otimes\sigma_\bu^{~ j}\Big(g^\perp_{\alpha i}g^\perp_{\beta j}+g^\perp_{\beta i}g^\perp_{\alpha j}
-g^\perp_{\alpha\beta}g_{ij}+g^\perp_{\alpha\beta}g_{ij}+{2p_{1\alpha}p_{2\beta}\over s}g_{ij}+{2p_{2\alpha}p_{1\beta}\over s}g_{ij}\Big)
\nonumber\\
&&\hspace{-1mm}
=~{2\over s^3}\sigma_\star^{~ i}\otimes\sigma_\bu^{~ j}\big(g^\perp_{\alpha i}g^\perp_{\beta j}+g^\perp_{\beta i}g^\perp_{\alpha j}
-g^\perp_{\alpha\beta}g_{ij}+g_{\alpha\beta}g_{ij}\big)
=~-{2\over s^3}\sigma_\star^{~ i}\otimes\sigma_\bu^{~ j}P_{\alpha\beta;ij}\nonumber
\end{eqnarray}
where 
\begin{equation}
P_{\alpha\beta;ij}~\equiv~g^\perp_{\alpha\beta}g_{ij}-g^\perp_{\alpha i}g^\perp_{\beta j}-g^\perp_{\beta i}g^\perp_{\alpha j}
\label{dob25}
\end{equation}
Note that in the last line in Eq. (\ref{dob24}) we dropped term $\sim g_{\alpha\beta}$ since it vanishes after multiplication by
$(\delta_\mu^\alpha\delta_\nu^\beta+\delta_\nu^\alpha\delta_\mu^\beta-\half g_{\mu\nu}g^{\alpha\beta})$. 
Thus, the second term in Eq. (\ref{dob18}) takes the form
\begin{eqnarray}
&&\hspace{-1mm}
\chU^{(1'c)}_{\alpha\beta(2)}(x)~
=~-{N_c\over 4s^2\alpha_q}\langle A,B|\big[\bsi_A(x)\sigma_{\alpha\xi}{1\over \beta}\slP_\perp\slB\psi_A(0)\big]
\label{dob26}\\
&&\hspace{16mm}
\times~\big[\bsi_B(0)\sigma_\beta^{~\xi}\psi_B(x)\big]|A,B\rangle
+\alpha\leftrightarrow\beta
+\big(x\leftrightarrow 0,\alpha_q\leftrightarrow -\alpha_q\big)
\nonumber\\
&&\hspace{-1mm}
=~{N_c\over 2s^3\alpha_q}P_{\alpha\beta;ij}
\langle A,B|\big[\bsi_A(x)\sigma_\star^{~ i}{1\over \beta}\slP_\perp\slB\psi_A(0)\big]
\nonumber\\
&&\hspace{16mm}
\times~\big[\bsi_B(0)\sigma_\bu^{~ j}\psi_B(x)\big]|A,B\rangle
+\big(x\leftrightarrow 0,\alpha_q\leftrightarrow -\alpha_q\big)
\nonumber\\
&&\hspace{-1mm}
=~{N_c\over 2s^3\alpha_q}P_{\alpha\beta;ij}\Big(\langle\bsi(x)\sigma_\star^{~ i}\slp_\perp\gamma_l
\bsi(0)\rangle_A\langle\bsi(0) \sigma_\bu^{~ j}\pizb^l\psi(x)\rangle_B
\nonumber\\
&&\hspace{-1mm}
+~\langle\bsi(x)\sigma_\star^{~ i}\gamma_k\gamma_l\psi(0)\rangle_A
\langle\bsi\sigma_\bu^{~ j}\big({1\over\beta}iD_kB^l\big)(0)\psi(x)\rangle_B\Big)
+\big(x\leftrightarrow 0,\alpha_q\leftrightarrow -\alpha_q\big)
\nonumber
\end{eqnarray}
Using formula (\ref{sigmasigma}) we get 
\begin{equation}
\sigma_{\star i}\gamma_k\gamma_l~=~g_{kl}\sigma_{\star i}+g_{ik}\sigma_{\star l}-g_{il}\sigma_{\star k}
\label{dob27}
\end{equation}
so the second contribution to Eq. (\ref{dob16}) is
\begin{eqnarray}
&&\hspace{-1mm}
\chU^{(1'c)}_{\alpha\beta(2)}(x)~=~{N_c\over 2s^3\alpha_q}P_{\alpha\beta;ij}\langle\bsi(x)\sigma_\star^{~ i}\slp_\perp\gamma_l
\psi(0)\rangle_A\langle\bsi(0) \sigma_\bu^{~ j}\pizb^l(0)\psi(x)\rangle_B
\nonumber\\
&&\hspace{17mm}
=~{N_c\over 2s^3\alpha_q}P_{\alpha\beta;ij}\langle\bsi(x)\big(\sigma_{\star i}p^l+\sigma_{\star l}p_i-g_{il}\sigma_{\star k}p^k\big)\psi(0)\rangle_A
\langle\bsi\pizb^l(0) \sigma_\bu^{~ j}\psi(x)\rangle_B
\nonumber\\
&&\hspace{17mm}+~\big(x\leftrightarrow 0,\alpha_q\leftrightarrow -\alpha_q\big)
\label{dob28}
\end{eqnarray}
and the corresponding Fourier transformation yields
\begin{eqnarray}
&&\hspace{-1mm}
U^{(1'c)}_{\alpha\beta(2)}(q)~=~{1\over 2\alpha_qsN_c}P_{\alpha\beta;ij}
\!\int\! d^2k_\perp
(2k_ik_l+k_\perp^2g_{il})
\nonumber\\
&&\hspace{7mm}
\times~\Big(h_1^\perp(\alpha_q,k_\perp)
\big[(q-k)^j(q-k)^l\barh_{1\pizg}+{g^{jl}\over 2}(q-k)_\perp^2(\barh_{1\pizg}+\barh_{2\pizg})(\beta_q,q_\perp-k_\perp)\big]
\nonumber\\
&&\hspace{7mm}
+~\barh_1^\perp(\alpha_q,k_\perp)
\big[(q-k)^j(q-k)^l h_{1\pizg}(\beta_q,q_\perp-k_\perp)+{g^{jl}\over 2}(h_{1\pizg}+h_{2\pizg})(\beta_q,q_\perp-k_\perp)\big]
\Big)
\nonumber\\
&&\hspace{-1mm}
=~{1\over 2\alpha_qsN_cm^2}\!\int\! d^2k_\perp
\Big(2\calW^\perp_{\alpha\beta}(q,k_\perp)\{h_1^\perp\barh_{1\pizg}+\barh_1^\perp h_{1\pizg}\}
\nonumber\\
&&\hspace{11mm}
-~[g^\perp_{\alpha\beta}k_\perp^2(q-k)_\perp^2+2k_\alpha k_\beta(q-k)_\perp^2]
\{h_1^\perp\barh_{2\pizg}+\barh_1^\perp h_{2\pizg}\}
\Big)
\label{dob29}
\end{eqnarray}
where we used parametrization (\ref{hrabs1}) and defined the notation
\begin{eqnarray}
&&\hspace{-1mm}
\calW^\perp_{\mu\nu}(q_\perp,k_\perp)~\equiv~g_{\mu\nu}^\perp(k,q-k)_\perp^2
-g_{\mu\nu}^\perp k_\perp^2(q-k_\perp)^2
\nonumber\\
&&\hspace{-1mm}
+~[k^\perp_\mu(q-k)^\perp_\nu+\mu\leftrightarrow\nu](k,q-k)_\perp
-k_\perp^2(q-k)^\perp_\mu(q-k)^\perp_\nu-~(q-k_\perp)^2k^\perp_\mu k^\perp_\nu
\label{calweperp}
\end{eqnarray}
It is easy to see that $q^\mu \calW^\perp_{\mu\nu}(q,k_\perp)=0$ and $\calW_i^{\perp i}(q,k_\perp)=0$.

The second term in Eq. (\ref{dob26}) is
\begin{eqnarray}
&&\hspace{-1mm}
\chU^{(1'c)}_{\alpha\beta(3)}(x)~=~{N_c\over 2s^3\alpha_q}P_{\alpha\beta;ij}
\langle\bsi(x)\sigma_\star^{~ i}\gamma_k\gamma_l\psi(0)\rangle_A
\langle\bsi\big({1\over\beta}iD_kB^l\big)(0)\sigma_\bu^{~ j}\psi(x)\rangle_B
\nonumber\\
&&\hspace{5mm}
=~{N_c\over 2s^3\alpha_q}P_{\alpha\beta;ij}\Big(\langle\bsi(x)\sigma_\star^{~ i}\psi(0)\rangle_A\langle\bsi\pizp\pizb(0)\sigma_\bu^{~ j}\psi(x)\rangle_B
\nonumber\\
&&\hspace{11mm}
+~i\langle\bsi(x)\sigma_{\star l}\psi(0)\rangle_A\langle\bsi\pizb^{il}(0)\sigma_\bu^{~ j}\psi(x)\rangle_B\Big)
+\big(x\leftrightarrow 0,\alpha_q\leftrightarrow -\alpha_q\big)
\label{dob30}
\end{eqnarray}
where we used Eq. (\ref{dob27}) and the notations (\ref{pizbez}).

For unpolarized hadrons, Eq. (\ref{dob30}) can re rewritten as 
\begin{eqnarray}
&&\hspace{-1mm}
\chU^{(1'c)}_{\alpha\beta(3)}(x)~
=~{N_c\over 2s^3\alpha_q}P_{\alpha\beta;ij}\langle\bsi(x)\sigma_\star^{~ i}\psi(0)\rangle_A
\label{dob31}\\
&&\hspace{-1mm}
\times~\Big(\langle\bsi(0)\sigma_\bu^{~ j}\pizp\pizb(0)\psi(x)\rangle_B
-i\bsi(0)\sigma_{\bu m}\pizb^{mj}(0)\psi(x)\rangle_B\Big)
+\big(x\leftrightarrow 0,\alpha_q\leftrightarrow -\alpha_q\big)
\nonumber
\end{eqnarray}
and, using  parametrizations (\ref{hrabs2}) and (\ref{hrabs3}), we obtain
\begin{eqnarray}
&&\hspace{-1mm}
U^{(1'c)}_{\alpha\beta(3)}(q)~=~-{1\over 2\alpha_qsN_c}P_{\alpha\beta;ij}\!\int\! d^2k_\perp
k^i(q-k)^j
\label{dob32}\\
&&\hspace{-1mm}
\times~\Big(h_1^\perp(\alpha_q,k_\perp)[\barh_{3\pizg}-i\barh_{4\pizg}](\beta_q,q_\perp-k_\perp)
+~\barh_1^\perp(\alpha_q,k_\perp)[h_{3\pizg}-ih_{4\pizg}](\beta_q,q_\perp-k_\perp)\Big)
\nonumber\\
&&\hspace{-1mm}
=~{1\over 2\alpha_qsN_c}\!\int\! d^2k_\perp\big[k^\perp_\alpha (q-k)^\perp_\beta
+k^\perp_\beta (q-k)^\perp_\alpha+(k,q-k)_\perp g^\perp_{\alpha\beta}\big]
\nonumber\\
&&\hspace{-1mm}
\times~\Big(h_1^\perp(\alpha_q,k_\perp)[\barh_{3\pizg}-i\barh_{4\pizg}](\beta_q,q_\perp-k_\perp)
+~\barh_1^\perp(\alpha_q,k_\perp)[h_{3\pizg}-ih_{4\pizg}](\beta_q,q_\perp-k_\perp)\Big)
\nonumber
\end{eqnarray}
Thus, from Eqs. (\ref{dob5}), (\ref{dob15}), (\ref{dob22}), (\ref{dob29}), and (\ref{dob32}) we get 
\begin{eqnarray}
&&\hspace{-1mm}
W_{\mu\nu}^{(1\Xi'_1)}(q)
=~{1\over(2\pi)^4N_c}\!\int\! dx_\bu dx_\star d^2x_\perp ~e^{-i\alpha_qx_\bu-i\beta_q x_\star+i(q,x)_\perp}
\cheW^{(1')}_{\mu\nu}(x)~=~{1\over \alpha_qsN_c}
\nonumber\\
&&\hspace{-1mm}
\times~\!\int\! d^2k_\perp\Big[-g^\perp_{\mu\nu}(k,q-k)_\perp\Big(
\{f_1\bar\acf_{1\pizg}+\barf_1\acf_{1\pizg}\}
+\alpha_q\{[f_\perp-ig_\perp]\bar\graf_{1\pizg}+[\barf_\perp+i\barg_\perp)]\graf_{1\pizg}\}\Big)
\nonumber\\
&&\hspace{-1mm}
-~g^\perp_{\mu\nu}m^2\{f_1[\barf_{2\pizg}+\barf_{3\pizg}]
+\barf_1[f_{2\pizg}+f_{3\pizg}]\}
\nonumber\\
&&\hspace{-1mm}
-~\alpha_q\big[2(q-k)^\perp_\mu (q-k)^\perp_\nu+g^\perp_{\mu\nu}(q-k)_\perp^2\big]\{[h+ie]\barh_{1\pizg}+[\barh-i\bare]h_{1\pizg}\}
\nonumber\\
&&\hspace{-1mm}
+~{1\over m^2}
\Big(2\calW^\perp_{\mu\nu}(q,k_\perp)\{h_1^\perp\barh_{1\pizg}+\barh_1^\perp h_{1\pizg}\}
\nonumber\\
&&\hspace{-1mm}
-~[g^\perp_{\mu\nu}k_\perp^2(q-k)_\perp^2+2k_\mu k_\nu(q-k)_\perp^2]
\{h_1^\perp\barh_{2\pizg}+\barh_1^\perp h_{2\pizg}\}
\Big)
\label{dob33}\\
&&\hspace{-1mm}
+~\big[k^\perp_\mu (q-k)^\perp_\nu
+k^\perp_\nu (q-k)^\perp_\mu+(k,q-k)_\perp g^\perp_{\mu\nu}\big]
\{h_1^\perp[\barh_{3\pizg}-i\barh_{4\pizg}]
+~\barh_1^\perp[h_{3\pizg}-ih_{4\pizg}]\}\Big]
\nonumber
\end{eqnarray}
%

\subsection{Terms coming from $\Bxi'_{1}$,  $\Xi'_2$, and $\Bxi'_2$}
Replacing  $\Xi_1\rightarrow \Xi'_1$ and $\Bxi_1\rightarrow \Bxi'_1$  in Eq. (\ref{procc}) 
we see that the contribution of terms with $\Bxi'_1(x)$ is a complex conjugate of the contribution (\ref{dob33}) of $\Xi'_1(0)$ so 
\begin{eqnarray}
&&\hspace{-1mm}
W^{(1\Xi'_1+1\Bxi'_1)}_{\mu\nu}(q)~=~{1\over(2\pi)^4N_c}\!\int\! dx_\bu dx_\star d^2x_\perp ~e^{-i\alpha_qx_\bu-i\beta_q x_\star+i(q,x)_\perp}
\big[\cheW^{(1\Xi'_1)}_{\mu\nu}(x)+\cheW^{(1\Bxi'_1)}_{\mu\nu}(x)\big]
\nonumber\\
&&\hspace{-1mm}
=~g_{\mu\nu}\big[\chU^{(1'a)}(q)+\chU^{(2'a)}(q)\big]
+(\delta_\mu^\alpha\delta_\nu^\beta+\delta_\nu^\alpha\delta_\mu^\beta-g_{\mu\nu}g^{\alpha\beta})
\big[\chU_{\alpha\beta}^{(1'b)}(q)+\chU_{\alpha\beta}^{(2'b)}(q)\big]
\nonumber\\
&&\hspace{-1mm}
+~(\delta_\mu^\alpha\delta_\nu^\beta+\delta_\nu^\alpha\delta_\mu^\beta-\half g_{\mu\nu}g^{\alpha\beta})
\big[\chU_{\alpha\beta}^{(1'c)}(q)+\chU_{\alpha\beta}^{(2'c)}(q)\big]
\nonumber\\
&&\hspace{-1mm}
~=~{2\over \alpha_qsN_c}\!\int\! d^2k_\perp\Big\{
-g^\perp_{\mu\nu}\Big[(k,q-k)_\perp\big(
\{f_1\Re\bar\acf_{1\pizg}+\barf_1\Re\acf_{1\pizg}\}
\label{dobresult}\\
&&\hspace{15mm}
+~\alpha_q\{f_\perp\Re\bar\graf_{1\pizg}+\barf_\perp\Re\graf_{1\pizg}\}+\alpha_q\{g_\perp\Im\bar\graf_{1\pizg}-\barg_\perp\Im\graf_{1\pizg}\} \big)
\nonumber\\
&&\hspace{-1mm}
+~m^2\big(\{f_1\Re\barf_{2\pizg}+\barf_1\Re f_{2\pizg}\}+\{f_1\Re\barf_{3\pizg}+\barf_1\Re f_{3\pizg}\}\Big]
\nonumber\\
&&\hspace{-1mm}
-~\alpha_q\big[2(q-k)^\perp_\mu (q-k)^\perp_\nu+g^\perp_{\mu\nu}(q-k)_\perp^2\big]
\big(\{h\Re\barh_{1\pizg}+\barh\Re h_{1\pizg}\}+\{\bare\Im h_{1\pizg}-e\Im \barh_{1\pizg}\}\big)
\nonumber\\
&&\hspace{-1mm}
+~
\Big({2\over m^2}\calW^\perp_{\mu\nu}\{h_1^\perp\Re\barh_{1\pizg}+\barh_1^\perp\Re h_{1\pizg}\}
-[g^\perp_{\mu\nu}k_\perp^2+2k_\mu k_\nu]{(q-k)_\perp^2\over m^2}
\{h_1^\perp\Re\barh_{2\pizg}+\barh_1^\perp\Re h_{2\pizg}\}
\Big)
\nonumber\\
&&\hspace{-1mm}
+~\big[(k,q-k)_\perp g^\perp_{\mu\nu}
+~k^\perp_\mu (q-k)^\perp_\nu
+k^\perp_\nu (q-k)^\perp_\mu\big]
\nonumber\\
&&\hspace{15mm}
\times~
\big(\{h_1^\perp\Re\barh_{3\pizg}+\barh_1^\perp\Re h_{3\pizg}\}+\{h_1^\perp\Im\barh_{4\pizg}
+\barh_1^\perp\Im h_{4\pizg}\}\big)\Big\}
\nonumber
\end{eqnarray}
where $\calW^\perp_{\mu\nu}(q,k_\perp)$ is defined in Eq. (\ref{calweperp}).

The corresponding contribution of terms  coming from $\Xi'_2$ and $\Bxi'_2$
\begin{eqnarray}
&&\hspace{-1mm}
\cheW_{\mu\nu}^{(1\Xi'_2)}(x)+\cheW_{\mu\nu}^{1(\Bxi'_2)}(x)~=~{N_c\over s}\langle A,B|
[\bar\psi_A(x)\gamma_\mu\psi_B(x)\big]\big[\Bxi'_2(0)\gamma_\nu\psi_A(0)\big]
\label{doba1}\\
&&\hspace{40mm}
+~[\bar\psi_A(x)\gamma_\mu \Xi'_2(x)\big]\big[\bar\psi_B(0)\gamma_\nu\psi_A(0)\big]|A,B\rangle~+~x\leftrightarrow 0
\nonumber
\end{eqnarray}
is obtained from Ea. (\ref{dobresult}) by the projectile$\leftrightarrow$target replacement (\ref{protareplace})
\begin{eqnarray}
&&\hspace{-3mm}
W^{(1\Xi'_2+1\Bxi'_2)}_{\mu\nu}(q)~=~{1\over(2\pi)^4N_c}\!\int\! dx_\bu dx_\star d^2x_\perp ~e^{-i\alpha_qx_\bu-i\beta_q x_\star+i(q,x)_\perp}
\big[\cheW^{(1\Xi'_2)}_{\mu\nu}(x)+\cheW^{(1\Bxi'_2)}_{\mu\nu}(x)\big]
\nonumber\\
&&\hspace{-1mm}
~=~{2\over \beta_qsN_c}\!\int\! d^2k_\perp\Big\{
-g^\perp_{\mu\nu}\Big[(k,q-k)_\perp\big(
\{\Re\bar\acf_{1\pizg}f_1+\Re\acf_{1\pizg}\barf_1\}+~\beta_q\{\Re\bar\graf_{1\pizg}f_\perp+\Re\graf_{1\pizg}\barf_\perp\}
\nonumber\\
&&\hspace{15mm}
+~\beta_q\{\Im\bar\graf_{1\pizg}g_\perp-\Im\graf_{1\pizg}\barg_\perp\} \big)
+~m^2\big(\{\Re\barf_{2\pizg}f_1+\Re f_{2\pizg}\barf_1\}+\{\Re\barf_{3\pizg}f_1+\Re f_{3\pizg}\barf_1\}\Big]
\nonumber\\
&&\hspace{-1mm}
-~\beta_q\big[2k^\perp_\mu k^\perp_\nu+g^\perp_{\mu\nu}k_\perp^2\big]
\big(\{\Re h_{1\pizg}\barh\Re\barh_{1\pizg}h\}+\{\Im h_{1\pizg}\bare-\Im \barh_{1\pizg}e\}\big)
\nonumber\\
&&\hspace{-1mm}
+~{1\over m^2}
\Big(2\calW^\perp_{\mu\nu}(q,k_\perp)\{\Re h_{1\pizg}\barh_1^\perp+\Re\barh_{1\pizg}h_1^\perp\}
\nonumber\\
&&\hspace{15mm}
-~[g^\perp_{\mu\nu}k_\perp^2(q-k)_\perp^2+2(q-k)_\mu(q-k)_\nu k_\perp^2]
\{\Re h_{2\pizg}\barh_1^\perp+\Re\barh_{2\pizg}h_1^\perp\}
\Big)
\nonumber\\
&&\hspace{-1mm}
+~\big[(k,q-k)_\perp g^\perp_{\mu\nu}+~k^\perp_\mu (q-k)^\perp_\nu
+k^\perp_\nu (q-k)^\perp_\mu\big]
\nonumber\\
&&\hspace{15mm}
\times~\big(\{\Re\barh_{3\pizg}h_1^\perp+\Re h_{3\pizg}\barh_1^\perp\}+\{\Im\barh_{4\pizg}h_1^\perp
+\Im h_{4\pizg}\barh_1^\perp\}\big)\Big\}
\label{dobaresult}
\end{eqnarray}

The final contribution of $\Xi'$ terms is a sum of Eq. (\ref{dobresult}) and (\ref{dobaresult})
\begin{eqnarray}
&&\hspace{-1mm}
W^{(1\Xi'_i+1\Bxi'_i)}_{\mu\nu}(q)~\equiv~W^{(1\Xi'_1+1\Bxi'_1)}_{\mu\nu}(q)
+W^{(1\Xi'_2+1\Bxi'_2)}_{\mu\nu}(q)
\label{dobavkaresult}\\
&&\hspace{-1mm}
~=~{2\over sN_c}\!\int\! d^2k_\perp\Big\{
-g^\perp_{\alpha\beta}\Big[(k,q-k)_\perp\Big({1\over\alpha_q}\{f_1\Re\bar\acf_{1\pizg}+\barf_1\Re\acf_{1\pizg}\}
+{1\over\beta_q}\{\Re\bar\acf_{1\pizg}f_1+\Re\acf_{1\pizg}\barf_1\}
\nonumber\\
&&\hspace{-1mm}
+~\{f_\perp\Re\bar\graf_{1\pizg}+\barf_\perp\Re\graf_{1\pizg}\}+\{g_\perp\Im\bar\graf_{1\pizg}-\barg_\perp\Im\graf_{1\pizg}\} 
+~\{\Re\bar\graf_{1\pizg}f_\perp+\Re\graf_{1\pizg}\barf_\perp\}\nonumber\\
&&\hspace{-1mm}
+\{\Im\bar\graf_{1\pizg}g_\perp-\Im\graf_{1\pizg}\barg_\perp\} \Big)
+~{m^2\over\alpha_q}\big(\{f_1\Re\barf_{2\pizg}+\barf_1\Re f_{2\pizg}\}+\{f_1\Re\barf_{3\pizg}+\barf_1\Re f_{3\pizg}\}
\nonumber\\
&&\hspace{40mm}
+~{m^2\over\beta_q}\big(\{\Re\barf_{2\pizg}f_1+\Re f_{2\pizg}\barf_1\}+\{\Re\barf_{3\pizg}f_1+\Re f_{3\pizg}\barf_1\}\Big]
\nonumber\\
&&\hspace{-1mm}
-~\big[2(q-k)^\perp_\mu (q-k)^\perp_\nu+g^\perp_{\mu\nu}(q-k)_\perp^2\big]
\big(\{h\Re\barh_{1\pizg}+\barh\Re h_{1\pizg}\}+\{\bare\Im h_{1\pizg}-e\Im \barh_{1\pizg}\}\big)
\nonumber\\
&&\hspace{-1mm}
-~\big[2k^\perp_\mu k^\perp_\nu+g^\perp_{\mu\nu}k_\perp^2\big]
\big(\{\Re h_{1\pizg}\barh+\Re\barh_{1\pizg}h\}+\{\Im h_{1\pizg}\bare-\Im \barh_{1\pizg}e\}\big)
\nonumber\\
&&\hspace{-1mm}
+~{2\over m^2}
\calW^\perp_{\mu\nu}(q,k_\perp)\Big({1\over\alpha_q}\{h_1^\perp\Re\barh_{1\pizg}+\barh_1^\perp\Re h_{1\pizg}\}
+{1\over\beta_q}\{\Re h_{1\pizg}\barh_1^\perp+\Re\barh_{1\pizg}h_1^\perp\}\Big)
\nonumber\\
&&\hspace{-1mm}
-~{(q-k)_\perp^2\over\alpha_qm^2}[g^\perp_{\mu\nu}k_\perp^2+2k_\mu k_\nu]
\{h_1^\perp\Re\barh_{2\pizg}+\barh_1^\perp\Re h_{2\pizg}\}
\nonumber\\
&&\hspace{-1mm}
-~{k_\perp^2\over\beta_qm^2}[g^\perp_{\mu\nu}(q-k)_\perp^2+2(q-k)_\mu(q-k)_\nu ]
\{\Re h_{2\pizg}\barh_1^\perp+\Re\barh_{2\pizg}h_1^\perp\}
\nonumber\\
&&\hspace{-1mm}
+~\big[(k,q-k)_\perp g^\perp_{\alpha\beta}
+~k^\perp_\alpha (q-k)^\perp_\beta
+k^\perp_\beta (q-k)^\perp_\alpha\big]
\big({1\over\alpha_q}\{h_1^\perp\Re\barh_{3\pizg}+\barh_1^\perp\Re h_{3\pizg}\}
\nonumber\\
&&\hspace{-1mm}
+~
{1\over\alpha_q}\{h_1^\perp\Im\barh_{4\pizg}
+\barh_1^\perp\Im h_{4\pizg}\}
+{1\over\beta_q}(\{\Re\barh_{3\pizg}h_1^\perp+\Re h_{3\pizg}\barh_1^\perp\}+
{1\over\beta_q}\{\Im\barh_{4\pizg}h_1^\perp
+\Im h_{4\pizg}\barh_1^\perp\}\big)\Big\}
\nonumber
\end{eqnarray}
%

\section{Terms  with two quark-quark-gluon operators \label{sec:twoperators}}

First, in Ref. \cite{Balitsky:2020jzt} it was demonstrated that after sorting out color-singlet matrix elements the contribution $W_{\mu\nu}^{(2c)}$ is $O\big({1\over N_c^2}\big)$ in comparison to 
$W_{\mu\nu}^{(2a)}$ (and $W_{\mu\nu}^{(2b)}$) so it will be neglected in accordance with our leading-$N_c$ accuracy.

\subsection{Terms  with two quark-quark-gluon operators coming from $\Xi_{1}$ and $\Xi_{2}$\label{sec:2qqGa}}
Let us start with the first term in the r.h.s. of Eq. (\ref{kalw2a}).
Performing Fierz transformation (\ref{fierz}) we obtain
\begin{equation}
\hspace{1mm}
{N_c\over s}\langle A,B|(\bsi_A^m(x)\gamma_\mu\Xi_{2}^m(x))(\bsi_B^n(0)\gamma_\nu\Xi_{1}^n(0))
+\mu\leftrightarrow\nu|A,B\rangle+x\leftrightarrow 0
~=~g_{\mu\nu}\cheV_1+\cheV_{2\mu\nu}+\cheV_{3\mu\nu}
\label{kalvs}
\end{equation}
where 
\begin{eqnarray}
&&\hspace{-3mm}
\cheV^{(1)}~
=~{N_c\over 2s}\langle A,B|-[\bsi_A^n(x)\Xi_{1}^m(0)][\bsi_B^n(0)\Xi_{2}^m(x)]
+[\bsi_A^m(x)\gamma_5\Xi_{1}^n(0)][\bsi_B^n(0)\gamma_5\Xi_{2}^m(x)]
\label{kalv1}\\
&&\hspace{-3mm}
+~[\bsi_A^m(x)\gamma_\alpha\Xi_{1}^m(0)][\bsi_B^n(0)\gamma^\alpha\Xi_{2}^n(x)]
+[\bsi_A^m(x)\gamma_\alpha\gamma_5\Xi_{1}^m(0)]
[\bsi_B^n(0)\gamma^\alpha\gamma_5\Xi_{2}^n(x)]|A,B\rangle+x\leftrightarrow 0,
\nonumber\end{eqnarray}
\begin{eqnarray}
&&\hspace{-1mm}
\cheV^{(2)}_{\mu\nu}~=~{N_c\over 2s}\langle A,B|
-~[\bsi_A^m(x)\gamma_\mu\Xi_{1}^n(0)][\bsi_B^n(0)\gamma_\nu\Xi_{2}^m(x)]
\nonumber\\
&&\hspace{11mm}
-~[\bsi_A^m(x)\gamma_\mu\gamma_5\Xi_{1}^n(0)][\bsi_B^n(0)\gamma_\nu\gamma_5\Xi_{2}^m(x)]
+\mu\leftrightarrow\nu|A,B\rangle+x\leftrightarrow 0,
\label{kalv2}
\end{eqnarray}
and
\begin{eqnarray}
&&\hspace{-1mm}
\cheV^{(3)}_{\mu\nu}~=~{N_c\over 2s}\langle A,B|
[(\bsi_A^m(x)\sigma_{\mu\alpha}\Xi_{1}^n(0)][\bsi_B^n(0)\sigma_\nu^{~\alpha}\Xi_{2}^m(x)]+\mu\leftrightarrow\nu
\nonumber\\
&&\hspace{11mm}
-~{g_{\mu\nu}\over 2}[\bsi_A^m(x)\sigma^{\alpha\beta}\Xi_{1}^n(0)][\bsi_B^n(0)\sigma_{\alpha\beta}\Xi_{2}^m(x)
|A,B\rangle+x\leftrightarrow 0
\label{kalv3}
\end{eqnarray}
It is convenient to define $\cheV^{(3)}_{\mu\nu}$ to be traceless. In next Sections, we will consider these terms in turn.

\subsubsection{Term propotional to $g_{\mu\nu}$}
Using $\Xi_{1}~=~-{g\slashed{p}_2\over s}\gamma^iB_i{1\over \alpha+i\epsilon}\psi_A$ and 
$\Xi_{2}~=~-{g\slashed{p}_1\over s}\gamma^iA_i{1\over\beta+i\epsilon}\psi_B$ from Eq. (\ref{fildz0}) and 
extracting color-singlet contributions one obtains
\begin{eqnarray}
&&\hspace{-1mm}
\cheV^{(1)}~=~{1\over 2s^3}
\nonumber\\
&&\hspace{-1mm}
\times~\Big\{-\Big[\langle \bsi A_i(x)\notp_2\gamma^j{1\over \alpha}\psi(0)\rangle_A
\langle\bsi B_j(0)\notp_1\gamma^i{1\over\beta}\psi(x)\rangle_B
-\psi(0)\otimes\psi(x)\leftrightarrow\gamma_5\psi(0)\otimes\gamma_5\psi(x)\Big]
\nonumber\\
&&\hspace{-1mm}
+~\Big[\langle\bsi A_i(x)\gamma_k\notp_2\gamma^j{1\over \alpha}\psi(0)\rangle_A
\langle\bsi B_j(0)\gamma^k\notp_1\gamma^i{1\over\beta}\psi(x)\rangle_B
+~\psi(0)\otimes\psi(x)\leftrightarrow\gamma_5\psi(0)\otimes\gamma_5\psi(x)\Big]
\nonumber\\
&&\hspace{14mm}
+~{2\over s}\Big[\langle\bsi A_i(x)\notp_1\notp_2\gamma^j{1\over \alpha}\psi(0)\rangle_A
\langle\bsi B_j(0)\notp_2\notp_1\gamma^i{1\over\beta}\psi(x)\rangle_B
\nonumber\\
&&\hspace{30mm}
+~\psi(0)\otimes\psi(x)\leftrightarrow\gamma_5\psi(0)\otimes\gamma_5\psi(x)\Big]\Big\}~+~x\leftrightarrow 0
\label{twoxia1}
\end{eqnarray}
From the power counting it is clear that the third term is $\sim {m_\perp^2\over s}$ with respect to the first two ones. 
By the same token,  if we replace $\Xi_{1}$ by $\Xi'_{1}$ or $\Xi_{2}$ by $\Xi'_{2}$ the contribution will be small.

Let us start with the first term in Eq. (\ref{twoxia1}). Using Eq. (\ref{gammas11}) and the fact that $\langle\bar\psi(x)\big[A_k\sigma_{\star j}-A_j(x)\sigma_{\star k}\big]\psi(0)\rangle_A=0$ (cf. Eq (\ref{flanpolarized})), 
we obtain
\begin{eqnarray}
&&\hspace{-1mm}
\cheV_{1a}^{(1)}~=~-{1\over 2s^3}\langle \bsi{\notA}(x)\notp_2{1\over \alpha}\psi(0)\rangle_A
\langle\bsi{\notB}(0)\notp_1{1\over\beta}\psi(x)\rangle_B
-{1\over 4s^2}\langle \bsi{\notA}(x)\gamma_i{1\over \alpha}\psi(0)\rangle_A
\nonumber\\
&&\hspace{11mm}
\times~\langle\bsi{\notB}(0)
\gamma^i{1\over\beta}\psi(x)\rangle_B
-~{1\over 8s^2}\langle \bsi A^i(x)\sigma_{jk}{1\over \alpha}\psi(0)\rangle_A
\langle\bsi B_i(0)\sigma^{jk}{1\over\beta}\psi(x)\rangle_B
\nonumber\\
&&\hspace{22mm}
=~-{1\over 2s^3}\langle \bsi{\notA}(x)\notp_2{1\over \alpha}\psi(0)\rangle_A
\langle\bsi{\notB}(0)\notp_1{1\over\beta}\psi(x)\rangle_B\Big[1+O\big({q_\perp^2\over s}\big)\Big]
\label{twoxia2}
\end{eqnarray}
where we used the fact that projectile and target matrix elements in the two last terms in the l.h.s. cannot produce factor of $s$. 
The corresponding contribution to $V^{(1)}(q)$ has the form
\begin{eqnarray}
&&\hspace{-1mm}
-{1\over \alpha_q\beta_qsN_c}{1\over 2m^2} \!\int\! d^2k_\perp\big[k_\perp^2h_1^\perp+m^2\alpha_q(h-ie)](\alpha_q,k_\perp)
\nonumber\\
&&\hspace{11mm}
\times~\big[(q-k)_\perp^2\barh_1^\perp+m^2\beta_q(\barh+i\bare)\big](\beta_q,q_\perp-k_\perp)
\end{eqnarray}
due to EOMs  (\ref{eqmp}), (\ref{eqmt}).

Next, consider second term in Eq. (\ref{twoxia1}). Using Eqs. (\ref{gammas6a}) and (\ref{gammas11}), one can rewrite is as
\begin{eqnarray}
&&\hspace{-1mm}
\cheV_{1b}^{(1)}~=~{1\over 2s^3}\Big[\langle\bsi A_i(x)\gamma_k\notp_2\gamma^j{1\over \alpha}\psi(0)\rangle_A
\langle\bsi B_j(0)\gamma^k\notp_1\gamma^i{1\over\beta}\psi(x)\rangle_B
\label{twoxia3}\\
&&\hspace{-1mm}
+~\psi(0)\otimes\psi(x)\leftrightarrow\gamma_5\psi(0)\otimes\gamma_5\psi(x)\Big]
=~{1\over s^3}\langle\bsi(x)\slA(x)\slp_2\gamma_i{1\over \alpha}\psi(0)\rangle_A
\langle\bsi(0)\slB(0)\slp_1\gamma^i{1\over\beta}\psi(x)\rangle_B
\nonumber
\end{eqnarray}
The corresponding contribution to $V^{(1)}(q)$ has the form
\beq
{1\over \alpha_q\beta_qsN_c}\!\int\! d^2k_\perp(k,q-k)_\perp\big[f_1-\alpha_q(f_\perp+ig_\perp)\big](\alpha_q,k_\perp)
\big[\barf_1-\beta_q(\barf_\perp-i\barg_\perp)\big](\beta_q,q_\perp-k_\perp)
\eeq
where again we used E)Ms (\ref{eqmp}), (\ref{eqmt}).

Similarly, from Eq. (\ref{gammas11}) we get the third term in the form
\begin{eqnarray}
&&\hspace{-1mm}
{1\over s^4}\Big[\langle\bsi A_i(x)\notp_1\notp_2\gamma^j{1\over \alpha}\psi(0)\rangle_A
\langle\bsi B_j(0)\notp_2\notp_1\gamma^i{1\over\beta}\psi(x)\rangle_B
+~\psi(0)\otimes\psi(x)\leftrightarrow\gamma_5\psi(0)\otimes\gamma_5\psi(x)\Big]
\nonumber\\
&&\hspace{-1mm}
=~{1\over 4s^2}\Big[\langle\bsi{\grA}_i(x)\gamma^j{1\over \alpha}\psi(0)\rangle_A
\langle\bsi{\graB}_j(0)\gamma^i{1\over\beta}\psi(x)\rangle_B
+~\psi(0)\otimes\psi(x)\leftrightarrow\gamma_5\psi(0)\otimes\gamma_5\psi(x)\Big]
\label{twoxia4}
\end{eqnarray}
Since both projectile and target matrix elements cannot give factor $s$ this contribution is $O\big({q_\perp^2\over s}\big)$ in comparison to that of 
the two  first  terms.
Using QCD equations of motion (\ref{eqmp}), (\ref{eqmt}),  we obtain the contribution to $W_{\mu\nu}$ in the form
\begin{eqnarray}
&&\hspace{-1mm}
g_{\mu\nu} V_1^{(1)}(q)
~=~{g_{\mu\nu}\over 16\pi^4N_c}\!\int\! dx_\bu dx_\star d^2x_\perp~e^{-i\alpha_qx_\bu-i\beta_qx_\star+i(q,x)_\perp}\cheV^{(1)}(x) 
\label{v11otvet}\\
&&\hspace{-1mm}
=~{g_{\mu\nu}\over Q^2N_c}\int\! d^2k_\perp\Big[(k,q-k)_\perp \big[f_1-\alpha_q(f_\perp+ig_\perp)\big](\alpha_q,k_\perp)
\big[\barf_1-\beta_q(\barf_\perp-i\barg_\perp)\big](\beta_q,q_\perp-k_\perp)
\nonumber\\
&&\hspace{-1mm}
-~{1\over 2m^2}\big[k_\perp^2h_1^\perp+m^2\alpha_q(h-ie)](\alpha_q,k_\perp\big]
\big[(q-k)_\perp^2\barh_1^\perp+m^2\beta_q(\barh+i\bare)\big](\beta_q,q_\perp-k_\perp)\Big]
\nonumber
\end{eqnarray}

Next, the $x\leftrightarrow 0$ term is
\begin{eqnarray}
&&\hspace{-1mm}
\cheV_2^{(1)}~=~
-{1\over 2s^3}\langle \bsi\slA(0)\slp_2{1\over \alpha}\psi(x)\rangle_A
\langle\bsi\slB(x)\slp_1{1\over\beta}\psi(0)\rangle_B
\nonumber\\
&&\hspace{5mm}
+~{1\over s^3}\langle\bsi\slA(0)\slp_2\gamma_i{1\over \alpha}\psi(x)\rangle_A
\langle\bsi\slB(x)\slp_1\gamma^{~i}{1\over\beta}\psi(0)\rangle_B
\label{chve1}
\end{eqnarray}

Similarly to Eq. (\ref{v11otvet}), we get
\begin{eqnarray}
&&\hspace{-1mm}
g_{\mu\nu} V_2^{(1)}(q)
~=~{g_{\mu\nu}\over 16\pi^4N_c}\!\int\! dx_\bu dx_\star d^2x_\perp~e^{-i\alpha_qx_\bu-i\beta_qx_\star+i(q,x)_\perp}\cheV_2^{(1)}(x) 
\label{v12otvet}\\
&&\hspace{-1mm}
=~{g_{\mu\nu}\over Q^2N_c}\int\! d^2k_\perp\Big[(k,q-k)_\perp \big[\barf_1-\alpha_q(\barf_\perp-i\barg_\perp)\big](\alpha_q,k_\perp)
\big[f_1-\beta_q(f_\perp+ig_\perp)\big](\beta_q,q_\perp-k_\perp)
\nonumber\\
&&\hspace{-1mm}
-~{1\over 2m^2}\big[k_\perp^2\barh_1^\perp+m^2\alpha_q(\barh+i\bare)](\alpha_q,k_\perp\big]
\big[(q-k)_\perp^2h_1^\perp+m^2\beta_q(h-i e)\big](\beta_q,q_\perp-k_\perp)\Big]
\nonumber
\end{eqnarray}
Sum of Eq. (\ref{v11otvet}) and (\ref{v12otvet}) is
\begin{eqnarray}
&&\hspace{-1mm}
g_{\mu\nu} V^{(1)}(q)
~=~{g_{\mu\nu}\over 16\pi^4N_c}\!\int\! dx_\bu dx_\star d^2x_\perp~e^{-i\alpha_qx_\bu-i\beta_qx_\star+i(q,x)_\perp}\cheV^{(1)}(x) 
\label{v1otvet}\\
&&\hspace{-1mm}
=~{g_{\mu\nu}\over Q^2N_c}\int\! d^2k_\perp\Big\{
(k,q-k)_\perp \Big(\{f_1\barf_1+\barf_1f_1\}-\alpha_q\{[f_\perp+ig_\perp]\barf_1+[\barf_\perp-i\barg_\perp]f_1\}
\nonumber\\
&&\hspace{-1mm}
-~\beta_q\{f_1[\barf_\perp-i\barg_\perp]+\barf_1[f_\perp+ig_\perp]\}
+\alpha_q\beta_q\{[f_\perp+ig_\perp][\barf_\perp-i\barg_\perp]+[\barf_\perp-i\barg_\perp][f_\perp+ig_\perp]\}\Big)
\nonumber\\
&&\hspace{-1mm}
-~{1\over 2}\Big(
{k_\perp^2(q-k)_\perp^2\over m^2}\{h_1^\perp\barh_1^\perp+\barh_1^\perp h_1^\perp\}
+\alpha_q(q-k)_\perp^2\{[(h-ie)]\barh_1^\perp+[\barh+i\bare]h_1^\perp\}
\nonumber\\
&&\hspace{-1mm}
+~\beta_qk_\perp^2\{h_1^\perp[\barh+i\bare]+\barh_1^\perp[h-i e]\}+m^2\alpha_q\beta_q\{[h-ie][\barh+i\bare]+[\barh+i\bare][h-ie]\}\Big)
\nonumber
\end{eqnarray}
%

\subsubsection{Term $\cheV^{(2)}_{\mu\nu}$ \label{sec:kalv2}}
Separating color-singlet contributions one can rewrite Eq. (\ref{kalv2}) as
\begin{eqnarray}
&&\hspace{-1mm}
\cheV^{(2)}_{\mu\nu}~=~-{1\over 2s^3}\big\{\langle\bsi A_i(x)\gamma_\mu\notp_2\gamma^j{1\over \alpha}\psi(0)\rangle_A
\langle\bsi B_j(0)\gamma_\nu\notp_1\gamma^i{1\over\beta}\psi(x)\rangle_B
\nonumber\\
&&\hspace{10mm}
+~\psi(0)\otimes\psi(x)\leftrightarrow\gamma_5\psi(0)\otimes\gamma_5\psi(x)+\mu\leftrightarrow\nu\}
~+~x\leftrightarrow 0
\label{kalve2}
\end{eqnarray}

As demonstrated in Ref. \cite{Balitsky:2020jzt}, only transverse  $\mu$ and $\nu$ contribute at ${1\over Q^2}$ level. 
In this case we can use formula (\ref{formula9}) and get
\begin{eqnarray}
&&\hspace{-1mm}
\cheV^{(2)}_{1\mu_\perp\nu_\perp}~
=~-{1\over 2s^3}\big\{\langle\bsi A_i(x)\gamma_{\mu_\perp}\notp_2\gamma^j{1\over \alpha}\psi(0)\rangle_A
\langle\bsi B_j(0)\gamma_{\nu_\perp}\notp_1\gamma^i{1\over\beta}\psi(x)\rangle_B
\nonumber\\
&&\hspace{10mm}
+~\psi(0)\otimes\psi(x)\leftrightarrow\gamma_5\psi(0)\otimes\gamma_5\psi(x)+\mu\leftrightarrow\nu\}
\nonumber\\
&&\hspace{10mm}
=~-{g_{\mu\nu}^\perp\over s^3}\langle\bsi\notA(x)\notp_2\gamma_i{1\over \alpha}\psi(0)\rangle_A
\langle\bsi\notB(0)\notp_1\gamma^i{1\over\beta}\psi(x)\rangle_B
\label{kalvedvaperp}
\end{eqnarray}
which gives the contribution to $W_{\mu\nu}$ in the form 
\begin{eqnarray}
&&\hspace{-11mm}
V^{(2)}_{1\mu\nu}(q)~
=~{1\over 16\pi^4N_c}\!\int\! dx_\bu dx_\star d^2x_\perp
~e^{-i\alpha_qx_\bu-i\beta_qx_\star+i(q,x)_\perp}\cheV^{(2)}_{1\mu_\perp\nu_\perp}(x) 
~=~-{g_{\mu\nu}^\perp\over \alpha_q\beta_qsN_c}
\nonumber\\
&&\hspace{-11mm}
\times \int\! d^2k_\perp (k,q-k)_\perp 
\big[f_1-\alpha_q(f^\perp+ig^\perp)\big](\alpha_q,k_\perp)
\big[\barf_1-\beta_q(\barf^\perp-i\barg^\perp)\big](\beta_q,q_\perp-k_\perp) 
\label{kalv2glav}
\end{eqnarray}
where we again used EOMs (\ref{eqmp}), (\ref{eqmt})

The corresponding $x\leftrightarrow 0$ contribution is
\begin{eqnarray}
&&\hspace{-1mm}
\cheV^{(2)}_{2\mu_\perp\nu_\perp}~
=~-{g_{\mu\nu}^\perp\over s^3}\langle\bsi\slA(0)\notp_2\gamma_i{1\over \alpha}\psi(x)\rangle_A
\langle\bsi\slB(x)\notp_1\gamma^i{1\over\beta}\psi(0)\rangle_A
\label{v22x}
\end{eqnarray}
which gives
\begin{eqnarray}
&&\hspace{-11mm}
V^{(2)}_{2\mu\nu}(q)~
=~-{g_{\mu\nu}^\perp\over Q^2N_c}\int\! d^2k_\perp (k,q-k)_\perp 
\nonumber\\
&&\hspace{3mm}
\times~\big[\barf_1-\alpha_q(\barf^\perp-i\barg^\perp)\big](\alpha_q,k_\perp)
\big[f_1-\beta_q(f^\perp+ig^\perp)\big](\beta_q,q_\perp-k_\perp) 
\label{v22q}
\end{eqnarray}

The  sum of Eq. (\ref{kalv2glav}) and (\ref{v22q}) has the form
\begin{eqnarray}
&&\hspace{-2mm}
V^{(2)}_{\mu\nu}(q)~
\label{v22qu}\\
&&\hspace{-2mm}
=~-{g_{\mu\nu}^\perp\over Q^2N_c}\int\! d^2k_\perp (k,q-k)_\perp \Big(\{f_1\barf_1+\barf_1f_1\}-\alpha_q\{[f^\perp+ig^\perp]\barf_1+[\barf^\perp-i\barg^\perp]f_1\}
\nonumber\\
&&\hspace{-2mm}
-~\beta_q\{f_1[\barf^\perp-i\barg^\perp]+\barf_1[f^\perp+ig^\perp]\}
+\alpha_q\beta_q\{[f_\perp+ig_\perp][\barf_\perp-i\barg_\perp]+[\barf_\perp-i\barg_\perp][f_\perp+ig_\perp]\}\Big)
\nonumber
\end{eqnarray}
%
\subsubsection{Term $\cheV^{(3)}_{\mu\nu}$ \label{sec:v3}}

Let us consider now 
\begin{eqnarray}
&&\hspace{-1mm}
\cheV'_{3\mu\nu}~=~{N_c\over 2s}\langle A,B|
[(\bsi_A^m(x)\sigma_{\mu\alpha}\Xi_{1}^m(0)][\bsi_B^n(0)\sigma_\nu^{~\alpha}\Xi_{2}^n(x)]+\mu\leftrightarrow\nu
+x\leftrightarrow 0
\label{kalv3a}
\end{eqnarray}
(the trace will be subtracted after the calculation). Separating color-singlet contributions, we get
\begin{equation}
\hspace{0mm}
\cheV'_{3\mu\nu}~=~{1\over 2s^3}\langle\bsi A_i(x)\sigma_{\mu\alpha}\notp_2\gamma^j{1\over \alpha}\psi(0)\rangle_A
\langle\bsi B_j(0)\sigma_\nu^{~\alpha}\notp_1\gamma^i{1\over\beta}\psi(x)\rangle_A+\mu\leftrightarrow\nu
+x\leftrightarrow 0
\label{kalv3b}
\end{equation}
From Eq. (\ref{mainfla4})
\begin{eqnarray}
&&\hspace{-1mm}
A^j\sigma_{\mu\xi}\sigma_{\star i}\otimes B^i\sigma_\nu^{~\xi}\sigma_{\bu j}~
=~-\big(A_\nu \sigma_{\star k}-{g_{\nu k}\over 2}A^j\sigma_{\star j}\big)\otimes\big(B_\mu\sigma_\bu^{~k}-\half\delta_\mu^kB^j\sigma_{\bu j}\big)
\nonumber\\
&&\hspace{35mm}
-~\big(A_k \sigma_{\star \mu_\perp}-{g_{\mu k}\over 2}A^j\sigma_{\star j}\big)\otimes\big(B^k\sigma_{\bu\nu_\perp}-\half\delta_\nu^kB^j\sigma_{\bu j}\big)
\nonumber\\
&&\hspace{35mm}
-~{2\over s}(p_{1\mu}p_{2\nu}+p_{2\mu}p_{1\nu})
\big(A^j\sigma_{\star i}-{\delta_i^j\over 2}A^k\sigma_{\star k}\big)
\otimes \big(B^i\sigma_{\bu j}-{\delta_i^j\over 2}B^l\sigma_{\bu l}\big)
\nonumber\\
&&\hspace{35mm}
-~{1\over s}(p_{1\mu}p_{2\nu}+p_{2\mu}p_{1\nu})\sigma_{\star i}A^i\otimes\sigma_{\bu j}B^j
+{g^\perp_{\mu\nu}\over 2}A^j\sigma_{\star j}\otimes B^k\sigma_{\bu k}
\label{mainfla4e}
\end{eqnarray}
Using EOMs (\ref{eqmp}), (\ref{eqmt}) and
parametrizations (\ref{1019}),  (\ref{maelsbe}) and similar one for projectile  matrix elements,  we get
\begin{eqnarray}
&&\hspace{-11mm}
V'_{3\mu\nu}~=~{g^\perp_{\mu\nu}-g^\parallel_{\mu\nu}\over 2\alpha_q\beta_qsN_c}\!\int d^2k_\perp 
\Big({1\over m^2}\big[k_\perp^2h_1^\perp+\alpha_qm^2(h-ie)\big]\big[(q-k)_\perp^2\barh_1^\perp+\beta_qm^2(\barh+i\bare)\big]
\nonumber\\
&&\hspace{3mm}
+~{1\over m^2}\big[k_\perp^2\barh_1^\perp+\alpha_qm^2(\barh+i\bare)\big]\big[(q-k)_\perp^2h_1^\perp+\beta_qm^2(h-ie)\big]\Big)
\label{kalvklad3perpmyy}\\
&&\hspace{-1mm}
+~{1\over \alpha_q\beta_qsN_c}\!\int d^2k_\perp {1\over m^2}\big\{[k^\perp_\mu(q-k)^\perp_\nu+\mu\leftrightarrow\nu](k,q-k)_\perp
-k_\perp^2(q-k)^\perp_\mu(q-k)^\perp_\nu
\nonumber\\
&&\hspace{-1mm}
-~(q-k_\perp)^2k^\perp_\mu k^\perp_\nu
-{g_{\mu\nu}^\perp\over 2}k_\perp^2(q-k_\perp)^2-g^\parallel_{\mu\nu}\big[(k,q-k)_\perp^2-\half k_\perp^2(q-k)_\perp^2\big]\big\}
\nonumber\\
&&\hspace{-1mm}
\times~\big[h_{1G}^\perp(\alpha_q,k_\perp)\barh_{1G}^\perp(\beta_q,q_\perp-k_\perp)
+\barh_{1G}^\perp(\alpha_q,k_\perp)h_{1G}^\perp(\beta_q,q_\perp-k_\perp)\big]
\nonumber
\end{eqnarray}
After subtracting trace we obtain
\begin{eqnarray}
&&\hspace{-2mm}
V^{(3)}_{\mu\nu}~=~V'_{3\mu\nu}-{g_{\mu\nu}\over 4}{V'_3}_\xi^{~\xi} 
\label{v3otvet}\\
&&\hspace{-2mm}
=~{g^\perp_{\mu\nu}-g^\parallel_{\mu\nu}\over 2Q^2N_c}\!\int d^2k_\perp
\Big({k_\perp^2(q-k)_\perp^2\over m^2}\{h_1^\perp\barh_1^\perp+\barh_1^\perp h_1^\perp\}
+\alpha_q(q-k)_\perp^2\{[h-ie]\barh_1^\perp+[\barh+i\bare]h_1^\perp\}
\nonumber\\
&&\hspace{-2mm}
+~\beta_qk_\perp^2\{h_1^\perp[\barh+i\bare]+\barh_1^\perp[h-ie]\}+\alpha_q\beta_qm^2\{[h-ie][\barh+i\bare]+[\barh+i\bare][h-ie]\}\Big)
\nonumber\\
&&\hspace{-1mm}
+~{1\over \alpha_q\beta_qsN_c}\!\int d^2k_\perp {1\over m^2}\calW^\perp_{\mu\nu}(q_\perp,k_\perp)\{h_{1G}^\perp\barh_{1G}^\perp+\barh_{1G}^\perp h_{1G}^\perp\}
\nonumber
\end{eqnarray}
where $\calw^\perp_{\mu\nu}(q_\perp,k)$ is defined in Eq. (\ref{calweperp}).

Let us now assemble the contribution of terms (\ref{kalvs}) to $W_{\mu\nu}$. Summing Eqs. (\ref{v1otvet}),  (\ref{v22qu}),  and (\ref{v3otvet}) we get
\begin{eqnarray}
&&\hspace{-1mm}
g_{\mu\nu}\cheV_1+\cheV_{2\mu\nu}+\cheV_{3\mu\nu}~=~{1\over 32\pi^4}
\!\int\!dx_\bu dx_\star d^2x_\perp~e^{-i\alpha x_\bu-i\beta x_\star+i(q,x)_\perp}
\nonumber\\
&&\hspace{22mm}
\big[\langle A,B|(\bsi_A^m(x)\gamma_\mu\Xi_{2}^m(x))(\bsi_B^n(0)\gamma_\nu\Xi_{1}^n(0))
+\mu\leftrightarrow\nu|A,B\rangle+x\leftrightarrow 0\big]
\nonumber\\
&&\hspace{-1mm}
=~{g_{\mu\nu}^\parallel\over Q^2N_c}\int\! d^2k_\perp\Big\{
(k,q-k)_\perp \Big[\{f_1\barf_1+\barf_1f_1\}-\alpha_q\{[f^\perp+ig^\perp]\barf_1+[\barf^\perp-i\barg^\perp]f_1\}
\nonumber\\
&&\hspace{-1mm}
-~\beta_q\{f_1[\barf^\perp-i\barg^\perp]+\barf_1[f^\perp+ig^\perp]\}
+\alpha_q\beta_q\{[f_\perp+ig_\perp][\barf_\perp-i\barg_\perp]+[\barf_\perp-i\barg_\perp][f_\perp+ig_\perp]\}\Big]
\nonumber\\
&&\hspace{-1mm}
-~\Big[{k_\perp^2(q-k)_\perp^2\over m^2}\{h_1^\perp\barh_1^\perp+\barh_1^\perp h_1^\perp\}
+\alpha_q(q-k)_\perp^2\{[h-ie]\barh_1^\perp+[\barh+i\bare]h_1^\perp\}
\nonumber\\
&&\hspace{-1mm}
+\beta_qk_\perp^2\{h_1^\perp[\barh+i\bare]+\barh_1^\perp[h-ie]\}+\alpha_q\beta_qm^2\{[h-ie][\barh+i\bare]+[\barh+i\bare][h-ie]\}\Big]\Big\}
\nonumber\\
&&\hspace{-1mm}
+~{1\over \alpha_q\beta_qsN_c}\!\int d^2k_\perp {1\over m^2}\calw^\perp_{\mu\nu}(q_\perp,k)\{h_{1G}^\perp\barh_{1G}^\perp+\barh_{1G}^\perp h_{1G}^\perp\}
\label{votvet}
\end{eqnarray}
Finally, to get $W_{\mu\nu}^{(2a)}(q)$ of Eq. (\ref{kalw2a}) we need to add 
the contribution of the term\\
 $[\Bxi_{1}(x)\gamma_\mu\psi_B(x)\big]\big[\Bxi_{2}(0)\gamma_\nu\psi_A(0)\big]$.
Similarly to the case of one quark-quark-gluon operator considered in Sect. \ref{1qqG}, it can be demonstrated that this contribution doubles the real part of the result (\ref{votvet}) so we get
\begin{eqnarray}
&&\hspace{-1mm}
W_{\mu\nu}^{(2a)}(q)~=~{1\over 32\pi^4}
\!\int\!dx_\bu dx_\star d^2x_\perp~e^{-i\alpha x_\bu-i\beta x_\star+i(q,x)_\perp}
\big[\langle A,B|(\bsi_A^m(x)\gamma_\mu\Xi_{2}^m(x))
\nonumber\\
&&\hspace{5mm}
\times(\bsi_B^n(0)\gamma_\nu\Xi_{1}^n(0))
+[\Bxi_{1}(x)\gamma_\mu\psi_B(x)\big]\big[\Bxi_{2}(0)\gamma_\nu\psi_A(0)\big]
+\mu\leftrightarrow\nu|A,B\rangle+x\leftrightarrow 0\big]
\nonumber\\
&&\hspace{-1mm}
=~{2g_{\mu\nu}^\parallel\over \alpha_q\beta_q sN_c}\int\! d^2k_\perp\Big\{
(k,q-k)_\perp \Big(\{f_1\barf_1+\barf_1 f_1\}-\alpha_q\{f_\perp\barf_1+\barf_\perp f_1\}-\beta_q\{f_1\barf_\perp+\barf_1 f_\perp\}
\nonumber\\
&&\hspace{-1mm}
+~\alpha_q\beta_q\{f_\perp\barf_\perp+\barf_\perp f_\perp\}+\alpha_q\beta_q\{g_\perp\barg_\perp+\barg_\perp g_\perp\}\Big)
\nonumber\\
&&\hspace{-1mm}
-~\big[{1\over m^2}k_\perp^2
(q-k)_\perp^2\{h_1^\perp\barh_1^\perp+\barh_1^\perp h_1^\perp\}
+\alpha_q(q-k)_\perp^2\{h\barh_1^\perp+\barh h_1^\perp\}+\beta_q k_\perp^2\{h_1^\perp\barh+\barh_1^\perp h\}
\nonumber\\
&&\hspace{-1mm}
+~\alpha_q\beta_qm^2\{h\barh+\barh h\}+\alpha_q\beta_qm^2\{e\bare+\bare e\}\big]\Big\}
\nonumber\\
&&\hspace{-1mm}
+~{2\over \alpha_q\beta_qsN_c}\!\int d^2k_\perp {1\over m^2}\calw^\perp_{\mu\nu}(q_\perp,k)
\Re\big(\{h_{1G}^\perp\barh_{1G}^\perp+\barh_{1G}^\perp h_{1G}^\perp\}\big)
\label{w2atvet}
\end{eqnarray}
%

\subsection{Terms  with two quark-quark-gluon operators coming from $\Bxi_{2}$ and  $\Xi_{2}$ \label{sec:2qqGb}}

Let us start with  the first term in Eq. (\ref{kalw2b}).
\begin{equation}
\hspace{-0mm}
\cheW^{(2b)}_{1\mu\nu}~=~{N_c\over s}\langle A,B|
\big[\bar\psi_A(x)\gamma_\mu\Xi_{2}(x)\big]\big[\Bxi_{2}(0)\gamma_\nu\psi_A(0)\big]
+\mu\leftrightarrow\nu|A,B\rangle~+~x\leftrightarrow 0
\label{w2b1munu}
\end{equation}
After Fierz transformation (\ref{fierz}) we obtain
\begin{eqnarray}
&&\hspace{-1mm}
\cheW^{(2b)}_{1\mu\nu}~=~-{N_c\over 2s}(\delta_\mu^\alpha\delta_\nu^\beta+\delta_\nu^\alpha\delta_\mu^\beta-g_{\mu\nu}g^{\alpha\beta})
\langle A,B|\big\{[\bar\psi_A^m(x)\gamma_\alpha\psi_A^n(0)][\Bxi_{2}^n(0)\gamma_\beta\Xi_{2}^m(x)]
\label{w2b1a}\\
&&\hspace{55mm}
+~\gamma_\alpha\otimes\gamma_\beta\leftrightarrow \gamma_\alpha\gamma_5\otimes\gamma_\beta\gamma_5\big\}|A,B\rangle
\nonumber\\
&&\hspace{-1mm}
+~{N_c\over 2s}(\delta_\mu^\alpha\delta_\nu^\beta+\delta_\nu^\alpha\delta_\mu^\beta-\half g_{\mu\nu}g^{\alpha\beta})
\langle A,B|[\bar\psi_A^m(x)\sigma_{\alpha\xi}\psi_A^n(0)][\Bxi_{2}^n(0)\sigma_\beta^{~\xi}\Xi_{2}^m(x)]|A,B\rangle
~+~x\leftrightarrow 0
\nonumber
\end{eqnarray}
(note that $\Bxi_{2}\Xi_{2}=\Bxi_{2}\gamma_5\Xi_{2}=0$). 
Using explicit expressions (\ref{fildz0}) for quark fields and separating color-singlet terms we get
\begin{equation}
\cheW^{(2b)}_{1\mu\nu}~=~\cheV^4_{\mu\nu}+\cheV^5_{\mu\nu}
\label{w2b1}
\end{equation}
where
\begin{eqnarray}
&&\hspace{-1mm}
\cheV^4_{\mu\nu}~=~-{1\over s^3}(\delta_\mu^\alpha p_{1\nu}+\delta_\nu^\alpha p_{1\mu}-g_{\mu\nu}p_1^\alpha)
\Big(\langle \bar\psi(x)A_j(x)\gamma_\alpha A_i(0)\psi(0)\rangle_A
\nonumber\\
&&\hspace{-1mm}
\times~\langle\big(\bar\psi{1\over \beta}\big)(0)\gamma^i\notp_1
\gamma^j{1\over\beta}\psi(x)\rangle_A
+\psi(0)\otimes\psi(x)\leftrightarrow \gamma_5\psi(0)\otimes\gamma_5\psi(x)\Big)~+~x\leftrightarrow 0
\label{v4}
\end{eqnarray}
and
\begin{eqnarray}
&&\hspace{-1mm}
\cheV^5_{\mu\nu}
=~
{1\over s^3}(\delta_\mu^\alpha\delta_\nu^\beta+\delta_\nu^\alpha\delta_\mu^\beta-\half g_{\mu\nu}g^{\alpha\beta})
\nonumber\\
&&\hspace{-1mm}
\times~\Big\{-p_{1\beta} \langle \bar\psi(x)A_j(x)\sigma_{\alpha k}A_i(0)\psi(0)\rangle_A
\langle\big(\bar\psi{1\over \beta}\big)(0)\gamma^i\sigma_\bu^{~k}
\gamma^j{1\over\beta}\psi(x)\rangle_A  
\nonumber\\
&&\hspace{-1mm} 
+~\langle \bar\psi(x)A_j(x)\sigma_{\alpha \bu}A_i(0)\psi(0)\rangle_A
\langle\big(\bar\psi{1\over \beta}\big)(0)\gamma^i\sigma_{\bu\beta_\perp}
\gamma^j{1\over\beta}\psi(x)\rangle_A \Big\}   ~+~x\leftrightarrow 0
\label{v5}
\end{eqnarray}
We will consider them in turn.

\subsubsection{Term  $\cheV^4_{\mu\nu}$ \label{sec:v4}}
First, as demonstrated in Ref. \cite{Balitsky:2020jzt}, the term $\sim g_{\mu\nu}$ is small,
so
\begin{eqnarray}
&&\hspace{-1mm}
\cheV^4_{\mu\nu}~=~-{p_{1\mu}\over s^3}\Big(
\langle \bar\psi A_j(x)\gamma_\nu A_i\psi(0)\rangle_A\langle\big(\bar\psi{1\over \beta}\big)(0)\gamma^i\notp_1
\gamma^j{1\over\beta}\psi(x)\rangle_A
\nonumber\\
&&\hspace{-1mm}
+~\psi(0)\otimes\psi(x)\leftrightarrow \gamma_5\psi(0)\otimes\gamma_5\psi(x)\Big)
~+~\mu\leftrightarrow \nu~+~x\leftrightarrow 0
\label{v4raz}
\end{eqnarray}
Also, it is demonstrated there that only longitudinal index $\nu$ gives ${1\over Q^2}$ power correction, so
\begin{eqnarray}
&&\hspace{-1mm}
\cheV^4_{\mu\nu}~=~-{4p_{1\mu} p_{1\nu}\over s^4}\Big(
\langle \bar\psi A_j(x)\notp_2A_i\psi(0)\rangle_A\langle\big(\bar\psi{1\over \beta}\big)(0)\gamma^i\notp_1
\gamma^j{1\over\beta}\psi(x)\rangle_A
\nonumber\\
&&\hspace{-1mm}
+~\psi(0)\otimes\psi(x)\leftrightarrow \gamma_5\psi(0)\otimes\gamma_5\psi(x)\Big)
\nonumber\\
&&\hspace{-1mm}
-~{g^\parallel_{\mu\nu}\over s^3}\Big(
\langle \bar\psi A_j(x)\notp_1A_i\psi(0)\rangle_A\langle\big(\bar\psi{1\over \beta}\big)(0)\gamma^i\notp_1
\gamma^j{1\over\beta}\psi(x)\rangle_A
\nonumber\\
&&\hspace{-1mm}
+~\psi(0)\otimes\psi(x)\leftrightarrow \gamma_5\psi(0)\otimes\gamma_5\psi(x)\Big)
~+~x\leftrightarrow 0
\label{v4long}
\end{eqnarray}
Moreover, the contribution of the second term to $W_{\mu\nu}$ is small \cite{Balitsky:2020jzt},
so we are left with the first term in the r.h.s. of Eq. (\ref{v4long}). Using Eq. (\ref{flagamma})
it can be rewritten as
\begin{eqnarray}
&&\hspace{-1mm}
\cheV^4_{\mu\nu}~=~-{4p_{1\mu} p_{1\nu}\over s^4}\Big(
\langle \bar\psi\slA(x)\slp_2\slA\psi(0)\rangle_A\langle\big(\bar\psi{1\over \beta}\big)(0)\notp_1{1\over\beta}\psi(x)\rangle_A
\nonumber\\
&&\hspace{-1mm}
+~\psi(0)\otimes\psi(x)\leftrightarrow \gamma_5\psi(0)\otimes\gamma_5\psi(x)\Big)
~+~x\leftrightarrow 0
\label{v4longa}
\end{eqnarray}

The corresponding contribution to $W_{\mu\nu}$ is obtained from formulas (\ref{maelqg2}) and 
 (\ref{projaa})  
\begin{eqnarray}
&&\hspace{-2mm}
V^4_{\mu\nu}(q)~=~{4p_{1\mu} p_{1\nu}\over \beta_q^2s^2N_c}\int\! d^2k_\perp\Big[
k_\perp^2\{f_1\barf_1+\barf_1f_1\}-2\alpha_q\{f_\perp\barf_1+\barf_\perp f_1\}
+2\alpha_q^2m^2\{f_3\barf_1+\barf_3f_1\}\Big]
\nonumber\\
\label{v4otvetrue}
\end{eqnarray}
%

\subsubsection{Term $\cheV^5_{\mu\nu}$ \label{sec:v5}}
Again, as demonstrated in Ref. \cite{Balitsky:2020jzt}, the term $\sim g_{\mu\nu}$ is small,
so
\begin{eqnarray}
&&\hspace{-7mm}
\cheV^5_{\mu\nu}
=~
{1\over s^3}
\Big\{-p_{1\mu} \langle \bar\psi A_j(x)\sigma_{\nu k}A_i\psi(0)\rangle_A
\langle\big(\bar\psi{1\over \beta}\big)(0)\gamma^i\sigma_\bu^{~k}
\gamma^j{1\over\beta}\psi(x)\rangle_A  
\nonumber\\
&&\hspace{-7mm} 
+~\langle \bar\psi A_j(x)\sigma_{\mu \bu}A_i\psi(0)\rangle_A
\langle\big(\bar\psi{1\over \beta}\big)(0)\gamma^i\sigma_{\bu\nu_\perp}
\gamma^j{1\over\beta}\psi(x)\rangle_A \Big\} +\mu\leftrightarrow\nu  ~+~x\leftrightarrow 0
\label{v5dva}
\end{eqnarray}
Moreover, the second term in the r.h.s. is also small \cite{Balitsky:2020jzt}, and therefore
\begin{eqnarray}
&&\hspace{-1mm}
\cheV^5_{\mu\nu}
=~
-{p_{1\mu}\over s^3}
\langle \bar\psi A_j(x)\sigma_{\nu k}A_i\psi(0)\rangle_A
\langle\big(\bar\psi{1\over \beta}\big)(0)\gamma^i\sigma_\bu^{~k}
\gamma^j{1\over\beta}\psi(x)\rangle_A  
 +\mu\leftrightarrow\nu+x\leftrightarrow 0
 \nonumber\\
&&\hspace{-1mm} 
=~
-{4p_{1\mu}p_{1\nu}\over s^4}
\langle \bar\psi A_j(x)\sigma_{\star k}A_i\psi(0)\rangle_A
\langle\big(\bar\psi{1\over \beta}\big)(0)\gamma^i\sigma_\bu^{~k}
\gamma^j{1\over\beta}\psi(x)\rangle_A  
\nonumber\\
&&\hspace{-1mm} 
-~{g^\parallel_{\mu\nu}\over s^4}
\langle \bar\psi A_j(x)\sigma_{\bu k}A_i\psi(0)\rangle_A
\langle\big(\bar\psi{1\over \beta}\big)(0)\gamma^i\sigma_\bu^{~k}
\gamma^j{1\over\beta}\psi(x)\rangle_A
\nonumber\\
&&\hspace{-1mm}
-~\Big({p_{1\mu}\over s^3}
\langle \bar\psi A_j(x)\sigma_{\nu k}A_i\psi(0)\rangle_A
\langle\big(\bar\psi{1\over \beta}\big)(0)\gamma^i\sigma_\bu^{~k}
\gamma^j{1\over\beta}\psi(x)\rangle_A  
 +\mu\leftrightarrow\nu\Big)+x\leftrightarrow 0
\label{v5raz}
\end{eqnarray}
As demonstrated in \cite{Balitsky:2020jzt}, the two last lines in the above equation are small.  As  to
the first term in r.h.s. of Eq. (\ref{v5raz}), using Eq. (\ref{sisigaga}) it can be rewritten as
\begin{equation}
\hspace{-0mm}
\cheV^5_{\mu\nu}
~=~
-{4p_{1\mu}p_{1\nu}\over s^4}
\langle \bar\psi \slA(x)\sigma_{\star j}\slA\psi(0)\rangle_A
\langle\big(\bar\psi{1\over \beta}\big)(0)\sigma_\bu^{~j}
{1\over\beta}\psi(x)\rangle_A  
~+~x\leftrightarrow 0
\label{v5razotvet}
\end{equation}
so the corresponding contribution to $W_{\mu\nu}$ takes the form
\begin{eqnarray}
&&\hspace{-1mm}
V^5_{\mu\nu}
=~
-{4p_{1\mu}p_{1\nu}\over \beta_q^2s^2N_c}
\label{v5otvetrue}\\
&&\hspace{-1mm}
\times~\!\int\! d^2k_\perp
(k,q-k)_\perp\Big( {k_\perp^2\over m^2} \{h_1^\perp\barh_1^\perp+\barh_1^\perp h_1^\perp\}+2\alpha_q \{h\barh_1^\perp+\barh h_1^\perp\}
-2\alpha_q^2\{h_3^\perp\barh_1^\perp+\barh_3^\perp h_1^\perp\}\Big)
\nonumber
\end{eqnarray}
where we used Eqs. (\ref{maelqg2}) and (\ref{projaa}).
The full result for $W_{\mu\nu}^{(2b)}$ is given by the sum of Eqs. (\ref{v4otvetrue}) and (\ref{v5otvetrue})
\begin{eqnarray}
&&\hspace{-1mm}
W_{1\mu\nu}^{(2b)}~=~
{4p_{1\mu}p_{1\nu}\over \beta_q^2s^2N_c}
\!\int\! d^2k_\perp\Big(
k_\perp^2\{f_1\barf_1+\barf_1f_1\}-2\alpha_q\{f_\perp\barf_1+\barf_\perp f_1\}
+2\alpha_q^2m^2\{f_3\barf_1+\barf_3f_1\}\Big)
\nonumber\\
&&\hspace{-1mm}
-~(k,q-k)_\perp\Big( {k_\perp^2\over m^2} \{h_1^\perp\barh_1^\perp+\barh_1^\perp h_1^\perp\}+2\alpha_q \{h\barh_1^\perp+\barh h_1^\perp\}
-2\alpha_q^2\{h_3^\perp\barh_1^\perp+\barh_3^\perp h_1^\perp\}\Big)
\Big]
\label{w2b1otvetrue}
\end{eqnarray}
%

\subsubsection{Second term in Eq. (\ref{kalw2b})}
Consider the second term in Eq. (\ref{kalw2b}).
\begin{equation}
\hspace{-0mm}
\cheW^{(2b)}_{2\mu\nu}~=~{N_c\over s}\langle A,B|
\big[\Bxi_{1}(x)\gamma_\mu\psi_B(x)\big]\big[\bar\psi_B(0)\gamma_\nu\Xi_{1}(0)\big]
+\mu\leftrightarrow\nu|A,B\rangle~+~x\leftrightarrow 0
\label{w2b2munu}
\end{equation}
After Fierz transformation (\ref{fierz}) we obtain
\begin{eqnarray}
&&\hspace{-1mm}
\cheW^{(2b)}_{2\mu\nu}~=~-{N_c\over 2s}(\delta_\mu^\alpha\delta_\nu^\beta+\delta_\nu^\alpha\delta_\mu^\beta-g_{\mu\nu}g^{\alpha\beta})
\langle A,B|\big\{[\Bxi_{1}^m(x)\gamma_\alpha\Xi_{1}^n(0)][\bsi_{B}^n(0)\gamma_\beta\psi_{B}^m(x)]
\\
&&\hspace{55mm}
+~\gamma_\alpha\otimes\gamma_\beta\leftrightarrow \gamma_\alpha\gamma_5\otimes\gamma_\beta\gamma_5\big\}|A,B\rangle
\nonumber\\
&&\hspace{-1mm}
+~{N_c\over 2s}(\delta_\mu^\alpha\delta_\nu^\beta+\delta_\nu^\alpha\delta_\mu^\beta-\half g_{\mu\nu}g^{\alpha\beta})
\langle A,B|[\Bxi_{1}^m(x)\sigma_{\alpha\xi}\Xi_{1}^n(0)][\psi_{1B}^n(0)\sigma_\beta^{~\xi}\psi_{1B}^m(x)]|A,B\rangle
~+~x\leftrightarrow 0
\nonumber
\end{eqnarray}
Sorting out color-singlet terms, we get similarly to sum of Eqs. (\ref{v4}) and  (\ref{v5})
\begin{eqnarray}
&&\hspace{-1mm}
\cheW^{(2b)}_{2\mu\nu}~=~-{1\over s^3}(\delta_\mu^\alpha p_{2\nu}+\delta_\nu^\alpha p_{2\mu}-g_{\mu\nu}p_2^\alpha)
\Big(\langle \bsi B_j(x)\gamma_\alpha B_i\psi(0)\rangle_A
\nonumber\\
&&\hspace{-1mm}
\times~\langle\big(\bar\psi{1\over \alpha}\big)(0)\gamma^i\notp_2
\gamma^j{1\over\alpha}\psi(x)\rangle_A
+\psi(0)\otimes\psi(x)\leftrightarrow \gamma_5\psi(0)\otimes\gamma_5\psi(x)\Big)~+~x\leftrightarrow 0
\nonumber\\
&&\hspace{-1mm}
+~
{1\over s^3}(\delta_\mu^\alpha\delta_\nu^\beta+\delta_\nu^\alpha\delta_\mu^\beta-\half g_{\mu\nu}g^{\alpha\beta})
\nonumber\\
&&\hspace{-1mm}
\times~\Big\{-p_{2\alpha}
\langle\big(\bar\psi{1\over \alpha}\big)(0)\gamma^i\sigma_\star^{~k}
\gamma^j{1\over\alpha}\psi(x)\rangle_A  
 \langle \bar\psi B_j(x)\sigma_{\beta k} B_i\psi(0)\rangle_A
\nonumber\\
&&\hspace{-1mm} 
+~
\langle\big(\bar\psi{1\over \alpha}\big)(0)\gamma^i\sigma_{\star\alpha_\perp}
\gamma^j{1\over\alpha}\psi(x)\rangle_A 
\langle \bar\psi B_j(x)\sigma_{\beta \ast} B_i\psi(0)\rangle_A \Big\}  ~+~x\leftrightarrow 0
\label{v45be}
\end{eqnarray}
Starting from this point, all calculations repeat those of Sections \ref{sec:v4} and \ref{sec:v5} with replacements of 
$p_1\leftrightarrow p_2$, $\alpha_q\leftrightarrow\beta_q$ and exchange of  projectile matrix elements and the target ones. The result is 
Eq. (\ref{w2b1otvetrue}) with these replacements so we get
\begin{eqnarray}
&&\hspace{-1mm}
W_{2\mu\nu}^{(2b)}~
\label{w2b2otvetrue}\\
&&\hspace{-1mm}
=~
{4p_{1\mu}p_{1\nu}\over \beta_q^2s^2N_c}
\!\int\! d^2k_\perp\Big(
(q-k)_\perp^2\{f_1\barf_1+\barf_1f_1\}-2\beta_q\{ f_1\barf_\perp+\barf_1f_\perp\}
+2\beta_q^2m^2\{f_1\barf_3+\barf_1f_3\}\Big)
\nonumber\\
&&\hspace{-1mm}
-~(k,q-k)_\perp\Big( {(q-k)_\perp^2\over m^2} \{h_1^\perp\barh_1^\perp+\barh_1^\perp h_1^\perp\}+2\beta_q \{h_1^\perp\barh+\barh_1^\perp h \}
-2\beta_q^2\{h_1^\perp\barh_3^\perp+\barh_1^\perp h_3^\perp\}\Big)
\Big]
\nonumber
\end{eqnarray}
and therefore the contribution of Eq. (\ref{kalw2b}) takes the form
\begin{eqnarray}
&&\hspace{-1mm}
W_{\mu\nu}^{(2b)}~
\label{w2botvetrue}\\
&&\hspace{-1mm}=~
{2\over sN_c}
\!\int\! d^2k_\perp\Big\{{2p_{1\mu}p_{1\nu}\over \beta_q^2s}\Big(k_\perp^2\{f_1\barf_1+\barf_1f_1\}
-2\alpha_qk_\perp^2\{f_\perp\barf_1+\barf_\perp f_1\}+2\alpha_q^2m^2\{f_3\barf_1+\barf_3 f_1\}
\nonumber\\
&&\hspace{-1mm}
-~{(k,q-k)_\perp\over m^2}\big[k_\perp^2 \{h_1^\perp\barh_1^\perp+\barh_1^\perp h_1^\perp\}+2\alpha_q m^2\{h\barh_1^\perp+\barh h_1^\perp\}
-2\alpha_q^2m^2\{h_3^\perp\barh_1^\perp+\barh_3^\perp h_1^\perp\}\big]\Big)
\nonumber\\
&&\hspace{-1mm}
+~{2p_{2\mu}p_{2\nu}\over \alpha_q^2s}\Big((q-k)_\perp^2\{f_1\barf_1+\barf_1f_1\}
-2\beta_q(q-k)_\perp^2\{f_1\barf_\perp+\barf_1f_\perp\}+2\beta_q^2m^2\{f_1\barf_3+\barf_1f_3\}
\nonumber\\
&&\hspace{-1mm}
-~(k,q-k)_\perp\big[{(q-k)_\perp^2\over m^2} \{h_1^\perp\barh_1^\perp+\barh_1^\perp h_1^\perp\}
+2\beta_q \{h_1^\perp \barh+\barh_1^\perp h\}
-2\beta_q^2\{h_1^\perp\barh_3^\perp+\barh_1^\perp h_3^\perp\}\big]\Big)
\nonumber
\end{eqnarray}
%

\section{Result \label{sec:result}}
The resulting power correction with ${1\over Q^2},~{1\over N_c}$ accuracy is a sum of equations
(\ref{ququ}), (\ref{w1qtru}),  (\ref{w2qtru}), (\ref{w2atvet}), (\ref{w2botvetrue}), and (\ref{dobavkaresult}).
It is convenient to  present it as a sum of three terms
\beq
W_{\mu\nu}(q)~=~{1\over N_c}\sum_f e_f^2~\big[W^{1f}_{\mu\nu}(q)+W^{2f}_{\mu\nu}(q)
+W^{3f}_{\mu\nu}(q)\big]~+~O\big({1\over Q^3}\big)~+~O\big({1\over N_c^2}\big)
\label{finalresult}
\eeq
The first part  was calculated in Ref. \cite{Balitsky:2020jzt} while
 the second and third parts are the result of this paper.
 
The first part has the form \cite{Balitsky:2020jzt}
\begin{eqnarray}
&&\hspace{-1mm}
W^1_{\mu\nu}(q)~=~{1\over N_c}\!\int\!d^2k_\perp \Big(\calW^F_{\mu\nu}(q,k_\perp)\{f_1\barf_1+\barf_1 f_1\}
+~\calW^H_{\mu\nu}(q,k_\perp)\{h_1^\perp\barh_1^\perp+\barh_1^\perp h_1^\perp\}
\Big)
\label{result2}
\end{eqnarray}
where

\begin{eqnarray}
&&\hspace{-1mm}
\calW^{F}_{\mu\nu}(q,k_\perp)~=~
-g_{\mu\nu}^\perp +{1\over Q^2}(q^\parallel_\mu q^\perp_\nu+q^\parallel_\nu q^\perp_\mu)
+{q_\perp^2\over Q^4}q^\parallel_\mu q^\parallel_\nu+{\tilq_\mu\tilq_\nu\over Q^4}[q_\perp^2-4(k,q-k)_\perp]
\nonumber\\
&&\hspace{20mm}
-~\Big[{\tilq_\mu\over  Q^2}\Big(g^\perp_{\nu i}-{q^\parallel_\nu q_i\over Q^2}\Big)(q-2k)_\perp^i
+\mu\leftrightarrow\nu\Big]
\label{WF}
\end{eqnarray}
%
\begin{eqnarray}
&&\hspace{-1mm}
\calW_{\mu\nu}^H(q,k_\perp)~
\label{WH}\\
&&\hspace{-1mm}
=~-{1\over m^2}\big[k^\perp_\mu(q-k)^\perp_\nu-k^\perp_\nu(q-k)^\perp_\mu-g_{\mu\nu}^\perp(k,q-k)_\perp\big]
+2{\tilq_\mu\tilq_\nu-q^\parallel_\mu q^\parallel_\nu \over Q^4m^2}k_\perp^2(q-k)_\perp^2
\nonumber\\
&&\hspace{-1mm}
-~{1\over m^2Q^2}\big(q^\parallel_\mu\big[k_\perp^2(q-k)^\perp_\nu+k^\perp_\nu(q-k)_\perp^2\big]
+~\tilq_\mu\big[k_\perp^2(q-k)^\perp_\nu-k^\perp_\nu(q-k)_\perp^2\big]+\mu\leftrightarrow\nu\big)
\nonumber\\
&&\hspace{-1mm}
-~{\tilq_\mu\tilq_\nu+q^\parallel_\mu q^\parallel_\nu \over Q^4m^2}\big[q_\perp^2-2(k,q-k)_\perp\big](k,q-k)_\perp
-~{q^\parallel_\mu\tilq_\nu+ \tilq_\mu q^\parallel_\nu \over Q^4m^2}(2k-q,q)_\perp(k,q-k)_\perp
\nonumber
\end{eqnarray}
Here $q^\parallel_\mu\equiv \alpha_qp_1+\beta_qp_2$ and $\tilq_\mu\equiv \alpha_qp_1-\beta_qp_2$.  Note also
that $\alpha_q\equiv x_A$ and $\beta_q\equiv x_B$ in the notations of conventional TMD factorization formula (\ref{TMDf}).

In Eq. (\ref{finalresult}) we need to sum over flavors. From the Fierz transformation (\ref{fierz}) 
it is easy to see that power corrections calculated in this paper are diagonal in flavor so 
the final result is a sum of $e_f^2$ multiplied by TMDs of corresponding flavor,
 for example $\{f_1\barf_1+\barf_1 f_1\}$ in Eq. (\ref{result2}) should be replaced by 
$\sum_f e_f^2\{f_1^f\barf_1^f+\barf_1^f f_1^f\}$.  To avoid cluttering of formulas, this summation over flavors is written only once in Eq. (\ref{finalresult}).

It is easy to see that $q^\mu \calW_{\mu\nu}^F=q^\mu \calW_{\mu\nu}^H$ 
so the first part is EM gauge invariant. Note that gauge invariance of the leading-twist part $\sim O(1)$ 
is restored by adding $\sim O\big(1/Q\big)$ and $\sim O\big(1/Q^2\big)$ contributions in Eqs. (\ref{WF}) and (\ref{WH}).
One may call $W^F$ a ``gauge-invariant completion''  of the leading-twist result (\ref{WLT}). 
It worth noting that if one takes only the $``f_1''$ part of the result (\ref{WF}) 
\footnote{Because  the $Z$-boson current has an additional weak current component 
there are additional power corrections to Z-boson cross sections calculated in Ref. \cite{Balitsky:2021fer}   so
the ``$f_1$'' part will be slightly more complicated than Eq. (\ref{WF})}  
and performs back-of-the-envelope estimation of $Z$-boson angular distribution coefficients one gets reasonable agreement with LHC data with expected ${1\over N_c}$ accuracy. Needless to say, the angular coefficients are determined by ``gauge completion'' terms in Eq. (\ref{WF}) rather than the leading-twist term. 

The second part of the result (\ref{finalresult}) is 
\begin{eqnarray}
&&\hspace{-1mm}
W^2_{\mu\nu}(q)~=~{2\over N_c Q^2}\!\int\!d^2k_\perp 
\nonumber\\
&&\hspace{-1mm}
\times~\Big\{
\Big[\tilq_\mu(q-k)_\nu+{2\over\beta_q s}\tilq_\mu p_{1\nu}(k,q-k)_\perp
+{2\over\alpha_q s}\tilq_\mu p_{2\nu}(q-k)_\perp^2~+~\mu\leftrightarrow\nu\Big]
\nonumber\\
&&\hspace{22mm}
\times~
\Big(\beta_q\{ f_1\barf_\perp+\barf_1f_\perp\}-\alpha_q\{h\barh_1^\perp+\barh h_1^\perp\}\Big)
\nonumber\\
&&\hspace{-1mm}
+~\Big[\tilq_\mu k^\perp_\nu+{2\over s\beta_q}k_\perp^2\tilq_\mu p_{1\nu}
+{2\over s\alpha_q}(k,q-k)_\perp\tilq_\mu p_{2\nu}~+~\mu\leftrightarrow\nu\Big]
\nonumber\\
&&\hspace{22mm}
\times~\Big(-\alpha_q\{f_\perp\barf_1+\barf_\perp f_1\}+\beta_q\{h^\perp_1\barh+\barh^\perp_1h\}\Big)
\nonumber\\
&&\hspace{-1mm}
+~{4\tilq_\mu\tilq_\nu\over Q^2}\Big[m^2\Big(
\alpha_q^2\{f_3\barf_1+\barf_3 f_1\}+\beta_q^2\{f_1\barf_3+\barf_1 f_3\}
+\alpha_q\beta_q\big[\{e\bare+\bare e\}+\{h\barh+\barh h \}\big]\Big)
\nonumber\\
&&\hspace{22mm}
+~(k,q-k)_\perp\Big(-\alpha_q\beta_q\big[\{f_\perp\barf_\perp+\barf_\perp f_\perp\}+\{g_\perp\barg_\perp+\barg_\perp g_\perp\}\big]
\nonumber\\
&&\hspace{22mm}
+~\beta_q^2\{h_1^\perp\barh_3^\perp+\barh_1^\perp h_3^\perp\}
+\alpha_q^2\{h_3^\perp\barh_1^\perp+\barh_3^\perp h_1^\perp\}\Big)\Big]
\nonumber\\
&&\hspace{-1mm}
+~{1\over m^2}
\calW^\perp_{\mu\nu}(q,k_\perp)\Big[{2\over\alpha_q}\{h_1^\perp\Re\barh_{1\pizg}+\barh_1^\perp\Re h_{1\pizg}\}
+{2\over\beta_q}\{\Re h_{1\pizg}\barh_1^\perp+\Re\barh_{1\pizg}h_1^\perp\}
\nonumber\\
&&\hspace{22mm}
+~\Re\big(\{h_{1G}^\perp\barh_{1G}^\perp+\barh_{1G}^\perp h_{1G}^\perp\}\big)\Big]
\label{wedva}
\end{eqnarray}
where the transverse structure $\calW^\perp_{\mu\nu}(q,k_\perp)$ was defined in Eq. (\ref{calweperp}):
\begin{eqnarray}
&&\hspace{-1mm}
\calW^\perp_{\mu\nu}(q_\perp,k_\perp)~\equiv~g_{\mu\nu}^\perp(k,q-k)_\perp^2
-g_{\mu\nu}^\perp k_\perp^2(q-k_\perp)^2
\nonumber\\
&&\hspace{-1mm}
+~[k^\perp_\mu(q-k)^\perp_\nu+\mu\leftrightarrow\nu](k,q-k)_\perp
-k_\perp^2(q-k)^\perp_\mu(q-k)^\perp_\nu-~(q-k_\perp)^2k^\perp_\mu k^\perp_\nu
\label{calweperpe}
\end{eqnarray}

This EM-gauge invariant part consists of two types of contributions: $\sim{1\over Q}$ (first terms
in the second and fourth lines) and $\sim{1\over Q^2}$ (all other terms). Note that, except for the last term,  it is determined by 
quark-antiquark TMDs.

The remaining third  part of the result (\ref{finalresult}) has the form
\begin{eqnarray}
&&\hspace{-1mm}
W^{3}_{\mu\nu}(q)~=~{2\over N_c Q^2}\!\int\!d^2k_\perp \bigg[g^\perp_{\mu\nu}\bigg\{
m^2\alpha_q\beta_q\big(\{ h\barh+ \barh h\}-\{e\bare+\bare e\}\big)
\nonumber\\
&&\hspace{13mm}
+~\alpha_q\beta_qm^2\big[
\Re f_D(\alpha_q,k_\perp)\barf'_1(\beta_q,q_\perp-k_\perp) 
+\Re \barf_D(\alpha_q,k_\perp)f'_1(\beta_q,q_\perp-k_\perp) 
\nonumber\\
&&\hspace{13mm}
+~f'_1(\alpha_q,k_\perp) \Re \barf_D(\beta_q,q_\perp-k_\perp)
+\barf'_1(\alpha_q,k_\perp) \Re f_D(\beta_q,q_\perp-k_\perp)
\big]
\nonumber\\
&&\hspace{13mm}
-~m^2\beta_q\big(\{f_1\Re\barf_{2\pizg}+\barf_1\Re f_{2\pizg}\}+\{f_1\Re\barf_{3\pizg}+\barf_1\Re f_{3\pizg}\}\big)
\nonumber\\
&&\hspace{13mm}
-~m^2\alpha_q\big(\{\Re\barf_{2\pizg}f_1+\Re f_{2\pizg}\barf_1\}+\{\Re\barf_{3\pizg}f_1+\Re f_{3\pizg}\barf_1\}\big)
\nonumber\\
&&\hspace{-1mm}
+~(k,q-k)_\perp\Big(-\beta_q\{h_1^\perp\barh+\barh_1^\perp h\}
-\alpha_q\{h\barh_1^\perp+\barh h_1^\perp\}+2\beta_q^2
\{h_1^\perp\barh_3^\perp+\barh_1^\perp h_3^\perp\}
\nonumber\\
&&\hspace{-1mm}
+~2\alpha_q^2\{h_3^\perp\barh_1^\perp+\barh_3^\perp h_1^\perp\}
+2\beta_q\{\Im \ace_G \barh_1^\perp+\Im \bar\ace_G h_1^\perp\}
+2\alpha_q\{h_1^\perp\Im \bar\ace_G+\barh_1^\perp\Im \ace_G \}
\nonumber\\
&&\hspace{-1mm}
-~\beta_q\{f_1\Re\bar\acf_{1\pizg}+\barf_1\Re\acf_{1\pizg}\}
-\alpha_q\{\Re\bar\acf_{1\pizg}f_1+\Re\acf_{1\pizg}\barf_1\}
-\alpha_q\beta_q\big[\{f_\perp\Re\bar\graf_{1\pizg}+\barf_\perp\Re\graf_{1\pizg}\}
\nonumber\\
&&\hspace{-1mm}
+~\{g_\perp\Im\bar\graf_{1\pizg}-\barg_\perp\Im\graf_{1\pizg}\} 
+\{\Re\bar\graf_{1\pizg}f_\perp+\Re\graf_{1\pizg}\barf_\perp\}
+\{\Im\bar\graf_{1\pizg}g_\perp-\Im\graf_{1\pizg}\barg_\perp\} \big]\bigg\}
\nonumber\\
&&\hspace{-1mm}
+~\big[g^\perp_{\mu\nu}(k,q-k)_\perp+k^\perp_\mu(q-k)^\perp_\nu+k^\perp_\nu(q-k)^\perp_\mu\big]
\nonumber\\
&&\hspace{-1mm}
\times~\bigg\{\alpha_q\beta_q\big[\{f_\perp\barf_\perp+\barf_\perp f_\perp\}-\{g_\perp\barg_\perp+\barg_\perp g_\perp\}\big]
\nonumber\\
&&\hspace{-1mm}
+~\alpha_q\beta_q\big[\Re h_D(\alpha_q,k_\perp)\bar{h'}_1^\perp(\beta_q,q_\perp-k_\perp)
+\Re\barh_D(\alpha_q,k_\perp){h'}^\perp_1(\beta_q,q_\perp-k_\perp)
\nonumber\\
&&\hspace{25mm}
+~{h'}^\perp_1(\alpha_q,k_\perp)\Re\barh_D(\beta_q,q_\perp-k_\perp)
+\bar{h'}_1^\perp(\alpha_q,k_\perp)\Re h_D(\beta_q,q_\perp-k_\perp)
\big]
\nonumber\\
&&\hspace{-1mm}
+~
\beta_q\{h_1^\perp\Re\barh_{3\pizg}+\barh_1^\perp\Re h_{3\pizg}\}+
\beta_q\{h_1^\perp\Im\barh_{4\pizg}
+\barh_1^\perp\Im h_{4\pizg}\}
\nonumber\\
&&\hspace{22mm}
+~\alpha_q\{\Re\barh_{3\pizg}h_1^\perp+\Re h_{3\pizg}\barh_1^\perp\}+
\alpha_q\{\Im\barh_{4\pizg}h_1^\perp
+\Im h_{4\pizg}\barh_1^\perp\}\bigg\}
\nonumber\\
&&\hspace{-1mm}
-~[g^\perp_{\mu\nu}k_\perp^2+2k_\mu k_\nu]
\nonumber\\
&&\hspace{-1mm}
\times~\Big({(q-k)_\perp^2\over \alpha_qm^2}\{h_1^\perp\Re\barh_{2\pizg}+\barh_1^\perp\Re h_{2\pizg}\}
+\{\Re h_{1\pizg}\barh+\Re\barh_{1\pizg}h\}+\{\Im h_{1\pizg}\bare-\Im \barh_{1\pizg}e\}\Big)
\nonumber\\
&&\hspace{-1mm}
-~\big[2(q-k)^\perp_\mu (q-k)^\perp_\nu+g^\perp_{\mu\nu}(q-k)_\perp^2\big]
\nonumber\\
&&\hspace{-1mm}
\times~\Big({k_\perp^2\over\beta_qm^2}
\{\Re h_{2\pizg}\barh_1^\perp+\Re\barh_{2\pizg}h_1^\perp\}
+(\{h\Re\barh_{1\pizg}+\barh\Re h_{1\pizg}\}+\{\bare\Im h_{1\pizg}-e\Im \barh_{1\pizg}\}\bigg]
\label{otvetadd}
\ega

This contribution is not gauge invariant: $q^\mu W^3_{\mu\nu}(q)\neq 0$. This is hardly surprising since 
from DVCS studies we know  that check of EM gauge invariance sometimes involves
cancellation of contributions of different twists (see e.g. Refs. \cite{Ji:1998pc,Guichon:1998xv, Anikin:2000em,Penttinen:2000dg,Belitsky:2000vx,Radyushkin:2000jy,Radyushkin:2000ap}).  
Still, 
the non-gauge-invariant contribution (\ref{otvetadd}) is proportional to transverse structures 
so the violation of gauge invariance is
$$
q^\mu W_{\mu\nu}(q)~=~q^\mu W^3_{\mu\nu}(q)~=~O\big({q_\perp\over Q^2}\big)
$$
Note that if, for example, we would have $g^\parallel_{\mu\nu}$ instead of $g^\perp_{\mu\nu}$ in Eq. (\ref{otvetadd}),
the violation of gauge invariance would be $\sim{1\over Q}$. The absence of such terms is a result of many cancellations involving
QCD equations of motion.
Thus, the EM gauge invariance of $W^3_{\mu\nu}(q)$ is restored by $\sim{1\over Q^3}$ and $\sim{1\over Q^4}$ corrections so one may say that at the $\sim{1\over Q^2}$ level our result (\ref{finalresult}) 
satisfies the requirement of EM gauge invariance.

Last but not least, let us discuss  the choice of basis of operators for ${1\over Q^2}$ corrections. Unlike ${1\over Q}$ corrections which are unique, one can represent  ${1\over Q^2}$ corrections in many different ways using QCD equations of motion
\footnote{As was mentioned in footnote \ref{oboz}, in this paper QCD  coupling constant $g$ is included in the definition of gluon field $A_\mu$.}
\begin{eqnarray}
&&\hspace{-1mm}
\bsi(x)\slA_\perp(x)~=~i\partial_i\bsi(x)\gamma_i+i{2\over s}\partial_\star\bsi(x)\slp_1
+i\bsi \stackrel{\leftarrow}D_\bu{2\over s}\slp_2,
\nonumber\\
&&\hspace{-1mm}
\slA_\perp(x)\psi(x)~=~-i\slpart_\perp\psi(x)-i{2\over s}\slp_1\partial_\star\psi(x)-i{2\over s}\slp_2D_\bu\psi(x)
\label{eqmp}
\ega
for the projectile operators in $A_\star=0$ gauge and
\begin{eqnarray}
&&\hspace{-1mm}
\bsi(x)\slB_\perp(x)~=~i\partial_i\bsi(x)\gamma_i+i{2\over s}\partial_\bu\bsi(x)\slp_2
+i\bsi \stackrel{\leftarrow}D_\star\!(x){2\over s}\slp_1,
\nonumber\\
&&\hspace{-1mm}
\slB_\perp(x)\psi(x)~=~-i\slpart_\perp\psi(x)-i{2\over s}\slp_2\partial_\bu\psi(x)-i{2\over s}\slp_1D_\star\psi(x)
\label{eqmt}
\ega
for operators in target matrix elements (in $B_\bu=0$ gauge). 
The choice in this paper is to reduce the l.h.s.'s of these EOMs to the r.h.s.'s whenever possible. This choice 
leads to the ``gauge completion'' (\ref{result2}) of the leading-twist result and helps with sorting out the ${1\over Q^2}$ corrections.
It should be mentioned that there is a different choice of basis of operators in the literature: in Refs. 
\cite{Vladimirov:2021hdn}, \cite{Rodini:2022wic}, \cite{Vladimirov:2023aot} the quark-antiquark TMDs of non-leading twist 
are expressed in terms of quark-antiquark-gluon TMDs using EOMs like Eq. (\ref{eqmp}). That choice is motivated by the fact that the evolution of $\barq q$ TMDs of non-leading twist involves $\barq q G$ TMDs anyway \cite{Rodini:2022wic}.
Ideally, to find the optimal basis of operators one should diagonalize the matrix of evolution equations of twist-4 TMDs and find those combinations which evolve like leading-twist TMDs which is a formidable task. 
\footnote{In the case of DVCS governed by the GPD evolution, there is a byway to find light-ray operators which evolve like 
the leading-twist ones using conformal $SL(2,R)$ invariance \cite{Braun:2011zr,Braun:2012hq,Braun:2014sta}. Unfortunately, this method is not applicable to TMD operators that are not   $SL(2,R)$ invariant.
}
In my opinion, it is useful  first to try
to assemble the   power corrections in gauge-invariant blocks like (\ref{result2}) and (\ref{wedva}) using QCD equations of motion. Of course, it is quite probable that among 
 the  ``leftovers'' entering Eq. (\ref{otvetadd}) there will be TMD combinations which evolve like, say, $f_1$ but 
 this is in agreement with our earlier statement that  the EM gauge invariance of $W^3_{\mu\nu}(q)$ is restored by $\sim{1\over Q^3}$ and $\sim{1\over Q^4}$ corrections.

\section{Conclusions and outlook \label{sec:coutlook}}

The result of this paper is a complete set of ${1\over Q^2}$ power corrections to TMD factorization for Drell-Yan process 
at the leading $N_c$ order. Let me emphasize that tree-level formulas of this paper are valid at both moderate  and small Bjorken $x$. The difference between
these two cases comes from different evolution of TMDs in the moderate-$x$ and small-$x$ regions, 
see the discussion in Ref. \cite{Balitsky:2023hmh}. Moreover, while the results of this paper were obtained using 
rapidity-only factorization, at the tree level they should be the same as  obtained by conventional CSS approach. Indeed, ${1\over Q}$ corrections
are the same as in Ref. \cite{Mulders:1995dh} and parts of ${1\over Q^2}$ corrections coincide with Ref.  \cite{Vladimirov:2023aot} after taking into account different choice of operator basis, see the discussion in previous Section.

One may wonder what can be a possible way to compare the result (\ref{finalresult}) with experimental data on DY process. There are phenomenological estimates of leading-twist TMDs \cite{Bacchetta:2017gcc,Scimemi:2019cmh, Moos:2023yfa,Bacchetta:2022awv}, but
due to the large number
of quark-antiquark-gluon TMDs involved, similar extraction of $\barq qG$ TMDs  from experiment  seems nearly impossible. On the other hand, there are attempts to 
calculate quark-antiquark TMDs on the lattice \cite{Musch:2011er,LPC:2022zci,Shu:2023cot} (see also the review \cite{Lin:2017snn})
 and one may expect to get lattice estimates of quark-quark-gluon TMDs in the future. 
It is well known that lattice calculations are not reliable at small $x$, so the moderate-$x$ result of this paper (\ref{finalresult}) may serve as a bridge between lattice 
calculations and experimental data.

An obvious outlook is to extend these results to the semi-inclusive deep inelastic scattering (SIDIS)  at 
EIC and elsewhere.  The study is in progress.

The author  is grateful to members of CERN TH department for kind hospitality
and to  A. Vladimirov for helpful discussions. This  work is
 supported by Jefferson Science Associates, LLC under the U.S. DOE contract \#DE-AC05-06OR23177
 and by U.S. DOE grant \#DE-FG02-97ER41028.

\section{Appendix}
\subsection{Formulas with Dirac matrices \label{diracs}}
\subsubsection{Fierz transformation}
First, let us write down Fierz transformation for symmetric hadronic tensor
\begin{eqnarray}
&&\hspace{-1mm}
\half[(\bsi\gamma_\mu\chi)(\bhi\gamma_\nu\psi)+\mu\leftrightarrow\nu]
\label{fierz}\\
&&\hspace{-1mm}
=~
-{1\over 4}\big(\delta_\mu^\alpha\delta_\nu^\beta+\delta_\nu^\alpha\delta_\mu^\beta-g_{\mu\nu}g^{\alpha\beta}\big)
\big[(\bsi\gamma_\alpha\psi)(\bhi\gamma_\beta\chi)
+(\bsi\gamma_\alpha\gamma_5\psi)(\bhi\gamma_\beta\gamma_5\chi)\big]
\nonumber\\
&&\hspace{-1mm}
+~{1\over 4}\big(\delta_\mu^\alpha\delta_\nu^\beta+\delta_\nu^\alpha\delta_\mu^\beta-\half g_{\mu\nu}g^{\alpha\beta}\big)
(\bsi\sigma_{\alpha\xi}\psi)(\bhi\sigma_\beta^{~\xi}\chi)-{g_{\mu\nu}\over 4}(\bsi\psi)(\bhi\chi)+{g_{\mu\nu}\over 4}(\bsi\gamma_5\psi)(\bhi\gamma_5\chi)
\nonumber
\end{eqnarray}
%

\subsubsection{Formulas with $\sigma$-matrices \label{sigmaflas}}
It is convenient to define
\footnote{We use conventions from {\it Bjorken \& Drell} where $\epsilon^{0123}=-1$ and
$
\gamma^\mu\gamma^\nu\gamma^\lambda=g^{\mu\nu}\gamma^\lambda +g^{\nu\lambda}\gamma^\mu-g^{\mu\lambda}\gamma^\nu
-i\epsilon^{\mu\nu\lambda\rho}\gamma_\rho\gamma_5
$.
Also, with this convention $\tigma_{\mu\nu}\equiv\half \epsilon_{\mu\nu\lambda\rho}\sigma^{\lambda\rho}=i\sigma_{\mu\nu}\gamma_5$.
}
\begin{equation}
\epsilon_{ij}~\equiv~ {2\over s}\epsilon_{\star\bu ij}~=~ {2\over s}p_2^\mu p_1^\nu\epsilon_{\mu\nu ij}
\label{eps2}
\end{equation}
such that $\epsilon_{12}~=~1$ and $\epsilon_{ij}\epsilon_{kl}~=~g_{ik}g_{jl}-g_{il}g_{jk}$. 
The frequently used formula is 
\begin{equation}
\hspace{-1mm}
\sigma_{\mu\nu}\sigma_{\alpha\beta}~=~(g_{\mu\alpha}g_{\nu\beta}-g_{\mu\beta}g_{\nu\alpha})-i\epsilon_{\mu\nu\alpha\beta}\gamma_5
-i(g_{\mu\alpha}\sigma_{\nu\beta}-g_{\mu\beta}\sigma_{\nu\alpha}-g_{\nu\alpha}\sigma_{\mu\beta}+g_{\nu\beta}\sigma_{\mu\alpha})
\label{sigmasigma}
\end{equation}

We need also the following formulas with $\sigma$-matrices in different matrix elements
\begin{eqnarray}
&&\hspace{-1mm}
\tigma_{\mu\nu}\otimes\tigma_{\alpha\beta}~
=~-\half(g_{\mu\alpha}g_{\nu\beta}-g_{\nu\alpha}g_{\mu\beta})\sigma_{\xi\eta}\otimes\sigma^{\xi\eta}
\nonumber\\
&&\hspace{-1mm}
+~g_{\mu\alpha}\sigma_{\beta\xi}\otimes\sigma_\nu^{~\xi}-g_{\nu\alpha}\sigma_{\beta\xi}\otimes\sigma_\mu^{~\xi}-g_{\mu\beta}\sigma_{\alpha\xi}\otimes\sigma_\nu^{~\xi}+g_{\nu\beta}\sigma_{\alpha\xi}\otimes\sigma_\mu^{~\xi}
-\sigma_{\alpha\beta}\otimes\sigma_{\mu\nu}
\label{tigmi}
\end{eqnarray}

and 
\begin{eqnarray}
&&\hspace{-1mm}
\tigma_{\mu\xi}\otimes\tigma_\nu^{~\xi}~=~-{g_{\mu\nu}\over 2}\sigma_{\xi\eta}\otimes\sigma^{\xi\eta}+\sigma_{\nu\xi}\otimes\sigma_\mu^{~\xi}
,~~~~\sigma_{\xi\eta}\otimes\tigma^{\xi\eta}~=~\tigma_{\xi\eta}\otimes\sigma^{\xi\eta}
\label{formulaxz}\\
&&\hspace{-1mm}
\sigma_{\mu\xi}\gamma_5\otimes\sigma_\nu^{~\xi}\gamma_5+\mu\leftrightarrow\nu-{g_{\mu\nu}\over 2}\sigma_{\xi\eta}\gamma_5\otimes\sigma^{\xi\eta}\gamma_5
~=~-[\sigma_{\mu\xi}\otimes\sigma_\nu^{~\xi}+\mu\leftrightarrow\nu-{g_{\mu\nu}\over 2}\sigma_{\xi\eta}\otimes\sigma^{\xi\eta}]
\nonumber
\end{eqnarray}
\begin{eqnarray}
&&\hspace{-1mm}
\sigma_\star^{~k}\otimes\gamma_i\sigma_{\bu k}\gamma_j~=~\hatp_2\gamma^k\otimes\notp_1\gamma_i\gamma_k\gamma_j
~=~\hatp_2\gamma^k\otimes\notp_1(g_{ik}\gamma_j+g_{jk}\gamma_i-g_{ij}\gamma_k)
\nonumber\\
&&\hspace{-1mm}
=~\hatp_2(g_{ik}\gamma_j+g_{jk}\gamma_i-g_{ij}\gamma_k)\otimes \notp_1\gamma^k~=~(\gamma_j\sigma_\star^{~k}\gamma_i)\otimes\sigma_{\bu k}
\label{sisigaga}
\end{eqnarray}
We will need also
\begin{eqnarray}
&&\hspace{-1mm}
\notp_2\otimes\gamma_i\notp_1\gamma_j+\notp_2\gamma_5\otimes\gamma_i\notp_1\gamma_j\gamma_5
~=~\gamma_j\notp_2\gamma_i\otimes\notp_1+\gamma_j\notp_2\gamma_i\gamma_5\otimes\notp_1\gamma_5
\label{flagamma}
\end{eqnarray}
and 
\begin{eqnarray}
&&\hspace{-1mm}
i\sigma_{\alpha\xi}\sigma_{\star i}\otimes\sigma_\beta^{~\xi}
\nonumber\\
&&\hspace{-1mm}
=~
g_{\alpha i}\sigma_{\star j}\otimes\sigma_{\beta_\perp}^{~j}+\sigma_{\alpha_\perp\ast}\otimes\sigma_{\beta_\perp i}  
-{2\over s}p_{2\alpha}\sigma_{\star i}\otimes\sigma_{\bu\beta_\perp}
+{2\over s}p_{2\beta}g_{\alpha i}\sigma_{\star j}\otimes\sigma_\bu^{~j}+{2\over s}p_{2\beta}\sigma_{\alpha_\perp\ast}\otimes\sigma_{\bu i} 
\nonumber\\
&&\hspace{-1mm}
+~{2\over s}p_{2\alpha}p_{2\beta}\big[i\otimes\sigma_{\bu i} -\sigma_{ij}\otimes\sigma_\bu^{~j}
\big]
+{4\over s^2}p_{2\alpha}p_{2\beta}\sigma_{\bu\ast}\otimes\sigma_{\bu i}
 -{2\over s}g^\parallel_{\alpha\beta}\sigma_{\star i}\otimes\sigma_{\bu\ast}+...,
\nonumber\\
&&\hspace{-1mm}
i\sigma_{\star i}\sigma_{\alpha\xi}\otimes\sigma_\beta^{~\xi}
\label{hformulas}\\
&&\hspace{-1mm}
=~
{2\over s}p_{2\alpha}\sigma_{\star i}\otimes\sigma_{\bu \beta_\perp}
-g_{i\alpha}\sigma_{\star j}\otimes\sigma_{\beta_\perp}^{~j}
+\sigma_{\star\alpha_\perp}\otimes\sigma_{\beta_\perp i}  
+{2\over s}p_{2\beta}\sigma_{\star\alpha_\perp}\otimes\sigma_{\bu i}
-{2\over s}g_{i\alpha}p_{2\beta}\sigma_{\star j}\otimes\sigma_\bu^{~j}
\nonumber\\
&&\hspace{-1mm}
+~{4\over s^2}p_{2\alpha}p_{2\beta}\sigma_{\star\bu}\otimes\sigma_{\bu i} 
+{2\over s}p_{2\alpha}p_{2\beta}\big[i\otimes\sigma_{\bu i} +\sigma_{ij}\otimes\sigma_\bu^{~j}
\big]
-{2\over s}g^\parallel_{\alpha\beta}\sigma_{\star i}\otimes\sigma_{\star\bu}~+~...
\nonumber
\end{eqnarray}
and
\begin{eqnarray}
&&\hspace{-1mm}
i\sigma_{\bu i}\sigma_{\alpha\xi}\otimes\sigma_\beta^{~\xi}
\nonumber\\
&&\hspace{-1mm}
=~
{2\over s}p_{1\alpha}\sigma_{\bu i}\otimes\sigma_{\star \beta_\perp}
-g_{i\alpha}\sigma_{\bu j}\otimes\sigma_{\beta_\perp}^{~j}
+\sigma_{\bu\alpha_\perp}\otimes\sigma_{\beta_\perp i}  
+{2\over s}p_{1\beta}\sigma_{\bu\alpha_\perp}\otimes\sigma_{\star i}
-{2\over s}g_{i\alpha}p_{1\beta}\sigma_{\bu j}\otimes\sigma_\star^{~j}
\nonumber\\
&&\hspace{-1mm}
+~{4\over s^2}p_{1\alpha}p_{1\beta}\sigma_{\bu\ast}\otimes\sigma_{\star i} 
+{2\over s}p_{1\alpha}p_{1\beta}\big[i\otimes\sigma_{\star i} +\sigma_{ij}\otimes\sigma_\star^{~j}
\big]
-{2\over s}g^\parallel_{\alpha\beta})\sigma_{\bu i}\otimes\sigma_{\bu\ast}+...,
\nonumber\\
&&\hspace{-1mm}
i\sigma_{\alpha\xi}\sigma_{\bu i}\otimes\sigma_\beta^{~\xi}
\label{mainfla3}\\
&&\hspace{-1mm}
=~
g_{\alpha i}\sigma_{\bu j}\otimes\sigma_{\beta_\perp}^{~j}+\sigma_{\alpha_\perp\bu}\otimes\sigma_{\beta_\perp i}  
-{2\over s}p_{1\alpha}\sigma_{\bu i}\otimes\sigma_{\star\beta_\perp}
+{2\over s}p_{1\beta}g_{\alpha i}\sigma_{\bu j}\otimes\sigma_\star^{~j}+{2\over s}p_{1\beta}\sigma_{\alpha_\perp\ast}\otimes\sigma_{\star i} 
\nonumber\\
&&\hspace{-1mm}
+~
{2\over s}p_{1\alpha}p_{1\beta}\big[i\otimes\sigma_{\star i}+\sigma_{ij}\otimes\sigma_{\star}^{~j}\big] +{4\over s^2}p_{1\alpha}p_{1\beta}\sigma_{\star\bu}\otimes\sigma_{\star i}
 -{2\over s}g^\parallel_{\alpha\beta}\sigma_{\bu i}\otimes\sigma_{\star\bu}
+...
\nonumber
\end{eqnarray}
From these equations after some algebra one obtains
\begin{eqnarray}
&&\hspace{-1mm}
\sigma_{\mu\xi}\sigma_{\star i}\otimes\sigma_\nu^{~\xi}\sigma_{\bu j}~
=~-g_{\mu i}g_{\nu j}\sigma_{\star k}\otimes\sigma_{\bu}^{~k}
+g_{\mu i}\sigma_{\star j}\otimes\sigma_{\bu \nu_\perp}
\label{mainfla4}\\
&&\hspace{-1mm}
+~g_{\nu_\perp j}\sigma_{\star\mu_\perp}\otimes\sigma_{\bu i} 
-g_{ij}\sigma_{\star\mu_\perp}\otimes\sigma_{\bu\nu_\perp}
-g^\parallel_{\mu\nu}\sigma_{\star i}\otimes\sigma_{\bu j}~+~...
\nonumber
\end{eqnarray}

The dots in the above formulas stand for the terms leading to contributions to $W_{\mu\nu}$ exceeding our $1/Q^2$accuracy.

\subsubsection{Formulas with $\gamma$-matrices and one gluon field}
In the gauge $A_\bu=0$ the field $A_i$ can be represented as 
\begin{equation}
A_i(x_\bu,x_\perp)~=~{2\over s}\!\int_{-\infty}^{x_\bu}\!dx'_\bu~F^{(A)}_{\star i}(x'_\bu,x_\perp)
\label{ai}
\end{equation}
Similarly, in the $B_\star=0$ gauge
\begin{equation}
B_i(x_\bu,x_\perp)~=~{2\over s}\!\int_{-\infty}^{x_\star}\!dx'_\star~F^{(B)}_{\bu i}(x'_\star,x_\perp)
\label{bi}
\end{equation}

We define ``dual'' fields by
\begin{equation}
\tilde{A}_i(x_\bu,x_\perp)~=~{2\over s}\!\int_{-\infty}^{x_\bu}\!dx'_\bu~\tilde{F}^{(A)}_{\star i}(x'_\bu,x_\perp),
~~~\tilde{B}_i(x_\star,x_\perp)~=~{2\over s}\!\int_{-\infty}^{x_\star}\!dx'_\star~\tilde{F}^{(B)}_{\bu i}(x'_\star,x_\perp),
\label{abefildz}
\end{equation}
 where $\tilF_{\mu\nu}=\half\epsilon_{\mu\nu\lambda\rho}F^{\lambda\rho}$ as usual. 
With this definition we have 
\beq
\tilA_i=-\epsilon_{ij}A^j,~~~\tilB_i=\epsilon_{ij}B^j,~~~
\epsilon_{ij}\tilA^j=A_i,~~~\epsilon_{ij}\tilB^j=-B_i
\label{deftildas}
\eeq
 and
\begin{equation}
\notp_2\grA_i~=~-\notA\notp_2\gamma_i,~~~~\grA_i\notp_2~=~-\gamma_i\notp_2\notA,~~~~
\notp_1\graB_i~=~-\notB\notp_1\gamma_i,~~~~\graB_i\notp_1~=~-\gamma_i\notp_1\notB
\label{glavla}
\end{equation}
where $\grA_i, \graB_i$ are defined in Eq. (\ref{agrebi}).

We also use
\begin{eqnarray}
&&\hspace{-1mm}
A^i\notp_2\otimes\gamma_n\notp_1\gamma_i+~A^i\notp_2\gamma_5\otimes\gamma_n\notp_1\gamma_i\gamma_5
~=~-\notp_2\grA_n\otimes \notp_1-\notp_2\grA_n\gamma_5\otimes \notp_1\gamma_5
\nonumber\\
&&\hspace{-1mm}
A^i\notp_2\otimes\gamma_i\notp_1\gamma_n+~A^i\notp_2\gamma_5\otimes\gamma_i\notp_1\gamma_n\gamma_5
~=~-\grA_n\notp_2\otimes \notp_1-\grA_n\notp_2\gamma_5\otimes \notp_1\gamma_5
\nonumber\\
&&\hspace{-1mm}
\gamma_n\slashed{p}_2\gamma^i\otimes\slashed{p}_1 B_i
~+~ \gamma_n\slashed{p}_2\gamma^i\gamma_5 \otimes\slashed{p}_1\gamma_5 B_i
~=~-\slashed{p}_2\otimes \slashed{p}_1\graB_n -
\slashed{p}_2 \gamma_5\otimes \slashed{p}_1\graB_n\gamma_5
\nonumber\\
&&\hspace{-1mm}
\gamma^i\slashed{p}_2\gamma_n \otimes\slashed{p}_1B_i
~+~\gamma^i \slashed{p}_2\gamma_n\gamma_5 \otimes\slashed{p}_1\gamma_5 B_i
~=~-\slashed{p}_2\otimes \graB_n\slashed{p}_1 -
\slashed{p}_2 \gamma_5\otimes \graB_n\slashed{p}_1\gamma_5
\label{gammas1fild}
\end{eqnarray}
and
\begin{eqnarray}
&&\hspace{-1mm}
{2\over s}\big[\notp_1\notp_2\gamma_i\otimes B^i\gamma_\nu
+\notp_1\notp_2\gamma_i\gamma_5\otimes B^i\gamma_\nu\gamma_5\big]
~=~\gamma_i\otimes\gamma_\nu\graB_i+\gamma_i\gamma_5\otimes\gamma_\nu\graB_i\gamma_5
\nonumber\\
&&\hspace{-1mm}
{2\over s}\big[\gamma_i\notp_2\notp_1\otimes B^i\gamma_\nu
+\gamma_i\notp_2\notp_1\gamma_5\otimes B^i\gamma_\nu\gamma_5\big]
~=~\gamma_i\otimes\graB_i\gamma_\nu+\gamma_i\gamma_5\otimes\graB_i\gamma_\nu\gamma_5
\nonumber\\
&&\hspace{-1mm}
{2\over s}\big[\notp_2\notp_1\gamma_i\otimes B^i\slp_1
+\notp_2\notp_1\gamma_i\gamma_5\otimes B^i\slp_1\gamma_5\big]
~=~\gamma_i\otimes\graB_i\slp_1+\gamma_i\gamma_5\otimes\graB_i\slp_1\gamma_5
\nonumber\\
&&\hspace{-1mm}
{2\over s}\big[\notp_1\notp_2\gamma_i\otimes B^i\slp_1
+\notp_1\notp_2\gamma_i\gamma_5\otimes B^i\slp_1\gamma_5\big]
~=~\gamma_i\otimes\slp_1\graB_i+\gamma_i\gamma_5\otimes\slp_1\graB_i\gamma_5
\label{14.19}
\end{eqnarray}
and
\begin{eqnarray}
&&\hspace{-1mm}
\gamma_\alpha\notp_2\gamma_i\otimes B_i\gamma_\beta+\gamma_\alpha\notp_2\gamma_i\gamma_5\otimes B_i\gamma_\beta\gamma_5
\label{fformula1}\\
&&\hspace{-1mm}
=~
-\slashed{p}_2\otimes\gamma_{\beta_\perp}\graB_\alpha
+{2\over s}p_{2\beta}\slashed{p}_2\otimes\slB\slashed{p}_1\gamma_{\alpha_\perp}
-{2\over s}p_{2\alpha}p_{2\beta}\big[\gamma_i\otimes\slB\slp_1\gamma_i+\gamma_i\gamma_5\otimes\slB\slp_1\gamma_i\gamma_5\big]
~+~...,
\nonumber\\
&&\hspace{-1mm}
\gamma_i\notp_2\gamma_\alpha\otimes B_i\gamma_\beta+\gamma_i\notp_2\gamma_\alpha\gamma_5\otimes B_i\gamma_\beta\gamma_5
\nonumber\\
&&\hspace{-1mm}
=~-\slashed{p}_2\otimes\graB_\alpha\gamma_{\beta_\perp}+{2\over s}p_{2\beta}\slashed{p}_2\otimes\gamma_\alpha\slp_1\slB
-{2\over s}p_{2\alpha}p_{2\beta}\big[\gamma_i\otimes\gamma_i\slp_1\slB
+\gamma_i\gamma_5\otimes\gamma_i\slp_1\slB \gamma_5\big]
~+~...
\nonumber
\end{eqnarray}
where the dots stand for the negligible terms as usual. Let us illustrate the derivation of the first of these equations.
After some algebra one obtains
\begin{eqnarray}
&&\hspace{-1mm}
\gamma_\mu\notp_2\gamma_i\otimes B_i\gamma_\nu+\gamma_\mu\notp_2\gamma_i\gamma_5\otimes B_i\gamma_\nu\gamma_5
\label{7.21}\\
&&\hspace{-1mm}
=~
-\slashed{p}_2\otimes\gamma_{\nu_\perp}\graB_\mu
-~{2\over s}p_{2\mu}p_{2\nu}\big[\gamma_i\otimes\slB\slp_1\gamma_i+\gamma_i\gamma_5\otimes\slB\slp_1\gamma_i\gamma_5\big]
+{2\over s}p_{2\nu}\slashed{p}_2\otimes\slB\slashed{p}_1\gamma_{\mu_\perp}
\nonumber\\
&&\hspace{-1mm}
-~\slashed{p}_2\gamma_5\otimes\gamma_\nu\graB_\mu\gamma_5+p_{2\mu}\gamma_i\otimes\gamma_{\nu_\perp}\graB_i+p_{2\mu}\gamma_i\gamma_5\otimes\gamma_{\nu_\perp}\graB_i\gamma_5
-~{2\over s}p_{1\nu}\slashed{p}_2\otimes\slashed{p}_2\graB_\mu
\nonumber\\
&&\hspace{-1mm}
+~{2\over s}p_{2\mu}p_{1\nu}\big[\gamma_i\otimes\slp_2\graB_i+\gamma_i\gamma_5\otimes\slp_2\graB_i\gamma_5\big]
\nonumber
\end{eqnarray}
Now, looking at Eq. (\ref{w1b1}) it is easy to see  that the last two lines in the r.h.s. do not give contributions of order of (\ref{pc}).

We need also
\bega
&&\hspace{-1mm}
\gamma_\mu\notp_1\gamma_i\otimes B_i\gamma_\nu+\gamma_\mu\notp_1\gamma_i\gamma_5
\otimes B_i\gamma_\nu\gamma_5
\nonumber\\
&&\hspace{-1mm}
=~p_{1\mu}\gamma^i\otimes\graB_i\gamma_{\nu}+p_{1\mu}\gamma^i\gamma_5\otimes\graB_i\gamma_{\nu}\gamma_5
-\slashed{p}_1\otimes\graB_\mu\gamma_{\nu}
-\slashed{p}_1\gamma_5\otimes\graB_\mu\gamma_\nu\gamma_5
\nonumber\\
&&\hspace{-1mm}
\gamma_i\notp_1\gamma_\mu\otimes B_i\gamma_\nu+\gamma_i\notp_1\gamma_\mu\gamma_5
\otimes B_i\gamma_\nu\gamma_5
\nonumber\\
&&\hspace{-1mm}
=~p_{1\mu}\gamma^i\otimes\gamma_{\nu}\graB_i+p_{1\mu}\gamma^i\gamma_5\otimes\gamma_{\nu}\graB_i\gamma_5
-\slashed{p}_1\otimes\gamma_{\nu}\graB_\mu
-\slashed{p}_1\gamma_5\otimes\gamma_\nu\graB_\mu\gamma_5
\label{14.20}
\end{eqnarray}
%

\subsubsection{Formulas with $\gamma$-matrices and two gluon fields}

With definition (\ref{abefildz}), we have the following formulas
\begin{eqnarray}
&&\hspace{-1mm}
A_i\otimes \tilB_j=g_{ij}\tilA_k\otimes B^k-\tilA_j\otimes B_i,~~~\tilde{A}_i\otimes B_j=g_{ij}A_k\otimes \tilB^k-A_j\otimes \tilB_i
\label{abeznaki}\\
&&\hspace{-1mm}
\tilde{A}_i\otimes \tilB_j=-g_{ij}A_k\otimes B^k+A_j\otimes B_i, ~~~\Rightarrow~~~\tilde{A}_i\otimes\tilde{B}^i=-A_i\otimes B^i,~~\tilde{A}_i\otimes B^i= A_i\otimes\tilde{B}^i
\nonumber
\end{eqnarray}
Using these formulas, after some algebra one obtains
\begin{eqnarray}
&&\hspace{-2mm}
\gamma_m\notp_2\gamma_jA^i\otimes\gamma_n\notp_1\gamma_iB^j
+\gamma_m\notp_2\gamma_jA^i\gamma_5\otimes\gamma_n\notp_1\gamma_iB^j\gamma_5
=\notp_2\grA_n\otimes\notp_1\graB_m+\notp_2\grA_n\gamma_5\otimes\notp_1\graB_m\gamma_5
\nonumber\\
&&\hspace{-2mm}
\gamma_j\notp_2\gamma_mA^i\otimes\gamma_n\notp_1\gamma_iB^j
+\gamma_j\notp_2\gamma_mA^i\gamma_5\otimes\gamma_n\notp_1\gamma_iB^j\gamma_5
=\notp_2\grA_n\otimes\graB_m\notp_1+\notp_2\grA_n\gamma_5\otimes\graB_m\notp_1\gamma_5
\nonumber\\
&&\hspace{-2mm}
\gamma_m\notp_2\gamma_jA^i\otimes\gamma_i\notp_1\gamma_nB^j
+\gamma_m\notp_2\gamma_jA^i\gamma_5\otimes\gamma_i\notp_1\gamma_nB^j\gamma_5
=\grA_n\notp_2\otimes\notp_1\graB_m+\grA_n\notp_2\gamma_5\otimes\notp_1\graB_m\gamma_5
\nonumber\\
&&\hspace{-2mm}
\gamma_j\notp_2\gamma_mA^i\otimes\gamma_i\notp_1\gamma_nB^j
+\gamma_j\notp_2\gamma_mA^i\gamma_5\otimes\gamma_i\notp_1\gamma_nB^j\gamma_5
=\grA_n\notp_2\otimes\graB_m\notp_1+\grA_n\notp_2\gamma_5\otimes\graB_m\notp_1\gamma_5
\nonumber\\
\label{gammas7}
\end{eqnarray}
and
\begin{eqnarray}
&&\hspace{-1mm}
\notp_2\grA_m\otimes\notp_1\graB_n+\notp_2\grA_n\gamma_5\otimes\notp_1\graB_m\gamma_5~=~g_{mn}\notp_2\grA_k\otimes\notp_1\graB^k
\nonumber\\
&&\hspace{-1mm}
\notp_2\grA_m\otimes\graB_n\notp_1+\notp_2\grA_n\gamma_5\otimes\gamma_5\graB_m\notp_1~\stackrel{\rm OK}=~g_{mn}\notp_2\grA_k\otimes\graB^k\notp_1
\nonumber\\
&&\hspace{-1mm}
\grA_m\notp_2\otimes\notp_1\graB_n+\gamma_5\grA_n\notp_2\otimes\notp_1\graB_m\gamma_5~=~g_{mn}\grA_k\notp_2\otimes\notp_1\graB^k
\nonumber\\
&&\hspace{-1mm}
\grA_m\notp_2\otimes\graB_n\notp_1+\gamma_5\grA_n\notp_2\otimes\gamma_5\graB_m\notp_1~=~g_{mn}\grA_k\notp_2\otimes\graB^k\notp_1
\label{gammas6}
\end{eqnarray}
The corollary of Eq. (\ref{gammas6}) is
\begin{eqnarray}
&&\hspace{-11mm}
\notp_2\grA_k\gamma_5\otimes\notp_1\graB^k\gamma_5~=~\notp_2\grA_k\otimes\notp_1\graB^k,~~~~~~~~
\notp_2\grA_k\gamma_5\otimes\gamma_5\graB^k\notp_1~=~\notp_2\grA_k\otimes\graB^k\notp_1
\nonumber\\
&&\hspace{-11mm}
\gamma_5\grA_k\notp_2\otimes\notp_1\graB^k\gamma_5~=~\grA_k\notp_2\otimes\notp_1\graB^k,~~~~~~~~
\gamma_5\grA_k\notp_2\otimes\gamma_5\graB^k\notp_1~=~\grA_k\notp_2\otimes\graB^k\notp_1
\label{gammas6a}
\end{eqnarray}

From Eqs. (\ref{gammas7}) and (\ref{gammas6}) one easily obtains
\begin{equation}
\gamma_m\notp_2\gamma_jA^i\otimes\gamma_n\notp_1\gamma_iB^j
+\gamma_m\notp_2\gamma_jA^i\gamma_5\otimes\gamma_n\notp_1\gamma_iB^j\gamma_5
~+~m\leftrightarrow n~=~2g_{mn}\notp_2\grA_k\otimes\notp_1\graB^k
\label{formula9}
\end{equation}
and
\begin{eqnarray}
&&\hspace{-1mm}
\gamma_m\notp_2\gamma_jA^i\otimes\gamma_n\notp_1\gamma_iB^j
+\gamma_m\notp_2\gamma_jA^i\gamma_5\otimes\gamma_n\notp_1\gamma_iB^j\gamma_5
~-~m\leftrightarrow n~
\nonumber\\
&&\hspace{-1mm}
=~2\notp_2\grA_n\otimes\notp_1\graB_m~-~m\leftrightarrow n,
\nonumber\\
&&\hspace{-1mm}
\gamma_j\notp_2\gamma_mA^i\otimes\gamma_i\notp_1\gamma_nB^j
+\gamma_j\notp_2\gamma_mA^i\gamma_5\otimes\gamma_i\notp_1\gamma_nB^j\gamma_5
~-~m\leftrightarrow n~
\nonumber\\
&&\hspace{-1mm}
=~2\grA_n\notp_2\otimes\graB_m\notp_1~-~m\leftrightarrow n
\label{formula9a}
\end{eqnarray}
We need also formulas
\begin{eqnarray}
&&\hspace{-1mm}
{4\over s^2}A^i\notp_1\notp_2\gamma_j\otimes B^j\notp_1\notp_2\gamma_i
\nonumber\\
&&\hspace{11mm}
=~
A^i\gamma_j\otimes B^j\gamma_i-iA^i\gamma_j\gamma_5\otimes \tilB^j\gamma_i
+i\tilA^i \gamma_j\otimes  B^j\gamma_i\gamma_5+\tilA^i\gamma_j\gamma_5\otimes \tilB^j\gamma_i\gamma_5,
\nonumber\\
&&\hspace{-1mm}
{4\over s^2}\big(A^i\notp_1\notp_2\gamma_j\otimes B^j\notp_1\notp_2\gamma_i
+A^i\notp_1\notp_2\gamma_j\gamma_5\otimes B^j\notp_1\notp_2\gamma_i\gamma_5\big)
\nonumber\\
&&\hspace{11mm}
=~\gamma^j\grA_i\otimes\gamma^i\graB_j+\gamma^j\grA_i\gamma_5\otimes\gamma^i\graB_j\gamma_5,
\nonumber\\
&&\hspace{-1mm}
\gamma_i\grA_j\gamma_5\otimes\gamma_j\grA_i\gamma_5~=~\gamma_i\grA_j\otimes\gamma^i\graB^j-\gamma_i\grA^i\otimes\gamma_j\graB^j
\label{gammas10}
\end{eqnarray}
and
\begin{eqnarray}
&&\hspace{-1mm}
A_k\gamma_i\slashed{p}_2\gamma^j\otimes B_j\gamma^i\slashed{p}_1\gamma^k~=~
\slashed{p}_2\grA_i\otimes \slashed{p}_1\graB^i
~=~\notA\notp_2\gamma_i\otimes\notB\notp_1\gamma^i,
\nonumber\\
&&\hspace{-1mm}
A_k\gamma^j\slashed{p}_2\gamma_i\otimes B_j\gamma^k\slashed{p}_1\gamma^i~=~
\grA_i\slashed{p}_2\otimes \graB_i\slashed{p}_1
~=~\gamma_i\notp_2\notA\otimes\gamma^i\notp_1\notB,
\nonumber\\
&&\hspace{-1mm}
A_k\gamma_i\slashed{p}_2\gamma_j\otimes B^j\gamma^k\slashed{p}_1\gamma^i
~=~\slashed{p}_2\grA_i\otimes\graB^i\slashed{p}_1
~=~~\notA\notp_2\gamma_i\otimes\gamma^i\notp_1\notB,
\nonumber\\
&&\hspace{-1mm}
A_k\gamma_j\slashed{p}_2\gamma_i\otimes B^j\gamma^i\slashed{p}_1\gamma^k
~=~\grA_i\slashed{p}_2\otimes\slashed{p}_1\graB^i
~=~\gamma_i\notp_2\notA\otimes\notB\notp_1\gamma^i,
\label{gammas11}
\end{eqnarray}
\begin{eqnarray}
&&\hspace{-1mm}
A^k\gamma_m\slashed{p}_2\gamma_j\otimes B^j\gamma_n\slashed{p}_1\gamma_k +m\leftrightarrow n
-g_{mn}A^k\gamma_i\slashed{p}_2\gamma_j\otimes B^j\gamma^i\slashed{p}_1\gamma_k
\nonumber\\
&&\hspace{-1mm}
=~\grA_m\notp_2\otimes\graB_n\notp_1+m\leftrightarrow n-g_{mn}\grA_k\notp_2\otimes\graB^k\notp_1,
\nonumber\\
&&\hspace{-1mm}
A^k\gamma_j\slashed{p}_2\gamma_m\otimes B^j\gamma_k\slashed{p}_1\gamma_n+m\leftrightarrow n
-g_{mn}A^k\gamma_j\slashed{p}_2\gamma_i\otimes B^j\gamma_k\slashed{p}_1\gamma^i
\nonumber\\
&&\hspace{-1mm}
=~\notp_2\grA_m\otimes\notp_1\graB_n+m\leftrightarrow n-g_{mn}\notp_2\grA_k\otimes\notp_1\graB^k,
\nonumber\\
&&\hspace{-1mm}
A^k\gamma_m\slashed{p}_2\gamma_j\otimes B^j\gamma_k\slashed{p}_1\gamma_n+m\leftrightarrow n
-g_{mn}A^k\gamma_j\slashed{p}_2\gamma_i\otimes B^j\gamma_k\slashed{p}_1\gamma^i
\nonumber\\
&&\hspace{-1mm}
=~\grA_m\notp_2\otimes\notp_1\graB_n+m\leftrightarrow n-g_{mn}\grA_k\notp_2\otimes\notp_1\graB^k,
\nonumber\\
&&\hspace{-1mm}
A^k\gamma_j\slashed{p}_2\gamma_m\otimes B^j\gamma_n\slashed{p}_1\gamma_k+m\leftrightarrow n
-g_{mn}A^k\gamma_j\slashed{p}_2\gamma_i\otimes B^j\gamma_k\slashed{p}_1\gamma^i
~=~
\nonumber\\
&&\hspace{-1mm}
=~\notp_2\grA_m\otimes\graB_n\notp_1+m\leftrightarrow n-g_{mn}\notp_2\grA_k\otimes\graB^k\notp_1,
\label{formulas67}
\end{eqnarray}
\begin{eqnarray}
&&\hspace{-1mm}
{2\over s}\big[A_i\notp_1\notp_2\gamma^j\otimes B_j\gamma_n\notp_1\gamma^i
+A_i\notp_1\notp_2\gamma^j\gamma_5\otimes B_j\gamma_{\nu_\perp}\notp_1\gamma^i\gamma_5\big]
\label{gammas16}\\
&&\hspace{-1mm}
=~-\gamma_i\grA_n\otimes \notp_1\graB^i - \gamma_i\grA_n\gamma_5\otimes \notp_1\graB^i\gamma_5
~=~\gamma_i\grA_n\otimes\notB\notp_1\gamma^i+\gamma_i\grA_n\gamma_5\otimes\notB\notp_1\gamma^i\gamma_5,
\nonumber\\
&&\hspace{-1mm} 
{2\over s}\big[A_i\gamma_n\notp_2\gamma^j\otimes B_j\notp_2\notp_1\gamma^i
+A_i\gamma_n\notp_2\gamma^j\gamma_5\otimes B_j\notp_2\notp_1\gamma^i\gamma_5\big]
\nonumber\\
&&\hspace{-1mm} 
=-\notp_2\grA_i\otimes\gamma^i\graB_n-\notp_2\grA_i\gamma_5\otimes\gamma^i\graB_n\gamma_5 
~=~\notA\notp_2\gamma_i\otimes\gamma^i\graB_n+\notA\notp_2\gamma_i\gamma_5\otimes\gamma^i\graB_n\gamma_5. 
\nonumber
\end{eqnarray}
%

 \subsection{TMD matrix elements \label{sec:tmdmat}}
 \subsubsection{Parametrization of leading-twist matrix elements \label{sec:paramlt}}
 Let us  first consider matrix elements of operators without $\gamma_5$. The standard parametrization of quark TMDs reads
 (see e.g. Ref. \cite{Arnold:2008kf,Boussarie:2023izj})
 \footnote{Our notations differ from ``TMD handbook'' \cite{Boussarie:2023izj}:  
 $g^{\rm QCD}_{\rm our}=-g^{\rm QCD}_{\rm hbook}$, 
 $\epsilon^{\mu\nu\lambda\rho}_{\rm our}=-\epsilon^{\mu\nu\lambda\rho}_{\rm hbook}$, but $\epsilon^{ij}_{\rm our}=\epsilon^{ij}_{\rm hbook}$}
\begin{eqnarray}
&&\hspace{-1mm}
{1\over 16\pi^3}\!\int\!dx_\bu d^2x_\perp~e^{-i\alpha x_\bu+i(k,x)_\perp}
~\langle A|\bsi(x_\bu,x_\perp)\gamma^\mu\psi(0)|A\rangle
\label{Amael}\\
&&\hspace{27mm}
=~p_1^\mu f_1(\alpha,k_\perp)
+k_\perp^\mu f_\perp(\alpha,k_\perp)+p_2^\mu{2m^2_N\over s}f_3(\alpha,k_\perp),
\nonumber\\
&&\hspace{-1mm}
{1\over 16\pi^3}\!\int\!dx_\bu d^2x_\perp~e^{-i\alpha x_\bu+i(k,x)_\perp}
~\langle A|\bsi(x_\bu,x_\perp)\psi(0)|A\rangle
~=~me(\alpha,k_\perp)
\nonumber
\end{eqnarray}
for quark distributions in the projectile and 
\begin{eqnarray}
&&\hspace{-1mm}
{1\over 16\pi^3}\!\int\!dx_\bu d^2x_\perp~e^{-i\alpha x_\bu+i(k,x)_\perp}
~\langle A|\bsi(0)\gamma^\mu\psi(x_\bu,x_\perp)|A\rangle
\label{baramael}\\
&&\hspace{27mm}
=~-p_1^\mu \barf_1(\alpha,k_\perp)
-k_\perp^\mu\barf _\perp(\alpha,k_\perp)-p_2^\mu{2m^2_N\over s}\barf_3(\alpha,k_\perp),
\nonumber\\
&&\hspace{-1mm}
{1\over 16\pi^3}\!\int\!dx_\bu d^2x_\perp~e^{-i\alpha x_\bu+i(k,x)_\perp}
~\langle A|\bsi(0)\psi(x_\bu,x_\perp)|A\rangle
~=~
m\bare(\alpha,k_\perp)
\nonumber
\end{eqnarray}
for the antiquark distributions. 
\footnote{In an arbitrary gauge, there are gauge links to $-\infty$ as displayed in  Eq. (\ref{gaugelinks}).}

The corresponding matrix elements for the target are obtained by trivial replacements $p_1\leftrightarrow p_2$, $x_\bu\leftrightarrow x_\star$
and $\alpha\leftrightarrow\beta$:
\begin{eqnarray}
&&\hspace{-1mm}
{1\over 16\pi^3}\!\int\!dx_\star d^2x_\perp~e^{-i\beta x_\star+i(k,x)_\perp}
~\langle B|\bsi(x_\star,x_\perp)\gamma^\mu\psi(0)|B\rangle
\label{Bmael}\\
&&\hspace{27mm}
=~p_2^\mu f_1(\beta,k_\perp)+k_\perp^\mu f_\perp(\beta,k_\perp)
+p_1^\mu{2m^2_N\over s}f_3(\beta,k_\perp),
\nonumber\\
&&\hspace{-1mm}
{1\over 16\pi^3}\!\int\!dx_\star d^2x_\perp~e^{-i\beta x_\star+i(k,x)_\perp}
~\langle B|\bsi(x_\star,x_\perp)\psi(0)|B\rangle
~=~me(\beta,k_\perp),
\nonumber
\end{eqnarray}
and
\begin{eqnarray}
&&\hspace{-1mm}
{1\over 16\pi^3}\!\int\!dx_\star d^2x_\perp~e^{-i\beta x_\star+i(k,x)_\perp}
~\langle B|\bsi(0)\gamma^\mu\psi(x_\star,x_\perp)|B\rangle
\label{barbmael}\\
&&\hspace{27mm}
=~-p_2^\mu \barf_1(\beta,k_\perp)
-k_\perp^\mu\barf _\perp(\beta,k_\perp)-p_1^\mu{2m^2_N\over s}\barf_3(\beta,k_\perp),
\nonumber\\
&&\hspace{-1mm}
{1\over 16\pi^3}\!\int\!dx_\star d^2x_\perp~e^{-i\beta x_\star+i(k,x)_\perp}
~\langle B|\bsi(0)\psi(x_\star,x_\perp)|B\rangle
~=~
m\bare(\beta,k_\perp).
\nonumber
\end{eqnarray}

Matrix elements of operators with $\gamma_5$ are parametrized as follows: 
\begin{eqnarray}
&&\hspace{-1mm}
{1\over 16\pi^3}\!\int\!dx_\bu d^2x_\perp~e^{-i\alpha x_\bu+i(k,x)_\perp}
~\langle A|\bsi(x_\bu,x_\perp)\gamma_\mu\gamma_5\psi(0)|A\rangle
~=~-\epsilon_{\mu_\perp i}k^ig^\perp(\alpha,k_\perp),
\nonumber\\
&&\hspace{-1mm}
{1\over 16\pi^3}\!\int\!dx_\bu d^2x_\perp~e^{-i\alpha x_\bu+i(k,x)_\perp}
~\langle A|\bsi(0)\gamma_\mu\gamma_5\psi(x_\bu,x_\perp)|A\rangle
~=~-\epsilon_{\mu_\perp i}k^i\barg^\perp(\beta,k_\perp)
\nonumber\\
\label{mael5p}
\end{eqnarray}
and
\begin{eqnarray}
&&\hspace{-1mm}
{1\over 16\pi^3}\!\int\!dx_\bu d^2x_\perp~e^{-i\alpha x_\bu+i(k,x)_\perp}
~\langle B|\bsi(x_\star,x_\perp)\gamma_\mu\gamma_5\psi(0)|B\rangle
~=~\epsilon_{\mu_\perp i}k^ig^\perp(\beta,k_\perp),
\nonumber\\
&&\hspace{-1mm}
{1\over 16\pi^3}\!\int\!dx_\bu d^2x_\perp~e^{-i\alpha x_\bu+i(k,x)_\perp}
~\langle B|\bsi(0)\gamma_\mu\gamma_5\psi(x_\star,x_\perp)|B\rangle
~=~\epsilon_{\mu_\perp i}k^i\barg^\perp(\beta,k_\perp)
\nonumber\\
\label{mael5t}
\end{eqnarray}

The parametrizations of time-odd Boer-Mulders TMDs are
\begin{eqnarray}
&&\hspace{-1mm}
{1\over 16\pi^3}\!\int\!dx_\bu d^2x_\perp~e^{-i\alpha x_\bu+i(k,x)_\perp}
~\langle A|\bsi(x_\bu,x_\perp)\sigma^{\mu \nu}\psi(0)|A\rangle
\nonumber\\
&&\hspace{11mm}
=~{1\over m}(k_\perp^\mu p_1^\nu -\mu\leftrightarrow\nu)h_{1}^\perp(\alpha,k_\perp)
+{2m\over s}(p_1^\mu p_2^\nu-\mu\leftrightarrow\nu)h(\alpha,k_\perp)
\nonumber\\
&&\hspace{33mm}
+~{2m\over s}(k_\perp^\mu p_2^\nu -\mu\leftrightarrow\nu)h_{3}^\perp(\alpha,k_\perp),
\nonumber\\
&&\hspace{-1mm}
{1\over 16\pi^3}\!\int\!dx_\bu d^2x_\perp~e^{-i\alpha x_\bu+i(k,x)_\perp}
~\langle A|\bsi(0)\sigma^{\mu \nu}\psi(x_\bu,x_\perp)|A\rangle
\nonumber\\
&&\hspace{11mm}
=~-{1\over m}(k_\perp^\mu p_1^\nu -\mu\leftrightarrow\nu)\barh_{1}^\perp(\alpha,k_\perp)
-{2m\over s}(p_1^\mu p_2^\nu-\mu\leftrightarrow\nu)\barh(\alpha,k_\perp)
\nonumber\\
&&\hspace{33mm}
-~{2m\over s}(k_\perp^\mu p_2^\nu -\mu\leftrightarrow\nu)\barh_{3}^\perp(\alpha,k_\perp)
\label{hmaelp}
\end{eqnarray}
and similarly for the target with usual replacements   $p_1\leftrightarrow p_2$, $x_\bu\leftrightarrow x_\star$
and $\alpha\leftrightarrow\beta$:
{\color{black}
\begin{eqnarray}
&&\hspace{-1mm}
{1\over 16\pi^3}\!\int\!dx_\star d^2x_\perp~e^{-i\beta x_\star+i(k,x)_\perp}
~\langle B|\bsi(x_\star,x_\perp)\sigma^{\mu \nu}\psi(0)|B\rangle
\nonumber\\
&&\hspace{11mm}
=~{1\over m}(k_\perp^\mu p_2^\nu -\mu\leftrightarrow\nu)h_{1}^\perp(\beta,k_\perp)
+{2m\over s}(p_2^\mu p_1^\nu-\mu\leftrightarrow\nu)h(\beta,k_\perp)
\nonumber\\
&&\hspace{33mm}
+~{2m\over s}(k_\perp^\mu p_1^\nu -\mu\leftrightarrow\nu)h_{3}^\perp(\beta,k_\perp),
\nonumber\\
&&\hspace{-1mm}
{1\over 16\pi^3}\!\int\!dx_\star d^2x_\perp~e^{-i\beta x_\star+i(k,x)_\perp}
~\langle B|\bsi(0)\sigma^{\mu \nu}\psi(x_\star,x_\perp)|B\rangle
\nonumber\\
&&\hspace{11mm}
=~-{1\over m}(k_\perp^\mu p_2^\nu -\mu\leftrightarrow\nu)\barh_{1}^\perp(\beta,k_\perp)
-{2m\over s}(p_2^\mu p_1^\nu-\mu\leftrightarrow\nu)\barh(\beta,k_\perp)
\nonumber\\
&&\hspace{33mm}
-~{2m\over s}(k_\perp^\mu p_1^\nu -\mu\leftrightarrow\nu)\barh_{3}^\perp(\beta,k_\perp)
\label{hmaelt}
\end{eqnarray}
}
Note that
the coefficients in front of $f_3$,  $g^\perp$, $h$ and $h_3^\perp$ in eqs. (\ref{Amael}),  (\ref{Bmael}), (\ref{mael5p}),  (\ref{mael5t}),  (\ref{hmaelp}),  and  (\ref{hmaelt}) 
 contain an extra ${1\over s}$ since $p_2^\mu$ enters only through the direction
of gauge link so the result should not depend on rescaling $p_2\rightarrow\lambda p_2$. 

\subsubsection{Matrix elements of quark-quark-gluon operators \label{sec:qqgparam}}

 First, let us demonstrate that operators ${1\over\alpha}$ and ${1\over\beta}$ in Eqs. (\ref{3.25}) are
 replaced by $\pm{1\over\alpha_q}$ and $\pm{1\over\beta_q}$ in forward matrix elements. Indeed,
\begin{eqnarray}
&&\hspace{-1mm}
\!\int\!dx_\bu ~e^{-i\alpha_q x_\bu}\langle\bar\Phi(x_\bu,x_\perp)\Gamma{1\over \alpha+\ie}\psi(0)\rangle_A
\label{maelqg1}\\
&&\hspace{-1mm}
=~{1\over i}\!\int\!dx_\bu \!\int_{-\infty}^0 \!\!\!dx'_\bu~e^{-i\alpha_q x_\bu}\langle\bar\Phi(x_\bu,x_\perp)\Gamma\psi(x'_\bu,0_\perp)\rangle_A
=~{1\over\alpha_q}\!\int\!dx_\bu ~e^{-i\alpha x_\bu}\langle\bar\Phi(x_\bu,x_\perp)\Gamma\psi(0)\rangle_A
\nonumber
\end{eqnarray}
where $\bar\Phi(x_\bu,x_\perp)$ can be $\bsi(x_\bu,x_\perp)$ or  $\bsi(x_\bu,x_\perp)A_i(x_\bu,x_\perp)$ and $\Gamma$ can be any $\gamma$-matrix.
Similarly,
\begin{eqnarray}
&&\hspace{-1mm}
\!\int\!dx_\bu ~e^{-i\alpha_q x_\bu}\langle\big(\bar\psi{1\over \alpha-\ie}\big)(x_\bu,x_\perp)\Gamma\Phi(0)\rangle_A
=~{1\over\alpha_q}\!\int\!dx_\bu ~e^{-i\alpha x_\bu}\langle\bar\psi(x_\bu,x_\perp)\Gamma\Phi(0)\rangle_A
\label{maelqg2}\\
&&\hspace{-1mm}
\!\int\!dx_\bu ~e^{-i\alpha_q x_\bu}\langle\big(\bar\psi{1\over \alpha-\ie}\big)(x_\bu,x_\perp)\Gamma{1\over \alpha+\ie}\psi(0)\rangle_A
=~{1\over\alpha_q^2}\!\int\!dx_\bu ~e^{-i\alpha_q x_\bu}\langle\bar\psi(x_\bu,x_\perp)\Gamma\psi(0)\rangle_A
\nonumber
\end{eqnarray}
where $\Phi(x_\bu,x_\perp)$ can be $\psi(x_\bu,x_\perp)$ or  $A_i(x_\bu,x_\perp)\psi(x_\bu,x_\perp)$. We need also
\begin{eqnarray}
&&\hspace{-7mm}
\!\int\!dx_\bu ~e^{-i\alpha_q x_\bu}\langle\big(\bar\psi{1\over \alpha-\ie}\big)(0)\Gamma\Phi(x_\bu,x_\perp)\rangle_A
=~-{1\over\alpha_q}\!\int\!dx_\bu ~e^{-i\alpha x_\bu}\langle\bar\psi(0)\Gamma\Phi(x_\bu,x_\perp)\rangle_A
\nonumber\\
&&\hspace{-7mm}
\!\int\!dx_\bu ~e^{-i\alpha_q x_\bu}\langle\bar\Phi(0)\Gamma{1\over \alpha+\ie}\psi(x_\bu,x_\perp)\rangle_A
=~-{1\over\alpha_q}\!\int\!dx_\bu ~e^{-i\alpha x_\bu}\langle\bar\Phi(0)\Gamma\psi(x_\bu,x_\perp)\rangle_A
\label{maelqg3}
\end{eqnarray}
The corresponding formulas for target matrix elements are obtained by substitution $\alpha\leftrightarrow\beta$ (and $x_\bu\leftrightarrow x_\star$).

\subsubsection{Matrix elements of quark-quark-gluon operators related to \\
quark-antiquark TMDs by QCD equations of motion\label{sec:qqgparame}}

Next, we will use QCD equation of motion to reduce quark-quark-gluon TMDs to leading-twist TMDs (see Ref. \cite{Mulders:1995dh}). 
For our quark fields QCD equations read
\footnote{As was mentioned in footnote \ref{oboz}, in this paper QCD  coupling constant $g$ is included in the definition of gluon field $A_\mu$.}
\begin{eqnarray}
&&\hspace{-1mm}
\bsi(x)\slA_\perp(x)~=~i\partial_i\bsi(x)\gamma_i+i{2\over s}\partial_\star\bsi(x)\slp_1
+i\bsi \stackrel{\leftarrow}D_\bu{2\over s}\slp_2,
\nonumber\\
&&\hspace{-1mm}
\slA_\perp(x)\psi(x)~=~-i\slpart_\perp\psi(x)-i{2\over s}\slp_1\partial_\star\psi(x)-i{2\over s}\slp_2D_\bu\psi(x)
\label{eqmp}
\ega
for the projectile operators in $A_\star=0$ gauge and
\begin{eqnarray}
&&\hspace{-1mm}
\bsi(x)\slB_\perp(x)~=~i\partial_i\bsi(x)\gamma_i+i{2\over s}\partial_\bu\bsi(x)\slp_2
+i\bsi \stackrel{\leftarrow}D_\star\!(x){2\over s}\slp_1,
\nonumber\\
&&\hspace{-1mm}
\slB_\perp(x)\psi(x)~=~-i\slpart_\perp\psi(x)-i{2\over s}\slp_2\partial_\bu\psi(x)-i{2\over s}\slp_1D_\star\psi(x)
\label{eqmt}
\ega
for operators in target matrix elements (in $B_\bu=0$ gauge). 
Our strategy is as follows: when we see an operator as in the left-hand sides of these equations, 
we rewrite it in terms
of the corresponding right-hand sides. For most of  the matrix elements listed in this Section, the result can be represented through quark-antiquark TMDs. Sometimes, however, one needs the last terms in the r.h.s.'s parametrized in the next Section.

Let us present the list of formulas derived in Ref. \cite{Balitsky:2020jzt}

\begin{eqnarray}
&&\hspace{-1mm}
{1\over 8\pi^3s}\!\int\! dx_\bu dx_\perp~e^{-i\alpha x_\bu+i(k,x)_\perp}\langle A|\bar\psi\slA(x)\slashed{p}_2\gamma_i\psi(0)|A\rangle
~=~k_i[f_1-\alpha (f_\perp+ig^\perp)\big](\alpha,k_\perp)
\nonumber\\
&&\hspace{0mm}
{1\over 8\pi^3s}\!\int\! dx_\bu dx_\perp~e^{-i\alpha x_\bu+i(k,x)_\perp}
\langle A|\bar\psi(x)\gamma_i\slashed{p}_2\slA\psi(0)|A\rangle
~=~k_i\big[f_1-\alpha(f_\perp-ig^\perp)\big](\alpha,k_\perp)
\nonumber\\
&&\hspace{-1mm}
{1\over 8\pi^3s}\!\int\! dx_\perp dx_\bu~e^{-i\alpha x_\bu+i(k,x)_\perp}
\langle A|\bsi(0)\gamma_i\slp_2\slA\psi(x)|A\rangle
~=~k_i\big[\barf_1-\alpha(\barf_\perp+i\barg^\perp)\big](\alpha,k_\perp)
\nonumber\\
&&\hspace{-1mm}
{1\over 8\pi^3s}\!\int\! dx_\perp dx_\bu~e^{-i\alpha x_\bu+i(k,x)_\perp}
\langle A|\bsi \slA(0)\slp_2\gamma_i\psi(x)|A\rangle
~=~
k_i\big[\barf_1-\alpha(\barf_\perp-i\barg_\perp)\big](\alpha_q,k_\perp)
\nonumber\\
\label{projmaels}
\end{eqnarray}
For brevity, hereafter in the projectile matrix elements $x=(x_\bu,0_\star,x_\perp)$ 

The target matrix elements are obtained by usual replacements (\ref{protareplace}):
\begin{eqnarray}
&&\hspace{-1mm}
{1\over 8\pi^3s}\!\int\! dx_\star dx_\perp~e^{-i\beta x_\star+i(k,x)_\perp}\langle B|\bar\psi\slB(x)\slashed{p}_1\gamma_i\psi(0)|B\rangle
~=~k_i[f_1-\beta (f_\perp+ig^\perp)\big](\beta,k_\perp)
\nonumber\\
&&\hspace{0mm}
{1\over 8\pi^3s}\!\int\! dx_\star dx_\perp~e^{-i\beta x_\star+i(k,x)_\perp}
\langle B|\bar\psi(x)\gamma_i\slashed{p}_1\slB\psi(0)|B\rangle
~=~k_i\big[f_1-\beta(f_\perp-ig^\perp)\big](\beta,k_\perp)
\nonumber\\
&&\hspace{-1mm}
{1\over 8\pi^3s}\!\int\! dx_\perp dx_\star~e^{-i\beta x_\star+i(k,x)_\perp}
\langle B|\bsi(0)\gamma_i\slp_1\slB\psi(x)|B\rangle
~=~k_i\big[\barf_1-\beta(\barf_\perp+i\barg^\perp)\big](\beta,k_\perp)
\nonumber\\
&&\hspace{-1mm}
{1\over 8\pi^3s}\!\int\! dx_\perp dx_\star~e^{-i\beta x_\star+i(k,x)_\perp}
\langle B|\bsi \slB(0)\slp_1\gamma_i\psi(x)|B\rangle
~=~
k_i\big[\barf_1-\beta(\barf_\perp-i\barg_\perp)\big](\beta_q,k_\perp)
\nonumber\\
\label{tarmaels}
\end{eqnarray}
Similarly,  in the target matrix elements $x\equiv(0_\bu,x_\star,x_\perp)$ 

Next, for the projectile matrix elements with an extra $\gamma_5$ one obtains
\bega
&&\hspace{-1mm}
\int\! {dx_\bu dx_\perp\over 8\pi^3s}~e^{-i\alpha x_\bu+i(k,x)_\perp}
\langle A|\bar\psi\slA(x)\slashed{p}_2\gamma_i\gamma_5\psi(0)|A\rangle~
=~i\epsilon_{ij}k^j\big[f_1-\alpha(f_\perp+ig^\perp)\big](\alpha,k_\perp),
\nonumber\\
&&\hspace{-1mm}
\int\! {dx_\bu dx_\perp\over 8\pi^3s}~e^{-i\alpha x_\bu+i(k,x)_\perp}
\langle A|\bar\psi(x)\gamma_i\gamma_5\slp_2\slA\psi(0)|A\rangle
=~-i\epsilon_{ij}k^j\big[f_1-\alpha(f_\perp-ig^\perp)\big](\alpha,k_\perp)
\nonumber\\
&&\hspace{-2mm}
\int\! {dx_\bu dx_\perp\over 8\pi^3s}~e^{-i\alpha x_\bu+i(k,x)_\perp}
\langle A|\bsi \slA(0)\slp_2\gamma_i\gamma_5\psi(x)|A\rangle
~=~i\epsilon_{ij}k^j\big[\barf_1-\alpha(\barf_\perp-i\barg_\perp)\big](\alpha,k_\perp)
\nonumber\\
&&\hspace{-2mm}
\int\! {dx_\bu dx_\perp\over 8\pi^3s}~e^{-i\alpha x_\bu+i(k,x)_\perp}
\langle A|\bsi _f(0)\gamma_i\gamma_5\slp_2\slA\psi(x)|A\rangle~
=~-i\epsilon_{ij}k^j
\big[\barf_{1}-\alpha(\barf_\perp+i\barg_\perp)\big](\alpha,k_\perp)
\nonumber\\
\label{projmaels5}
\end{eqnarray}
and for the target
\begin{eqnarray}
&&\hspace{-1mm}
\int\! {dx_\star dx_\perp\over 8\pi^3s}~e^{-i\beta x_\star+i(k,x)_\perp}
\langle B|\bar\psi\slB(x)\slashed{p}_1\gamma_i\gamma_5\psi(0)|B\rangle
~=~-i\epsilon_{ij}k^j\big[f_1-\beta (f_\perp+ig^\perp)\big](\beta,k_\perp)
\nonumber\\
&&\hspace{-1mm}
\int\! {dx_\star dx_\perp\over 8\pi^3s}~e^{-i\beta x_\star+i(k,x)_\perp}
\langle B|\bar\psi(x)\gamma_i\gamma_5\slp_1\slB(0)\psi(0)|B\rangle
=~i\epsilon_{ij}k^j\big[f_1-\beta(f_\perp-ig_\perp)\big](\beta,k_\perp)
\nonumber\\
&&\hspace{-2mm}
\int\! {dx_\star dx_\perp\over 8\pi^3s}~e^{-i\beta x_\star+i(k,x)_\perp}
\langle B|\bsi \slB(0)\slp_1\gamma_i\gamma_5\psi(x)|B\rangle
~=~-i\epsilon_{ij}k^j\big[\barf_1-\beta(\barf_\perp-i\barg_\perp)\big](\beta,k_\perp)
\nonumber\\
&&\hspace{-2mm}
\int\! {dx_\star dx_\perp\over 8\pi^3s}~e^{-i\beta x_\star+i(k,x)_\perp}
\langle B|\bsi _f(0)\gamma_i\gamma_5\slp_1\slB\psi(x)|B\rangle
=~i\epsilon_{ij}k^j
\big[\barf_{1}-\beta(\barf_\perp+i\barg_\perp)\big](\beta,k_\perp)
\nonumber\\
\label{tarmaels5}
\end{eqnarray}
The different sign  in  projectile$\leftrightarrow$target replacement of matrix elements with $\gamma_5$ is due to the difference in the definitions (\ref{deftildas}).

\begin{eqnarray}
&&\hspace{-1mm}
\int\! {dx_\bu dx_\perp\over 8\pi^3s}~e^{-i\alpha_qx_\bu+i(k,x)_\perp}
\langle A|\bsi\slA(x)\slashed{p}_2\psi(0)|A\rangle
=~\big[-i{k_\perp^2\over m}h_1^\perp-\alpha m(e+ih)\big](\alpha,k_\perp)
\nonumber\\
&&\hspace{-1mm}
\int\! {dx_\bu dx_\perp\over 8\pi^3s}~e^{-i\alpha x_\bu+i(k,x)_\perp}
\langle A|\bsi(x)\slashed{p}_2\slA\psi(0)|A\rangle
=~\big[i{k_\perp^2\over m}h_1^\perp-\alpha m(e-ih)\big](\alpha,k_\perp)
\label{1056}\\
&&\hspace{-1mm}
\!\int\! {dx_\bu dx_\perp\over 8\pi^3s}~e^{-i\alpha x_\bu+i(k,x)_\perp}
\langle A|\bsi(0)\slashed{p}_2\slA\psi(x_\bu,x_\perp)|A\rangle
=~\big[i{k_\perp^2\over m}\barh_{1}^\perp
+\alpha m(\bare+i\barh)\big](\alpha,k_\perp),
\nonumber\\
&&\hspace{-1mm}
\!\int\! {dx_\bu dx_\perp\over 8\pi^3s}~e^{-i\alpha x_\bu+i(k,x)_\perp}
\langle A|\bsi\slA(0)\slp_2\psi(x_\bu,x_\perp)|A\rangle
=~\big[-i{k_\perp^2\over m}\barh_{1}^\perp
+\alpha m(\bare-i\barh)\big](\alpha,k_\perp)
\nonumber
\end{eqnarray}
The target matrix elements are obtained by usual replacements (\ref{protareplace}) (without sign change).
\begin{eqnarray}
&&\hspace{-1mm}
\!\int\! {dx_\star dx_\perp\over 8\pi^3s}~e^{-i\beta_qx_\star+i(k,x)_\perp}
\langle B|\bsi\slB(x)\slashed{p}_1\psi(0)|B\rangle
=~\big[-i{k_\perp^2\over m}h_{1}^\perp-\beta_q m(e+ih)\big](\beta,k_\perp),
\nonumber\\
&&\hspace{-1mm}
\!\int\! {dx_\star dx_\perp\over 8\pi^3s}~e^{-i\beta x_\star+i(k,x)_\perp}
\langle B|\bsi\slB(0)\slashed{p}_1\psi(x)|B\rangle
=~\big[-i{k_\perp^2\over m}\barh_{1}^\perp+\beta m(\bare-i\barh)\big](\beta,k_\perp).
\nonumber\\
&&\hspace{-1mm}
\!\int\! {dx_\star dx_\perp\over 8\pi^3s}~e^{-i\beta x_\star+i(k,x)_\perp}
\langle B|\bsi(0)\slashed{p}_1\slB\psi(x)|B\rangle
=~\big[i{k_\perp^2\over m}\barh_{1}^\perp
+\beta m(\bare+i\barh)\big](\beta,k_\perp),
\label{1019}\\
&&\hspace{-1mm}
\!\int\! {dx_\star dx_\perp\over 8\pi^3s}~e^{-i\beta x_\star+i(k,x)_\perp}
\langle B|\bsi\slB(0)\slp_2\psi(x_\star,x_\perp)|B\rangle
=~\big[-i{k_\perp^2\over m}\barh_1^\perp
+\beta m(\bare-i\barh)\big](\beta,k_\perp)
\nonumber
\end{eqnarray}
Next, we need
\begin{eqnarray}
&&\hspace{-2mm}
\!\int\! {dx_\bu dx_\perp\over 16\pi^3}~e^{-i\alpha x_\bu+i(k,x)_\perp}
\langle A|\bsi\slA(x)\gamma_i\psi(0)|A\rangle
=~k_im\big[-e-i\alpha h_3+ih_D\big](\alpha ,k_\perp),
\nonumber\\
&&\hspace{-2mm}
\!\int\! {dx_\bu dx_\perp\over 16\pi^3}~e^{-i\alpha x_\bu+i(k,x)_\perp}
\langle A|\bsi(x)\gamma_i\slA\psi(0)|A\rangle
=~k_im\big[-e+i\alpha h_3+ih_D^\ast\big](\alpha ,k_\perp),
\nonumber\\
&&\hspace{-2mm}
\!\int\! {dx_\bu dx_\perp\over 16\pi^3}~e^{-i\alpha x_\bu+i(k,x)_\perp}
\langle A|\bsi\slA(0)\gamma_i
\psi(x)|A\rangle
=~mk_i\big[\bare-i\alpha \barh_3-i\barh_D\big](\alpha ,k_\perp),
\nonumber\\
&&\hspace{-2mm}
\!\int\! {dx_\bu dx_\perp\over 16\pi^3}~e^{-i\alpha x_\bu+i(k,x)_\perp}
\langle A|\bsi(0)\gamma_i\slA\psi(x)|A\rangle
=~mk_i\big[\bare+i\alpha \barh_3+i\barh_D^\ast\big](\alpha ,k_\perp)
\nonumber\\
\label{projagammai}
\end{eqnarray}
The matrix elements $h_D$ are $\barq q$ TMDs with an extra longitudinal derivative of the quark field. They are 
defined in the next Section. It is worth noting that contributions of these terms in the r.h.s.'s  cancel in the final result.

For the target we get
\begin{eqnarray}
&&\hspace{-2mm}
\!\int\! {dx_\star dx_\perp\over 16\pi^3}~e^{-i\beta x_\star+i(k,x)_\perp}
\langle B|\bsi\slB(x)\gamma_i\psi(0)|B\rangle
=~
mk_i[-e-i\beta h_3^\perp+ih_D](\beta,k_\perp),
\nonumber\\
&&\hspace{-1mm}
\!\int\! {dx_\star dx_\perp\over 16\pi^3}~e^{-i\beta x_\star+i(k,x)_\perp}
\langle B|\bsi(x)\gamma_i\slB\psi(0)|B\rangle
=~
mk_i[-e+i\beta h_3^\perp-ih_D^\star](\beta,k_\perp),
\nonumber\\
&&\hspace{-1mm}
\!\int\! {dx_\star dx_\perp\over 16\pi^3}~e^{-i\beta_qx_\star+i(k,x)_\perp}
\langle B|\bsi\notB(0)\gamma_i\psi(x)|B\rangle
=~mk_i[\bare-i\beta\barh_3^\perp-i\barh_D](\beta,k_\perp),
\nonumber\\
&&\hspace{-1mm}
\!\int\! {dx_\star dx_\perp\over 16\pi^3}~e^{-i\beta x_\star+i(k,x)_\perp}
\langle B|\bsi(0)\gamma_i\slB\psi(x))|B\rangle
=~mk_i[e+i\beta \barh_3^\perp+i\barh_D^\star](\beta,k_\perp).
\label{protbegammai}
\end{eqnarray}

Finally, we need
\begin{eqnarray}
&&\hspace{-2mm}
\int\! {dx_\bu dx_\perp\over 8\pi^3s}~e^{-i\alpha x_\bu+i(k,x)_\perp}
\langle A|\bsi\slA(x)\slashed{p}_2\slA\psi(0)|A\rangle~
=~\big[k_\perp^2(f_1-2\alpha f_\perp)+2\alpha^2m^2f_3\big](\alpha,k_\perp)
\nonumber\\
&&\hspace{-2mm}
\int\! {dx_\bu dx_\perp\over 8\pi^3s}~e^{-i\alpha x_\bu+i(k,x)_\perp}
\langle A|\bsi\slA(0)\slashed{p}_2\slA\psi(x)|A\rangle
~=~-\big[k_\perp^2(\barf_1-2\alpha\barf_\perp)+2\alpha^2m^2\barf_3\big](\alpha,k_\perp)
\nonumber\\\
&&\hspace{-2mm}
\int\! {dx_\bu dx_\perp\over 8\pi^3s}~e^{-i\alpha x_\bu+i(k,x)_\perp}
\langle A|\bsi\slA(x)\sigma_{\star i}\slA\psi(0)|A\rangle
=-k_i\big[{k_\perp^2\over m}h_1^\perp+2\alpha mh-2\alpha^2mh_3^\perp\big](\alpha,k_\perp)
\nonumber\\
&&\hspace{-2mm}
\int\! {dx_\bu dx_\perp\over 8\pi^3s}~e^{-i\alpha x_\bu+i(k,x)_\perp}
\langle A|\bsi\slA(x)\sigma_{\star i}\slA\psi(0)|A\rangle
={k_i\over m}\big[k_\perp^2 \barh_1^\perp+2\alpha m^2\barh-2\alpha^2m^2\barh_3^\perp\big](\alpha,k_\perp)
\nonumber\\
\label{projaa}
\end{eqnarray}
and for the target
\begin{eqnarray}
&&\hspace{-3mm}
\int\! {dx_\star dx_\perp\over 8\pi^3s}~e^{-i\beta x_\star+i(k,x)_\perp}
\langle B|\bsi\slB(x)\slashed{p}_2\slB\psi(0)|B\rangle~
=~\big[k_\perp^2(f_1-2\beta f_\perp)+2\beta^2m^2f_3\big](\beta,k_\perp)
\nonumber\\
&&\hspace{-3mm}
\int\! {dx_\star dx_\perp\over 8\pi^3s}~e^{-i\beta x_\star+i(k,x)_\perp}
\langle B|\bsi\slB(0)\slashed{p}_2\slB\psi(x)|B\rangle
~=-\big[k_\perp^2(\barf_1-2\beta\barf_\perp)+2\beta^2m^2\barf_3\big](\beta,k_\perp)
\nonumber\\\
&&\hspace{-3mm}
\int\! {dx_\star dx_\perp\over 8\pi^3s}~e^{-i\beta x_\star+i(k,x)_\perp}
\langle B|\bsi\slB(x)\sigma_{\bu i}\slB\psi(0)|B\rangle
=-k_i\big[{k_\perp^2\over m}h_1^\perp+2\beta mh-2\beta^2mh_3^\perp\big](\beta,k_\perp)
\nonumber\\
&&\hspace{-3mm}
\int\! {dx_\star dx_\perp\over 8\pi^3s}~e^{-i\beta x_\star+i(k,x)_\perp}
\langle B|\bsi\slB(x)\sigma_{\bu i}\slB\psi(0)|B\rangle
={k_i\over m}\big[k_\perp^2 \barh_1^\perp+2\beta m^2\barh-2\beta^2m^2\barh_3^\perp\big](\beta,k_\perp)
\nonumber\\
\end{eqnarray}
%

\subsubsection{Parametrization of other quark-quark-gluon TMDs \label{sec:nonreduce}}
First, let us parametrize matrix elements from Sect. \ref{sec:qqpc}. 
\begin{eqnarray}
&&\hspace{-1mm}
{2\over 8\pi^3s}\!\int\!dx_\bu d^2x_\perp~e^{-i\alpha x_\bu+i(k,x)_\perp}
~\langle A|\bsi i\stackrel{\leftarrow}D_\bu(x_\bu,x_\perp)\slp_2\psi(0)|A\rangle~=~-m^2f_D(\alpha,k_\perp),
\label{Dmaelp}\\
&&\hspace{-1mm}
{2\over 8\pi^3s}\!\int\!dx_\bu d^2x_\perp~e^{-i\alpha x_\bu+i(k,x)_\perp}
~\langle A|\bsi(x_\bu,x_\perp)\slp_2iD_\bu\psi(0)|A\rangle~=~m^2f^\star_D(\alpha,k_\perp),
\nonumber\\
&&\hspace{-1mm}
{2\over 8\pi^3s}\!\int\!dx_\bu d^2x_\perp~e^{-i\alpha x_\bu+i(k,x)_\perp}
~\langle A|\bsi i\stackrel{\leftarrow}D_\bu(0)\slp_2\psi(x_\bu,x_\perp)|A\rangle~=~m^2\barf_D(\alpha,k_\perp),
\nonumber\\
&&\hspace{-1mm}
{2\over 8\pi^3s}\!\int\!dx_\bu d^2x_\perp~e^{-i\alpha x_\bu+i(k,x)_\perp}
~\langle A|\bsi(0)\slp_2iD_\bu\psi(x_\bu,x_\perp)|A\rangle~=~-m^2\barf^\star_D(\alpha,k_\perp),
\nonumber\\
&&\hspace{-1mm}
{2\over 8\pi^3s}\!\int\!dx_\star d^2x_\perp~e^{-i\alpha x_\bu+i(k,x)_\perp}
~\langle A|\bsi i\stackrel{\leftarrow}D_\bu(x_\bu,x_\perp)\sigma_{\star i}\psi(0)|A\rangle~=~-mk_ih_D(\alpha,k_\perp),
\nonumber\\
&&\hspace{-1mm}
{2\over 8\pi^3s}\!\int\!dx_\star d^2x_\perp~e^{-i\alpha x_\bu+i(k,x)_\perp}
~\langle A|\bsi(x_\bu,x_\perp)\sigma_{\star i}iD_\bu\psi(0)|A\rangle~=~mk_ih^\star_D(\alpha,k_\perp)
\nonumber\\
&&\hspace{-1mm}
{2\over 8\pi^3s}\!\int\!dx_\star d^2x_\perp~e^{-i\alpha x_\bu+i(k,x)_\perp}
~\langle A|\bsi i\stackrel{\leftarrow}D_\bu(0)\sigma_{\star i}\psi(x_\bu,x_\perp)|A\rangle~=~mk_i\barh_D(\alpha,k_\perp),
\nonumber\\
&&\hspace{-1mm}
{2\over 8\pi^3s}\!\int\!dx_\star d^2x_\perp~e^{-i\alpha x_\bu+i(k,x)_\perp}
~\langle A|\bsi(0)\sigma_{\star i}iD_\bu\psi(x_\bu,x_\perp)|A\rangle~=~-mk_i\barh^\star_D(\alpha,k_\perp)
\nonumber
\end{eqnarray}
and
\begin{eqnarray}
&&\hspace{-5mm}
{2\over 8\pi^3s}\!\int\!dx_\star d^2x_\perp~e^{-i\beta x_\star+i(k,x)_\perp}
~\langle B|\bsi i\stackrel{\leftarrow}D_\star(x_\star,x_\perp)\slp_1\psi(0)|B\rangle~=~-m^2f_D(\beta,k_\perp),
\label{Dmaelt}\\
&&\hspace{-5mm}
{2\over 8\pi^3s}\!\int\!dx_\star d^2x_\perp~e^{-i\beta x_\star+i(k,x)_\perp}
~\langle B|\bsi(x_\star,x_\perp)\slp_1iD_\star\psi(0)|B\rangle~=~m^2f^\ast_D(\beta,k_\perp),
\nonumber\\
&&\hspace{-5mm}
{2\over 8\pi^3s}\!\int\!dx_\star d^2x_\perp~e^{-i\beta x_\star+i(k,x)_\perp}
~\langle B|\bsi i\stackrel{\leftarrow}D_\star(0)\slp_1\psi(x_\star,x_\perp)|B\rangle~=~m^2\barf_D(\beta,k_\perp),
\nonumber\\
&&\hspace{-5mm}
{2\over 8\pi^3s}\!\int\!dx_\star d^2x_\perp~e^{-i\beta x_\star+i(k,x)_\perp}
~\langle B|\bsi(0)\slp_1iD_\star\psi(x_\star,x_\perp)|B\rangle~=~-m^2\barf^\ast_D(\beta,k_\perp),
\nonumber\\
&&\hspace{-5mm}
{2\over 8\pi^3s}\!\int\!dx_\star d^2x_\perp~e^{-i\beta x_\star+i(k,x)_\perp}
~\langle B|\bsi i\stackrel{\leftarrow}D_\star(x_\star,x_\perp)\sigma_{\bu i}\psi(0)|A\rangle~=~-mk_ih_D(\beta,k_\perp),
\nonumber\\
&&\hspace{-5mm}
{2\over 8\pi^3s}\!\int\!dx_\star d^2x_\perp~e^{-i\beta x_\bu+i(k,x)_\perp}
~\langle B|\bsi(x_\star,x_\perp)\sigma_{\bu i}iD_\star\psi(0)|B\rangle~=~mk_ih^\ast_D(\alpha,k_\perp)
\nonumber\\
&&\hspace{-5mm}
{2\over 8\pi^3s}\!\int\!dx_\star d^2x_\perp~e^{-i\beta x_\star+i(k,x)_\perp}
~\langle B|\bsi i\stackrel{\leftarrow}D_\star(0)\sigma_{\bu i}\psi(x_\star,x_\perp)|B\rangle~=~mk_i\barh_D(\beta,k_\perp),
\nonumber\\
&&\hspace{-5mm}
{2\over 8\pi^3s}\!\int\!dx_\star d^2x_\perp~e^{-i\beta x_\bu+i(k,x)_\perp}
~\langle B|\bsi(0)\sigma_{\bu i}iD_\star\psi(x_\star,x_\perp)|B\rangle~=~-mk_i\barh^\ast_D(\alpha,k_\perp)
\nonumber
\end{eqnarray}

Next, in Sect. \ref{1qqG} and \ref{sec:primed} we calculate target $\barq Gq$  matrix elements and restore the corresponding contributions
with projectile $\barq Gq$ ones by trivial replacements (\ref{protareplace}). Consequently, we will list only parametrizations of target 
$\barq Gq$  matrix elements. The projectile matrix elements can be obtained by the usual (\ref{protareplace}) replacements.
\footnote{One shold be careful with $\tilA_i\leftrightarrow \tilB_i$ replacement due to Eq. (\ref{deftildas}), see e.g. Eq. (\ref{tarmaels5}).  }

We parametrize the quark-quark-gluon TMDs with matrices $1$ or $\gamma_5$ as follows
\begin{eqnarray}
&&\hspace{-1mm}
{1\over 16\pi^3}\!\int\!dx_\star d^2x_\perp~e^{-i\beta x_\star+i(k,x)_\perp}
~\langle B|\bsi(x_\star,x_\perp)B_i(x_\star,x_\perp)\psi(0)|B\rangle~=~k_ime_G(\beta,k_\perp)
\nonumber\\
&&\hspace{-1mm}
{1\over 16\pi^3}\!\int\!dx_\star d^2x_\perp~e^{-i\beta x_\star+i(k,x)_\perp}
~\langle B|\bsi(x_\star,x_\perp)B_i(0)\psi(0)|B\rangle~=~k_ime_G^\ast(\beta,k_\perp)
\nonumber\\
&&\hspace{-1mm}
{1\over 16\pi^3}\!\int\!dx_\star d^2x_\perp~e^{-i\beta x_\star+i(k,x)_\perp}
~\langle B|\bsi(0)B_i(0)\psi(x_\star,x_\perp)|B\rangle~=~k_i\bare_Gm(\beta,k_\perp)
\label{bes}\\
&&\hspace{-1mm}
{1\over 16\pi^3}\!\int\!dx_\star d^2x_\perp~e^{-i\beta x_\star+i(k,x)_\perp}
~\langle B|\bsi(0)B_i(x)\psi(x_\star,x_\perp)|B\rangle~=~k_i\bare_G^\ast m(\beta,k_\perp)
\nonumber
\end{eqnarray}
\begin{eqnarray}
&&\hspace{-1mm}
{1\over 16\pi^3}\!\int\!dx_\star d^2x_\perp~e^{-i\beta x_\star+i(k,x)_\perp}
~\langle B|\bsi(x_\star,x_\perp)i\tilB_i(x_\star,x_\perp)\gamma_5\psi(0)|B\rangle~=~k_im\tile_G(\beta,k_\perp)
\nonumber\\
&&\hspace{-1mm}
{1\over 16\pi^3}\!\int\!dx_\star d^2x_\perp~e^{-i\beta x_\star+i(k,x)_\perp}
~\langle B|\bsi(x_\star,x_\perp)i\tilB_i(0)\gamma_5\psi(0)|B\rangle~=~k_im\tile_G^\ast(\beta,k_\perp)
\nonumber\\
&&\hspace{-1mm}
{1\over 16\pi^3}\!\int\!dx_\star d^2x_\perp~e^{-i\beta x_\star+i(k,x)_\perp}
~\langle B|\bsi(0)i\tilB_i(0)\gamma_5\psi(x_\star,x_\perp)|B\rangle~=~k_i\bar\tile_Gm(\beta,k_\perp)
\label{besg5}\\
&&\hspace{-1mm}
{1\over 16\pi^3}\!\int\!dx_\star d^2x_\perp~e^{-i\beta x_\star+i(k,x)_\perp}
~\langle B|\bsi(0)i\tilB_i(x)\gamma_5\psi(x_\star,x_\perp)|B\rangle~=~k_i\bar\tile_G^\ast m(\beta,k_\perp)
\nonumber
\end{eqnarray}
and accordingly%
\begin{eqnarray}
&&\hspace{-1mm}
{1\over 16\pi^3}\!\int\!dx_\star d^2x_\perp~e^{-i\beta x_\star+i(k,x)_\perp}
~\langle B|\bsi(x_\star,x_\perp)\acB_i(x_\star,x_\perp)\psi(0)|B\rangle~=~k_im\ace_G(\beta,k_\perp)
\nonumber\\
&&\hspace{-1mm}
{1\over 16\pi^3}\!\int\!dx_\star d^2x_\perp~e^{-i\beta x_\star+i(k,x)_\perp}
~\langle B|\bsi(x_\star,x_\perp)\acB_i(0)\psi(0)|B\rangle
~=~k_im\ace_G^\ast(\beta,k_\perp)
\nonumber\\
&&\hspace{-1mm}
{1\over 16\pi^3}\!\int\!dx_\star d^2x_\perp~e^{-i\beta x_\star+i(k,x)_\perp}
~\langle B|\bsi(0)\acB_i(0)\psi(x_\star,x_\perp)|B\rangle~=~k_im\bar\ace_G(\beta,k_\perp)
\nonumber\\
&&\hspace{-1mm}
{1\over 16\pi^3}\!\int\!dx_\star d^2x_\perp~e^{-i\beta x_\star+i(k,x)_\perp}
~\langle B|\bsi(0)\acB_i(x)\psi(x_\star,x_\perp)|B\rangle~=~k_im\bar\ace^\ast_B(\beta,k_\perp)
\label{bres}
\end{eqnarray}
Next, we turn to quark-quark-gluon TMDs with matrices $\sigma_{\mu\nu}$. First, 
consider the case of $\sigma_{\bu\ast}$. We get
\begin{eqnarray}
&&\hspace{-1mm}
\int\!{dx_\star d^2x_\perp\over 8\pi^3s}~e^{-i\beta x_\star+i(k,x)_\perp}
~\langle B|\bsi B_i(x)\sigma_{\bu\ast}\psi(0)|B\rangle~=~k_im[\beta h_3^\perp-h+h_D-i\tile_G](\beta,k_\perp)
\nonumber\\
&&\hspace{-1mm}
\int\!{dx_\star d^2x_\perp\over 8\pi^3s}~e^{-i\beta x_\star+i(k,x)_\perp}
~\langle B|\bsi(x)\sigma_{\bu\ast}B_i\psi(0)|B\rangle~=~k_im[\beta h_3^\perp-h+h_D^\ast+i\tile_G^\ast](\beta,k_\perp)
\nonumber\\
&&\hspace{-1mm}
\int\!{dx_\star d^2x_\perp\over 8\pi^3s}~e^{-i\beta x_\star+i(k,x)_\perp}
~\langle B|\bsi(0)\sigma_{\bu\ast}\psi B_i(x)|B\rangle~=~k_im[\beta \barh_3^\perp-\barh-\barh_D^\ast+i\bar\tile_G^\ast](\beta,k_\perp)
\nonumber\\
&&\hspace{-1mm}
\int\!{dx_\star d^2x_\perp\over 8\pi^3s}~e^{-i\beta x_\star+i(k,x)_\perp}
~\langle B|\bsi B_i(0)\sigma_{\bu\ast}\psi(x)|B\rangle~=~k_im[\beta \barh_3^\perp-\barh-\barh_D-i\bar\tile_G](\beta,k_\perp)
\nonumber\\
\label{hbes}
\end{eqnarray}
Let us illustrate the derivation of these equations.
From equations of motion (\ref{eqmt}) and   Eq. (\ref{sigmasigma}) we see that
\bega
&&\hspace{-1mm}
{1\over 8\pi^3s}\!\int\! dx_\star dx_\perp~e^{-i\beta_qx_\star+i(k,x)_\perp}
\langle B|\bsi \slB(x)\gamma_i\sigma_{\star\bu}\psi(0)|B\rangle
~
=~{1\over 16\pi^3}\!\int\! dx_\star dx_\perp~e^{-i\beta_qx_\star+i(k,x)_\perp}
\nonumber\\
&&\hspace{-1mm}
\times~\langle B|\bsi(x)[\beta\sigma_{\star i}-{2\over s}k_i\sigma_{\star\bu}-\epsilon_{ij}k^j\gamma_5
+i\stackrel{\leftarrow}D_\ast\sigma_{\bu i}]\psi(0)|B\rangle
~=~-mk_i[\beta h_3^\perp-h+h_D](\beta,k_\perp)
\nonumber\\
\ega
On the other hand
\begin{eqnarray}
&&\hspace{-1mm}
{2\over s}\langle B|\bsi \slB(x)\gamma_i\sigma_{\star\bu}\psi(0)|B\rangle
~=~{2\over s}\langle B|\bsi B^j(x)(\delta_{ij}+i\sigma_{ij})\sigma_{\star\bu}\psi(0)|B\rangle
\\
&&\hspace{11mm}
~=~{2\over s}\langle B|\bsi B_i(x)\sigma_{\star\bu}\psi(0)|B\rangle+\langle B|\bsi \tilB_i(x)\gamma_5\psi(0)|B\rangle
\nonumber\\
&&\hspace{-1mm}
\Rightarrow~
{1\over 8\pi^3s}\!\int\! dx_\star dx_\perp~e^{-i\beta_qx_\star+i(k,x)_\perp}
\langle B|\bsi B_i(x)\sigma_{\star\bu}\psi(0)|B\rangle
\nonumber\\
&&\hspace{11mm}
=~{1\over 8\pi^3s}\!\int\! dx_\star dx_\perp~e^{-i\beta_qx_\star+i(k,x)_\perp}
\langle B|\bsi \slB(x)\gamma_i\sigma_{\star\bu}\psi(0)|B\rangle-i\tile_G(\beta,k_\perp)
\nonumber
\end{eqnarray}
where we used  parametrization (\ref{bes}). From the above two equations
 we easily get the first of Eqs (\ref{hbes}).
In a similar way one can obtain the rest of formulas (\ref{hbes}).

Second, for transverse $\sigma$'s we get
\begin{eqnarray}
&&\hspace{-1mm}
\int\! {dx_\star dx_\perp\over 16\pi^3}~e^{-i\beta x_\star+i(k,x)_\perp}
\langle B|\bsi B^{\mu}(x)\sigma_{\nu_\perp j}\psi(0)|B\rangle
\label{besigmat}\\
&&\hspace{11mm}
=~m(\delta_{\nu_\perp}^{\mu_\perp} k_j-\delta^\mu_j k^\perp_\nu) \big[-ie(\beta,k_\perp)-ie_G(\beta,k_\perp)
+\beta h_3^\perp(\beta,k_\perp)-h_D(\beta,k_\perp)\big],
\nonumber\\
&&\hspace{-1mm}
\int\! {dx_\star dx_\perp\over 16\pi^3}~e^{-i\beta x_\star+i(k,x)_\perp}
\langle B|\bsi(x) B^{\mu}(0)\sigma_{\nu_\perp j}\psi(0)|B\rangle
\nonumber\\
&&\hspace{11mm}
=~m(\delta_{\nu_\perp}^{\mu_\perp} k_j-\delta^\mu_j k^\perp_\nu)
\big[ie(\beta,k_\perp)+ie_G^\ast(\beta,k_\perp)+\beta h_3^\perp(\beta,k_\perp)-h_D^\ast(\beta,k_\perp)\big],
\nonumber\\
&&\hspace{-1mm}
\int\! {dx_\star dx_\perp\over 16\pi^3}~e^{-i\beta x_\star+i(k,x)_\perp}
\langle B|\bsi B^{\mu}(0)\sigma_{\nu_\perp j}\psi(x)|B\rangle
\nonumber\\
&&\hspace{11mm}
=~m(\delta_{\nu_\perp}^{\mu_\perp} k_j-\delta^\mu_j k^\perp_\nu) \big[i\bare(\beta,k_\perp)
-i\bare_G(\beta,k_\perp+\beta \barh_3^\perp(\beta,k_\perp)+\barh_D(\beta,k_\perp)\big],
\nonumber\\
&&\hspace{-1mm}
\int\! {dx_\star dx_\perp\over 16\pi^3}~e^{-i\beta x_\star+i(k,x)_\perp}
\langle B|\bsi (0)\sigma_{\nu_\perp j}B^{\mu}\psi(x)|B\rangle
\nonumber\\
&&\hspace{11mm}
=~m(\delta_{\nu_\perp}^{\mu_\perp} k_j-\delta^\mu_j k^\perp_\nu) \big[-i\bare(\beta,k_\perp)
+i\bare_G^\ast(\beta,k_\perp)+\beta \barh_3^\perp(\beta,k_\perp)+\barh_D^\ast(\beta,k_\perp)\big]
\nonumber
\end{eqnarray}

Again, let us illustrate the derivation of the first of the above equation. After convoluting $\mu$ and $\nu$, we need to prove that
\begin{eqnarray}
&&\hspace{-1mm}
{1\over 16\pi^3}\!\int\! dx_\star dx_\perp~e^{-i\beta_qx_\star+i(k,x)_\perp}
\langle B|\bsi B^{\nu}(x)\sigma_{\nu_\perp j}\psi(0)|B\rangle
\nonumber\\
&&\hspace{-1mm}
=~{1\over 16\pi^3}\!\int\! dx_\star dx_\perp~e^{-i\beta_qx_\star+i(k,x)_\perp}
\langle B|\bsi [i\slB_\perp(x)\gamma_j-iB_j(x)\psi(0)|B\rangle
\nonumber\\
&&\hspace{11mm}
=~-imk_j\big[e(\beta_q,k_\perp)+e_G(\beta_q,k_\perp)+i\beta_q h_3^\perp(\beta,k_\perp)\big],
\nonumber\\
&&\hspace{11mm}
=~mk_j\big[-ie-ie_G+\beta_q h_3^\perp-h_D\big](\beta,k_\perp)
\label{10.88}
\ega
which easily follows from equations of motion (\ref{eqmt}) and parametrizations (\ref{bes}).The rest of the equations (\ref{besigmat}) is proved  in a  similar way.

For $\sigma_{\mu\nu}$ with one longitudinal and one transverse indices we define
\begin{eqnarray}
&&\hspace{-1mm}
{1\over 16\pi^3}{2\over s}\!\int\!dx_\star d^2x_\perp~e^{-i\beta x_\star+i(k,x)_\perp}
~\langle B|\bsi(x_\star,x_\perp)\big[B_i(x)\sigma_{\bu j}-\half g_{ij}B^k\sigma_{\bu k}(x)\big]\psi(0)|B\rangle~
\nonumber\\
&&\hspace{33mm}
=~-(k_ik_j+\half g_{ij}k_\perp^2){1\over m}h_{1G}^{\perp f}(\beta,k_\perp)
\nonumber\\
&&\hspace{-1mm}
{1\over 16\pi^3}{2\over s}\!\int\!dx_\star d^2x_\perp~e^{-i\beta x_\star+i(k,x)_\perp}
~\langle B|\bsi(0)[B_i(0)\sigma_{\bu j}-\half g_{ij}B^k\sigma_{\bu k}(0)]\psi(x_\star,x_\perp)|B\rangle~
\nonumber\\
&&\hspace{33mm}
=~-(k_ik_j+\half g_{ij}k_\perp^2){1\over m}\barh_{1G}^{\perp f}(\beta,k_\perp)
\label{maelsbe}
\end{eqnarray}

Next, we parametrize
\begin{eqnarray}
&&\hspace{-1mm}
{1\over 16\pi^3}\!\int\!dx_\star d^2x_\perp~e^{-i\beta x_\star+i(k,x)_\perp}
~\langle B|\bsi(x_\star,x_\perp)
\gamma_j\graB_i(x_\star,x_\perp)\psi(0)|B\rangle
\label{bresgi}\\
&&\hspace{12mm}
=~\big[k_ik_j+k_\perp^2{g_{ij}\over 2}\big]\graf_{1G}(\beta,k_\perp)
+{g_{ij}\over 2}[k_\perp^2 (f_\perp-ig_\perp)-2\beta m^2f_3](\beta,k_\perp),
\nonumber\\
&&\hspace{-1mm}
{1\over 16\pi^3}\!\int\!dx_\star d^2x_\perp~e^{-i\beta x_\star+i(k,x)_\perp}
~\langle B|\bsi(x_\star,x_\perp)
\graB_i(0)\gamma_j\psi(0)|B\rangle
\nonumber\\
&&\hspace{12mm}
=~\big[k_ik_j+k_\perp^2{g_{ij}\over 2}\big]\graf_{1G}^\ast(\beta,k_\perp)
+{g_{ij}\over 2}[k_\perp^2 (f_\perp+ig_\perp)-2\beta m^2f_3t](\beta,k_\perp),
\nonumber\\
&&\hspace{-1mm}
{1\over 16\pi^3}\!\int\!dx_\star d^2x_\perp~e^{-i\beta x_\star+i(k,x)_\perp}
~\langle B|\bsi(0)\gamma_j\graB_i(0)\psi(x_\star,x_\perp)|B\rangle
\nonumber\\
&&\hspace{12mm}
=~\big[k_ik_j
+k_\perp^2{g_{ij}\over 2}\big]\bar\graf_{1G} 
(\beta,k_\perp)+{g_{ij}\over 2}[k_\perp^2(\barf_\perp+i\barg_\perp)-2\beta m^2\barf_3](\beta,k_\perp),
\nonumber\\
&&\hspace{-1mm}
{1\over 16\pi^3}\!\int\!dx_\star d^2x_\perp~e^{-i\beta x_\star+i(k,x)_\perp}
~\langle B|\bsi(0)
\graB_i(x_\star,x_\perp)\gamma_j\psi(x_\star,x_\perp)|B\rangle
\nonumber\\
&&\hspace{12mm}
=~\big[k_ik_j
+k_\perp^2{g_{ij}\over 2}\big]\bar\graf_{1G}^\ast 
(\beta,k_\perp)+{g_{ij}\over 2}[k_\perp^2(\barf_\perp-i\barg_\perp)-2m^2\barf_3](\beta,k_\perp)
\nonumber
\end{eqnarray}
Let us prove the first of the above equations.
Consider
\begin{eqnarray}
&&\hspace{-1mm}
{1\over 16\pi^3}\!\int\!dx_\star d^2x_\perp~e^{-i\beta x_\star+i(k,x)_\perp}{2\over s}\langle B|\bsi(x)\slB(x)\slp_2\slp_1\psi(0)|B\rangle
\nonumber\\
&&\hspace{11mm}
=~{k_i\over 16\pi^3}\!\int\!dx_\star d^2x_\perp~e^{-i\beta x_\star+i(k,x)_\perp}\langle B|\bsi(x)\big[-\gamma_i+i\epsilon_{ij}\gamma^j\gamma_5
+2i\stackrel{\leftarrow}D_\ast \slp_1x\big]\psi(0)|B\rangle
\nonumber\\
&&\hspace{22mm}
=~k_\perp^2 [f_\perp(\beta,k_\perp)+ig_\perp(\beta,k_\perp)]-2m^2f_D(\beta,k_\perp)
\label{formul1}
\end{eqnarray}
where we again used QCD equations (\ref{eqmt})

On the other hand,
\begin{eqnarray}
&&\hspace{-1mm}
{2\over s}\langle B|\bsi(x)\slB(x)\slp_2\slp_1\psi(0)|B\rangle~=~\bsi(x)\gamma^i(B_i+i\tilB_i\gamma_5)(x)\psi(0)\rangle~=~
\bsi(x)\graB^i(x)\gamma_i\psi(0)\rangle~\Rightarrow~
\nonumber\\
&&\hspace{-1mm}
{1\over 16\pi^3}\!\int\!dx_\star d^2x_\perp~e^{-i\beta x_\star+i(k,x)_\perp}
~\langle B|\bsi\graB^i(x)\gamma_i\psi(0)|B\rangle~=~[k_\perp^2 (f_\perp+ig_\perp)-2m^2f_D](\beta,k_\perp)
\nonumber\\
\end{eqnarray}
Next, $\gamma_i\graB^i=2\slB-\graB^i\gamma_i$ so from the equation of motion (\ref{eqmt}) 
\beq
{1\over 16\pi^3}\!\int\!dx_\star d^2x_\perp~e^{-i\beta x_\star+i(k,x)_\perp}
~\langle B|\bsi(x)\slB(x)\psi(0)|B\rangle
~=~[k_\perp^2 f_\perp-\beta m^2f_3-m^2f_D](\beta,k_\perp)
\eeq
and we easily get
\begin{equation}
\int\!{dx_\star d^2x_\perp\over 16\pi^3}~e^{-i\beta x_\star+i(k,x)_\perp}
~\langle B|\bsi(x)\gamma_i\graB^i(x)\psi(0)|B\rangle
~=~[k_\perp^2 (f_\perp-ig_\perp)-2\beta m^2f_3](\beta,k_\perp)
\label{chekbe1}
\end{equation}
The rest of the convolutions in Eqs. (\ref{bresgi}) are obtained in a similar way.

Finally, we parametrize TMDs with integrated gluon fields as in Eq. (\ref{pizbez}) as follows
\begin{eqnarray}
&&\hspace{-1mm}
{1\over 16\pi^3}{2\over s}\!\int\!dx_\star d^2x_\perp~e^{-i\beta x_\star+i(k,x)_\perp}
~\langle B|\bsi(x)\grave\pizb_i(x_\star,x_\perp)\slp_1\psi(0)|B\rangle~=~k_i\graf_{1\pizg}(\beta,k_\perp)
\nonumber\\
&&\hspace{-1mm}
{1\over 16\pi^3}{2\over s}\!\int\!dx_\star d^2x_\perp~e^{-i\beta x_\star+i(k,x)_\perp}
~\langle B|\bsi(0)\grave\pizb_i(0)\slp_1\psi(x_\star,x_\perp)|B\rangle~=~-k_i\bar\graf_{1\pizg}(\beta,k_\perp)
\nonumber\\
&&\hspace{-1mm}
{1\over 16\pi^3}{2\over s}\!\int\!dx_\star d^2x_\perp~e^{-i\beta x_\star+i(k,x)_\perp}
~\langle B|\bsi(x)\slp_1\grave\pizb_i(0)\psi(0)|B\rangle~=~-k_i\graf_{1\pizg}^\ast(\beta,k_\perp)
\label{frabs1}\\
&&\hspace{-1mm}
{1\over 16\pi^3}{2\over s}\!\int\!dx_\star d^2x_\perp~e^{-i\beta x_\star+i(k,x)_\perp}
~\langle B|\bsi(0)\slp_1\grave\pizb_i(x)\psi(x_\star,x_\perp)|B\rangle~=~k_i\bar\graf_{1\pizg}^\ast(\beta,k_\perp)
\nonumber
\end{eqnarray}
\begin{eqnarray}
&&\hspace{-1mm}
\int\!{dx_\star d^2x_\perp\over 8\pi^3s}~e^{-i\beta x_\star+i(k,x)_\perp}
~\langle B|\bsi(x)\slp_1\pizp\pizb(x_\star,x_\perp)\psi(0)|B\rangle~=~m^2 f_{2\pizg}(\beta,k_\perp)
\nonumber\\
&&\hspace{-1mm}
\int\!{dx_\star d^2x_\perp\over 8\pi^3s}~e^{-i\beta x_\star+i(k,x)_\perp}
~\langle B|\bsi(0)\slp_1\pizp\pizb(0)\psi(x_\star,x_\perp)|B\rangle~=~m^2\barf_{2\pizg}(\beta,k_\perp)
\nonumber\\
&&\hspace{-1mm}
\int\!{dx_\star d^2x_\perp\over 8\pi^3s}~e^{-i\beta x_\star+i(k,x)_\perp}
~\langle B|\bsi(x)(\pizp\pizb)^\ast(0)\slp_1\psi(0)|B\rangle~=~m^2f_{2\pizg}^\ast(\beta,k_\perp)
\label{frabs2}\\
&&\hspace{-1mm}
\int\!{dx_\star d^2x_\perp\over 8\pi^3s}~e^{-i\beta x_\star+i(k,x)_\perp}
~\langle B|\bsi(0)(\pizp\pizb)^\ast(x)\slp_1\psi(x_\star,x_\perp)|B\rangle~=~m^2\barf_{2\pizg}^\ast(\beta,k_\perp)
\nonumber
\end{eqnarray}
\begin{eqnarray}
&&\hspace{-1mm}
\int\!{dx_\star d^2x_\perp\over 8\pi^3s}~e^{-i\beta x_\star+i(k,x)_\perp}
~\langle B|\bsi(x)\slp_1\gamma_5{\epsilon^{ij}\over 2}\pizb_{ij}(x_\star,x_\perp)\psi(0)|B\rangle~=~m^2f_{3\pizg}(\beta,k_\perp)
\nonumber\\
&&\hspace{-1mm}. 
\int\!{dx_\star d^2x_\perp\over 8\pi^3s}~e^{-i\beta x_\star+i(k,x)_\perp}
~\langle B|\bsi(0)\slp_1\gamma_5{\epsilon^{ij}\over 2}\pizb_{ij}(0)\psi(x_\star,x_\perp)|B\rangle~=~m^2\barf_{3\pizg}(\beta,k_\perp)
\nonumber\\
&&\hspace{-1mm}
\int\!{dx_\star d^2x_\perp\over 8\pi^3s}~e^{-i\beta x_\star+i(k,x)_\perp}
~\langle B|\bsi(x)\slp_1\gamma_5{\epsilon^{ij}\over 2}\pizb_{ij}(0)\psi(0)|B\rangle~=~m^2f_{3\pizg}^\ast(\beta,k_\perp)
\label{frabs3}\\
&&\hspace{-1mm}
\int\!{dx_\star d^2x_\perp\over 8\pi^3s}~e^{-i\beta x_\star+i(k,x)_\perp}
~\langle B|\bsi(0)\slp_1\gamma_5{\epsilon^{ij}\over 2}\pizb_{ij}(x_\star,x_\perp)\psi(x_\star,x_\perp)|B\rangle~=~m^2\barf_{3\pizg}^\ast(\beta,k_\perp)
\nonumber
\end{eqnarray}
%

\begin{eqnarray}
&&\hspace{-1mm}
\int\!{dx_\star d^2x_\perp\over 8\pi^3s}~e^{-i\beta x_\star+i(k,x)_\perp}
~\langle B|\bsi(x)\
\sigma_{\bu\beta_\perp} \pizb_\alpha(x)
\psi(0)|B\rangle~
\nonumber\\
&&\hspace{22mm}
=~k^\perp_\alpha k^\perp_\beta {1\over m} h_{1\pizg}(\beta,k_\perp)
+{k_\perp^2\over m}{g_{\alpha\beta}^\perp\over 2}[h_{1\pizg}+h_{2\pizg}](\beta,k_\perp),
\nonumber\\
&&\hspace{-1mm}
\int\!{dx_\star d^2x_\perp\over 8\pi^3s}~e^{-i\beta x_\star+i(k,x)_\perp}
~\langle B|\bsi(0)\sigma_{\bu\beta_\perp}\pizb_\alpha(0)
\psi(x)|B\rangle~
\nonumber\\
&&\hspace{22mm}
=~-k^\perp_\alpha k^\perp_\beta {1\over m} \barh_{1\pizg}(\beta,k_\perp)
-{k_\perp^2\over m}{g_{\alpha\beta}^\perp\over 2}[\barh_{1\pizg}+\barh_{2\pizg}](\beta,k_\perp)
\label{hrabs1}
\nonumber\\
&&\hspace{-1mm}
\int\!{dx_\star d^2x_\perp\over 8\pi^3s}~e^{-i\beta x_\star+i(k,x)_\perp}
~\langle B|\bsi(x)\
\sigma_{\bu\beta_\perp} \pizb_\alpha(0)
\psi(0)|B\rangle~
\nonumber\\
&&\hspace{22mm}
=~k^\perp_\alpha k^\perp_\beta {1\over m} h_{1\pizg}^\ast(\beta,k_\perp)
+{k_\perp^2\over m}{g_{\alpha\beta}^\perp\over 2}[h_{1\pizg}^\ast+h_{2\pizg}^\ast](\beta,k_\perp),
\nonumber\\
&&\hspace{-1mm}
\int\!{dx_\star d^2x_\perp\over 8\pi^3s}~e^{-i\beta x_\star+i(k,x)_\perp}
~\langle B|\bsi(0)\sigma_{\bu\beta_\perp}\pizb_\alpha(x)
\psi(x)|B\rangle~
\nonumber\\
&&\hspace{22mm}
=~-k^\perp_\alpha k^\perp_\beta {1\over m} \barh_{1\pizg}^\ast(\beta,k_\perp)
-{k_\perp^2\over m}{g_{\alpha\beta}^\perp\over 2}[\barh_{1\pizg}^\ast+\barh_{2\pizg}^\ast](\beta,k_\perp)
\end{eqnarray}
and
\begin{eqnarray}
&&\hspace{-1mm}
\int\!{dx_\star d^2x_\perp\over 8\pi^3s}~e^{-i\beta x_\star+i(k,x)_\perp}
~\langle B|\bsi(x)\sigma_{\bu i}\pizp\pizb(x_\star,x_\perp)\psi(0)|B\rangle~=~k_imh_{3\pizg}(\beta,k_\perp)
\nonumber\\
&&\hspace{-1mm}. 
\int\!{dx_\star d^2x_\perp\over 8\pi^3s}~e^{-i\beta x_\star+i(k,x)_\perp}
~\langle B|\bsi(0)\sigma_{\bu i}\pizp\pizb(0)\psi(x_\star,x_\perp)|B\rangle~=~k_im\barh_{3\pizg}(\beta,k_\perp)
\nonumber\\
&&\hspace{-1mm}
\int\!{dx_\star d^2x_\perp\over 8\pi^3s}~e^{-i\beta x_\star+i(k,x)_\perp}
~\langle B|\bsi(x)\sigma_{\bu i}\pizb\pizp(0)\psi(0)|B\rangle~=~k_imh_{3\pizg}^\ast(\beta,k_\perp)
\label{hrabs2}\\
&&\hspace{-1mm}
\int\!{dx_\star d^2x_\perp\over 8\pi^3s}~e^{-i\beta x_\star+i(k,x)_\perp}
~\langle B|\bsi(0)\sigma_{\bu i}\pizb\pizp(x_\star,x_\perp)\psi(x_\star,x_\perp)|B\rangle~=~k_im\barh_{3\pizg}^\ast(\beta,k_\perp)
\nonumber
\end{eqnarray}
\begin{eqnarray}
&&\hspace{-1mm}
\int\!{dx_\star d^2x_\perp\over 8\pi^3s}~e^{-i\beta x_\star+i(k,x)_\perp}
~\langle B|\bsi(x)\sigma_{\bu i}\pizb_{ij}(x_\star,x_\perp)\psi(0)|B\rangle~=~k_jmh_{4\pizg}(\beta,k_\perp)
\nonumber\\
&&\hspace{-1mm}. 
\int\!{dx_\star d^2x_\perp\over 8\pi^3s}~e^{-i\beta x_\star+i(k,x)_\perp}
~\langle B|\bsi(0)\sigma_{\bu i}\pizb_{ij}(0)\psi(x_\star,x_\perp)|B\rangle~=~k_jm\barh_{4\pizg}(\beta,k_\perp)
\nonumber\\
&&\hspace{-1mm}
\int\!{dx_\star d^2x_\perp\over 8\pi^3s}~e^{-i\beta x_\star+i(k,x)_\perp}
~\langle B|\bsi(x)\sigma_{\bu i}\pizb_{ij}(0)\psi(0)|B\rangle~=~k_jmh_{4\pizg}^\ast\beta,k_\perp)
\label{hrabs3}\\
&&\hspace{-1mm}
\int\!{dx_\star d^2x_\perp\over 8\pi^3s}~e^{-i\beta x_\star+i(k,x)_\perp}
~\langle B|\bsi(0)\sigma_{\bu i}\pizb_{ij}(x_\star,x_\perp)\psi(x_\star,x_\perp)|B\rangle~=~k_jm\barh_{4\pizg}^\ast(\beta,k_\perp)
\nonumber
\end{eqnarray}
As usual, the corresponding matrix elements for the projectile are obtained by trivial replacements
$x_\star\leftrightarrow x_\bu$, $\alpha_q\leftrightarrow\beta_q$ and
$\slashed{p}_2\leftrightarrow\slashed{p}_1$.

\bibliography{dysi}
\bibliographystyle{JHEP}

\end{document}